\RequirePackage{ifpdf}
\documentclass[hyper,letterpaper]{JHEP3}
\pdfoutput=1
\usepackage{amssymb,amsmath,amsfonts}
\usepackage{mathtools}
\usepackage{graphicx}
\usepackage{cite}
\usepackage{multirow}
\usepackage{verbatim}
\usepackage{appendix}
\usepackage{fancybox}
\usepackage{array}
\usepackage{url}
\usepackage{float}
\usepackage[mathscr]{euscript}
\usepackage{bbm} 

\newcommand{\bea}{\begin{eqnarray}}
\newcommand{\eea}{\end{eqnarray}}
\newcommand{\be}{\begin{equation}}
\newcommand{\ee}{\end{equation}}

\newcommand{\beq}{\begin{equation}\begin{aligned}}
\newcommand{\eeq}{\end{aligned}\end{equation}}

\newcommand{\Z}{{\mathbb Z}}
\newcommand{\R}{{\mathbb R}}
\newcommand{\C}{{\mathbb C}}

\newcommand{\Q}{{\mathbb Q}}

\newcommand{\cp}{{\mathbb{C}}{\mathbf{P}}}

\def\frak{\mathfrak}

\def\tilde{\widetilde}
\renewcommand{\bar}{\overline}
\renewcommand{\hat}{\widehat}
\def\^{{\wedge}}
\def\*{{\star}}

\def\CA{{\mathcal A}}
\def\CB{{\mathcal B}}
\def\CC{{\mathcal C}}

\def\CF{{\mathcal F}}
\def\CG{{\mathcal G}}
\def\CH{{\mathcal H}}
\def\CI{{\mathcal I}}

\def\CL{{\mathcal L}}
\def\CM{{\mathcal M}}
\def\CN{{\mathcal N}}
\def\CO{{\mathcal O}}
\def\CP{{\mathcal P}}

\def\CR{{\mathcal R}}

\def\CT{{\mathcal T}}

\def\CW{{\mathcal W}}

\def\g{\gamma}

\def\la{\lambda}

\def\eg{{\textit{e.g.}}}
\def\ie{{\textit{i.e.}}}

\newcommand{\fq}{{\frak q}}

\newcommand{\Tr}{{\rm Tr}}

\def\q{{\mathbbm{q}}}
\def\lk{{\ell k}}
\def\Coker{{\mathrm{Coker}\,}}
\def\Tor{{\mathrm{Tor}\,}}
\def\top{{\text{top}}}

\newtheorem{conj}{Conjecture}[section]

\title{BPS spectra and 3-manifold invariants}

\author{Sergei Gukov$^{1}$, Du Pei$^{1,2}$, Pavel Putrov$^{3}$, Cumrun Vafa$^{4}$
\\
$^1$ Walter Burke Institute for Theoretical Physics, California Institute of Technology, Pasadena, CA 91125, USA \\
$^2$ Center for Quantum Geometry of Moduli Spaces, Deparment of Mathematics, University
of Aarhus, DK-8000, Denmark\\
$^3$ School of Natural Sciences, Institute for Advanced Study, Princeton, NJ 08540, USA \\
$^4$ Jefferson Physical Laboratory, Harvard University, Cambridge, MA 02138, USA}

\abstract{We provide a physical definition of new homological invariants $\CH_a (M_3)$ of 3-manifolds (possibly, with knots)
labeled by abelian flat connections.
The physical system in question involves a 6d fivebrane theory on $M_3$ times a 2-disk, $D^2$, whose Hilbert space of BPS states
plays the role of a basic building block in categorification of various partition functions of 3d $\CN=2$ theory $T[M_3]$:
$D^2\times S^1$ half-index, $S^2\times S^1$ superconformal index, and $S^2\times S^1$ topologically twisted index.
The first partition function is labeled by a choice of boundary condition and provides a refinement of Chern-Simons (WRT) invariant. A linear combination of them in the unrefined limit gives
the analytically continued WRT invariant of $M_3$. The last two can be factorized into the product of half-indices.
We show how this works explicitly for many examples, including Lens spaces, circle fibrations over Riemann surfaces, and plumbed 3-manifolds.

\vspace{2cm}

\vspace{2cm}

\texttt{CALT-TH-2016-039}
}

\begin{document}

\section{Introduction}

Can we hear the shape of a drum?
Much like harmonics of a musical instrument, spectra of quantum systems contain wealth of useful information.
Of particular interest are supersymmetric or the so-called BPS states which, depending on the problem at hand,
can manifest themselves either as minimal surfaces, or solutions to partial differential equations, or other ``extremal'' objects.
Thus, a spectrum of BPS states in Calabi-Yau compactifications can be used to reconstruct the geometry
of the Calabi-Yau space itself and, as we explain in this paper, spectra of BPS states play a similar role in low-dimensional topology.

The old approach to constructing numerical invariants of 3- and 4-manifolds, as well as the corresponding homological invariants of 3-manifolds,
is based on gauge theory.
The famous examples are Donaldson-Witten and Seiberg-Witten (SW) invariants of 4- and 3-manifolds,
and corresponding instanton and monopole \cite{kronheimer2007monopoles} Floer homologies of 3-manifolds.
All of them were extensively studied in mathematical literature and have appropriate rigorous definitions that go back to the previous century.
The numerical invariants are realized in terms of counting solutions to certain partial differential equations,
while the homological invariants build on the ideas of Andreas Floer \cite{MR956166}.
In particular, Seiberg-Witten invariants of 4-manifolds had great success distinguishing some
homeomorphic but non-diffeomorphic 4-manifolds.
And, in the world of 3-manifolds, the so-called Heegaard Floer homology constructed
by Ozsvath and Szabo \cite{ozsvath2004holomorphic} gives a much simpler and non-gauge theoretic definition
of a homological invariant, which is believed to be equivalent to the monopole Floer homology.

A seemingly different class of 3-manifold invariants, the so-called Witten-Reshetikhin-Turaev (WRT) invariants \cite{Witten:1988hf,MR1091619},
comes from a different type of TQFT, which sometimes is called ``of Schwarz type'' to distinguish it from
the TQFTs ``of cohomological type'' mentioned in the previous paragraph \cite{Birmingham:1991ty}.
At the turn of the century, however, the distinction between the two types started to blur and
the ideas of the present paper suggest it may even go away completely in the future.
In fact, the first hints for this go back to
the early work \cite{Rozansky:1992td,Chang:1991df,Rozansky:1992zt,meng1996sw}
that relates Seiberg-Witten theory in three dimensions to Chern-Simons theory with $U(1|1)$ super gauge group.
The latter provides a much simpler invariant compared to the usual Chern-Simons theory, say, with $SU(2)$ gauge group,
due to cancelations between bosonic and fermionic contributions.
Therefore, if one can find a 4d TQFT that categorifies $SU(2)$ Chern-Simons theory in 3d,
similar to how 4d SW theory categorifies 3d SW theory, it would help a great deal with the classification problem
of smooth 4-manifolds.
The first step in constructing such categorification is, of course,
to find a homological invariant of 3-manifolds whose (equivariant) Euler characteristic gives the WRT invariant.

The existence of such 4d TQFT was envisioned by Crane and Frenkel \cite{MR1295461} more than 20 years ago,
and the first evidence came with the advent of knot homology \cite{khovanov1999categorification,khovanov2008matrix,MR2901970}
which categorifies WRT invariants of knots and links (realized by Wilson lines in CS theory) in $S^3$, also known as the colored Jones polynomial.
The physical understanding of a homological approach to HOMFLY polynomials was independently initiated in the physics literature in \cite{Ooguri:1999bv}, which later led to the physical interpretation of Khovanov-Rozansky homology
as certain BPS Hilbert spaces \cite{Gukov:2004hz,Gukov:2007ck,Witten:2011zz,Aganagic:2011sg}
(see {\it e.g.} \cite{Chun:2015gda,Nawata:2015wya} for an overview and an extensive list of references).

The physical construction suggests that there should be a homological invariant that categorifies (in a certain sense)
the WRT invariant of general 3-manifolds with knots inside.
Namely, such homological invariant can be understood as the BPS sector of the Hilbert space of
the 6d $\CN=(2,0)$ theory (that is a theory describing dynamics of coincident M5-branes in M-theory)
on $M_3\times D^2\times \R$ with a certain supersymmetry preserving background along $M_3\times D^2$.
Equivalently, if one first reduces the 6d theory on $M_3$,
it can be understood as the BPS Hilbert space of the effective 3d $\CN=2$ theory $T[M_3]$ on $D^2\times \R$.
On the other hand, if one first compactifies on $D^2$, one does not get an ordinary 4d gauge theory on $M_3\times \R$ like 4d SW gauge theory\footnote{Roughly speaking, the effective 4d theory is an infinite KK-like tower on 4d gauge theories. However one needs to appropriately sum it up. The decategorified counterpart of such summation was studied in \cite{Gukov:2016njj}.}:
\beq
\begin{array}{ccccc}
\; & \;\text{6d $(2,0)$ theory} &\text{on}  &\text{$\R\times D^2\times M_3$}  \; & \; \\
\; & \swarrow \qquad\qquad\qquad& \; &\qquad\qquad\qquad \searrow & \; \\
\text{space of BPS states $\CH_{\text{BPS}}$  } & \; & \; & \; & \text{Hilbert space $\CH_{M_3}$ of } \\
\text{of 3d $\CN=2$ theory $T[M_3]$} & \; & \; & \; & \text{4d ``Crane-Frenkel TQFT''}\\
\text{on $D^2$} & \; & \; & \; & \text{on $M_3$}
\end{array}
\label{CatGeom0}
\eeq

There is another natural SUSY-preserving background on which one can quantize $T[M_3]$.
Since the IR physics of $T[M_3]$ is governed by a non-trivial 3d $\CN=2$ SCFT,
one can consider its radial quantization and study its Hilbert space on $S^2\times \R$.
This should provide us with another non-trivial homological invariant of $M_3$ which should
have roughly the same level of complexity as the BPS Hilbert space on $D^2\times \R$ which categorifies the WRT invariant,
but with several advantages due to the presence of operator-state correspondence and no need to specify a boundary condition at $\partial D^2=S^1$.

The set of boundary conditions that one can put at $\partial D^2\cong S^1$ in the path integral can be understood as follows. Let us represent $D^2$ as an elongated cigar, which asymptoically looks like $S^1\times \R$. After compactification of the stack of fivebranes on $S^1$ we obtain 5d maximally supersymmetric gauge theory. The supersymmetric vacua of such theory on $M_3\times \R$ (where $\R$ is the original time direction) are given by flat connections\footnote{The choice of such flat connections should not be confused with the choice of flat connections in CS theory on $M_3$. As will be explained later in detail they are related by S-transform.} on $M_3$. The number of such supersymetric vacua is the same as the number of boundary conditions (\textit{cf.} \cite{Cecotti:1991me,Witten:2011zz,Pasquetti:2011fj,Beem:2012mb,Cecotti:2013mba}). As we will see later, the subset of such boundary conditions that corresponds to abelian flat connections plays an important role; for convenience, here we summarize various clues that all point to a special role of abelian flat connections:

\begin{itemize}

\item
charges of BPS states in the target space theory \cite{Ooguri:1999bv};

\item
relation between homological invariants of different rank \cite{Gukov:2016gkn};

\item
resurgent analysis of complex Chern-Simons theory \cite{Gukov:2016njj};

\item
similar analysis of 3d $\CN=2$ theories (section \ref{sec:summary} below);

\item
Langlands duality for flat connections on 3-manifolds (section \ref{sec:Langlands}).

\end{itemize}
\noindent
Moreover, in this paper we give many examples of $q$-series invariants for various 3-manifolds
which exhibit integrality, and in many cases we also write the corresponding homological invariants.
All of these invariants are labeled by (connected components of) abelian flat connections;
we do not have any such example associated with a non-abelian flat connection.

In this paper we study the relation between such homological invariants and their decategorified counterparts --- superconformal indices. The structure described above, of course, will also manifest itself at the level of partition functions, {\it i.e.} indices, once we compactify the time on $S^1$. For some of the examples in this paper for there are no cancellations among states in computing the refined index. In those cases, the homological invariants are faithfully captured by the refined index computation.

The organization of the paper is as follows. In Section 2 we review some results of \cite{Gukov:2016gkn} and summarize general conjectures about homological invariants of closed 3-manifolds mentioned here and their decategorified versions. In Section 3 we consider various examples for which we explicitly verify these conjectures. In Section 4 we extend it to the case of 3-manifolds with knots.
Note, the main part of Section 4 is written in jargonish shorthand, using the concepts and notations introduced earlier.
Various details and generalizations of this work can be found in the appendices.
Thus, Appendix~\ref{app:Zq-from-WRT} explains how the WRT invariants of general negative definite plumbed 3-manifolds
can be analytically continued away from roots of unity to produce power series in $q$ with integer powers and integer coefficients,
required for categorification.
In Appendix~\ref{app:cabling}, we compare the ordinary Khovanov homology of the $n$-th cabling of the unknot in a 3-sphere
to the refined partition function of 3d $\CN=2$ theory $T[S^3]$ in the presence of line operators, {\it cf.} Figure~\ref{fig:impurity}.
Finally, in Appendix~\ref{sec:TV} we explain how categorification of the index of $T[M_3]$
relates to categorification of the Turaev-Viro invariants.

\begin{figure}[ht]
\centering
\includegraphics[trim={0 0.5in 0 0.5in},clip,width=4.0in]{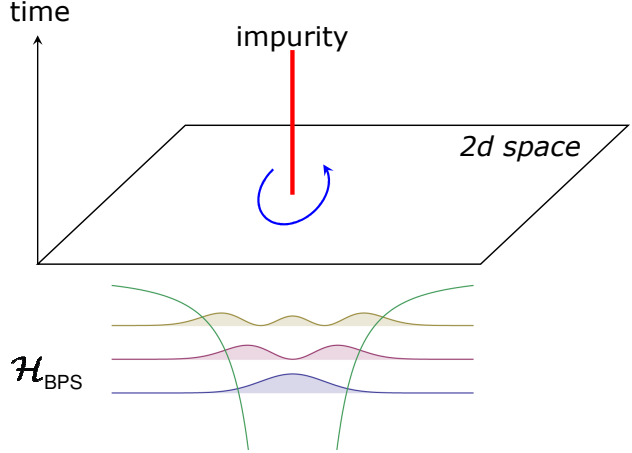}
\caption{The space of BPS states in 3d $\CN=2$ theory on $\R \times D^2$ with an impurity,
relevant to the physical realization of Heegaard Floer homology $HF(M_3)$,
monopole Floer homology $HM(M_3)$, as well as categorification of WRT invariants of 3-manifolds with knots.}
\label{fig:impurity}
\end{figure}

\section{Fivebranes on 3-manifolds and categorification of WRT invariant}
\label{sec:summary}

The goal of this section is to introduce the key players and their interrelation.
To keep the discussion simple and concrete, we choose the gauge group to be $G=SU(2)$ for most of it,
and then in section \ref{sec:higherrk} briefly comment how everything can be generalized to higher ranks.

\subsection{Preliminaries}
\label{sec:preliminaries}

Before we present a mathematically-friendly summary of our proposal and the physics behind it, we need to introduce
some notations, especially those relevant to abelian flat connections that will be central in our discussion.

Consider a closed and connected 3-manifold $M_3$, with $\partial M_3=\emptyset$. In order to present the results in full generality, it will be useful to consider the linking pairing on the torsion part of $H_1(M_3,\Z)$:
\begin{equation}
	\begin{array}{cccc}
		\lk: & \Tor H_1(M_3)\otimes \Tor H_1(M_3) & \longrightarrow & \Q/\Z \\
		& [a]\otimes [b] & \longmapsto & {\#(a\cap b')}/{n} \\
	\end{array}
\end{equation}
where $b'$ is a 2-chain such that $\partial b'=nb$ for an integer $n$. Such $b'$ and $n$ exist because $[b]$ is torsion. As usual, $\#(a\cap b')$ denotes the number of intersection points counted with signs determined by the orientation.  Note that the linking form provides an isomorphism between $\Tor H_1(M_3)$ and its Pontryagin dual $(\Tor H_1(M_3))^*\equiv\mathrm{Hom}(\Tor H_1(M_3),U(1))$ via the pairing $e^{2\pi i\lk(\cdot,\cdot)}$.

The $\Z_2$ Weyl group acts on the elements $a\in\Tor H_1(M_3,\Z)$ via $a\mapsto -a$. The set of orbits is the set of connected components of abelian flat $SU(2)$ connections on $M_3$ (\ie, connections in the image of $\rho: \CM_\text{flat}\left(M_3,U(1)\right)\rightarrow \CM_\text{flat}\left(M_3,SU(2)\right)$ from the embedding $U(1)\subset SU(2)$),\footnote{It is in fact $\left(\Tor H_1(M_3,\Z)\right)^*/\Z_2$ that is canonically identified with components of abelian flat connections. However, as the distinction between $\Tor H_1(M_3,\Z)$ and its dual is only important in section~\ref{sec:summaryConj}, we will use the same set of labels $\{a, b, \ldots\}$ for elements in both groups. }
\begin{equation}
	\Tor H_1(M_3,\Z)/\Z_2 \; \cong \; \pi_0\CM_\text{flat}^\text{ab}(M_3,SU(2)) \,.
\end{equation}
 It is also useful to introduce a shorthand notation for the stabilizer subgroup:
\begin{equation}
	\CW_a \; \equiv \; \text{Stab}_{\Z_2}(a) \; = \; \left\{
	\begin{array}{cl}
		\Z_2, & a=-a \,, \\
		1, & \text{otherwise\,.}
	\end{array}
	\right.
\end{equation}

\subsection{${D^2\times S^1}$ partition function of ${T[M_3]}$ and WRT invariant}
\label{sec:summaryConj}

Now we are ready to present a slightly generalized and improved version of the results from \cite[sec.~6]{Gukov:2016gkn}.

\subsubsection*{Categorification of WRT invaraiant}

Let $Z_{SU(2)_k}[M_3]$ be the partition function of $SU(2)$ Chern-Simons theory
with ``bare'' level $(k-2)$ on $M_3$, also known as the WRT invariant.
We use the standard ``physics'' normalization where
\be
Z_{SU(2)_k}\left[S^2\times S^1\right]=1,
\ee
and
\be
Z_{SU(2)_k}\left[S^3\right]=\sqrt{\frac2k}\sin\left(\frac\pi k\right).
\ee
The following conjecture was proposed in \cite{Gukov:2016gkn}.\footnote{A related conjecture was made in \cite{hikami2011decomposition}. However it did not include the $S$-transform, which is crucial for restoring integrality and categorification.}

\begin{conj}
The WRT invariant can be decomposed into the following form:
\begin{equation}
	Z_{SU(2)_k}[M_3]=(i\sqrt{2k})^{b_1(M_3)-1}\sum_{a,b\;\in \;\atop\Tor H_1(M_3,\Z)/\Z_2}e^{2\pi ik\lk(a,a)}\,S_{ab}
	\, \hat{Z}_b(q)|_{q\rightarrow e^{\frac{2\pi i }{k}}}
	\label{WRT-decomposition}
\end{equation}
with
\footnote{The constant positive integer $c$ depends only on $M_3$ and in a certain sense measures its ``complexity''. In many simple examples $c=0$, and the reader is welcome to ignore $2^{-c}$ factor which arises from some technical subtleties. Its physical origin will be explained later in the paper.}\footnote{Later in the text we will sometimes use slightly redefined quantities, $\hat{Z}_a(q)\rightarrow q^{\Delta}\hat{Z}_a(q)$, where $\Delta$ is a common, $a$ independent rational number.}
\begin{equation}
	\hat{Z}_b(q) \in \, 2^{-c} q^{\Delta_b} \Z[[q]]\qquad \Delta_b\in \Q,\qquad c\in\Z_+
	\label{Block-qSeries}
\end{equation}
convergent in $|q|<1$ and
\begin{equation}
	S_{ab}=\frac{e^{4\pi i\lk(a,b)}+e^{-4\pi i\lk(a,b)}}{|\CW_a|\sqrt{|\Tor H_1(M_3,\Z)|}}.
\end{equation}
\label{Conjecture1}
\end{conj}
In other words, we claim the existence of new 3-manifold invariants $\hat{Z}_a$, which admit $q$-series
expansion with integer powers and integer coefficients (hence, more suitable for categorification)
and from which the WRT invariant can be reconstructed via \eqref{WRT-decomposition}.
While the formal mathematical definition of the invariants $\hat{Z}_a$ is waiting to be discovered,
they admit a physics definition that will be reviewed below and
can be independently computed via techniques of resurgent analysis.
In particular, each term
\begin{equation}
	\sum_{b\in\Tor H_1(M_3,\Z)}\left.e^{2\pi ik\lk(a,a)}\,S_{ab}
		\, \hat{Z}_b(q)\right|_{q\rightarrow e^{\frac{2\pi i }{k}}}
		\equiv
			e^{2\pi ik\lk(a,a)}\,Z_a(q)
\label{WRT-ab-contrib}
\end{equation}
in the sum (\ref{WRT-decomposition}) is a certain resummation of the perturbative (in $\frac{2\pi i}k$ or, equivalently, in $(1-q)$) expansions of the WRT invariant around the corresponding abelian flat connection $a$ \cite{Gukov:2016njj}.

In order to avoid unnecessary technical complications, in the rest of this paper
we assume that $\Tor H_1(M_3,\Z)$ has no $\Z_2$ factors.\footnote{Recall that $\Tor H_1(M_3,\Z)$, as a finitely generated abelian group, can be decomposed into $\Tor H_1(M_3,\Z)= \prod_{i}\Z_{p_i}$. We ask for a fairly weak condition that $\Z_2$ doesn't appear in this decomposition. In other words, $M_3$ is a $\Z_2$-homology sphere. Equivalently, there is a unique Spin structure on $M_3$, so that there is no ambiguity in specifying Nahm-pole boundary condition for $\CN=4$ $SU(2)$ SYM on $M_3\times\R_+$ \cite{Witten:2011zz}. The general case, in principle, could also be worked out. We leave it as an exercise to an interested reader.} Under this assumption, the $S$-matrix satisfies
\begin{equation}
	\sum_{b}S_{ab}S_{bc}=\delta_{ac}.
\label{S-squared}
\end{equation}

Moreover, as will be discussed in detail below, physics predicts the existence of a $\Z \times \Z \times \Tor H_1(M_3,\Z)/\Z_2$ graded homological invariant of $M_3$:
\begin{equation}
	\CH_{D^2}[M_3] \; =\bigoplus_{a\in \Tor H_1(M_3,\Z)/\Z_2}\CH_a[M_3],\qquad \CH_a[M_3] \; =\bigoplus_{i\in \Z+\Delta_a,\atop j\in\Z}\CH^{i,j}_a
\end{equation}
 which categorifies the $q$-series $\hat{Z}_a(q)$. Namely\footnote{ The presence of $1/2$ factors that produce $2^{-c}$ overall factor in (\ref{Block-qSeries}) can be interpreted as presence of factors $\cong \C[x]$ with $\text{deg}_qx=0$ in $\CH_a[M_3]$. The $q$-graded Euler charecteristic of $\C[x]$ is naively divergent: $1+1+1+\ldots$, but its zeta-regularization gives $1/2$. },
\begin{equation}
	\hat{Z}_a(q)=\sum_{i\in \Z+\Delta_a,\atop j\in\Z}q^{i}(-1)^j\dim\CH^{i,j}_a.
	\label{BlockEuler}
\end{equation}
Because of their close relation to homological invariants, we usually refer to $\hat{Z}_a(q)$ as \textit{homological blocks}. The vector space $\CH_{D^2}[M_3]$ can be interpreted as the closed 3-manifold analog of Khovanov-Rozansky knot homology. From this point of view, $e^{4\pi i\lk(a,\cdot)}$ can be understood as the variable associated with $\Tor H_1(M_3,\Z)/\Z_2$ grading and enters into the decomposition \eqref{WRT-decomposition} much like $q$, the variable for one of the $\Z$-gradings (the ``$q$-grading'').
Note that the label $a$ in $\CH^{i,j}_a$ is reminiscent of the Spin$^c$ structure in Heegaard/monopole Floer homologies of 3-manifolds.
This fact, of course, is not an accident and plays an important role in the relation between Heegaard/monopole Floer homologies of 3-manifolds
and the categorification of WRT invariants \cite{Gukov:2016gkn}.

Next, we describe the physics behind the Conjecture~\ref{Conjecture1}.
(A mathematically inclined reader may skip directly to Conjecture~\ref{Conjecture2}.)

\begin{figure}[ht]
\centering
\includegraphics[width=3.5in]{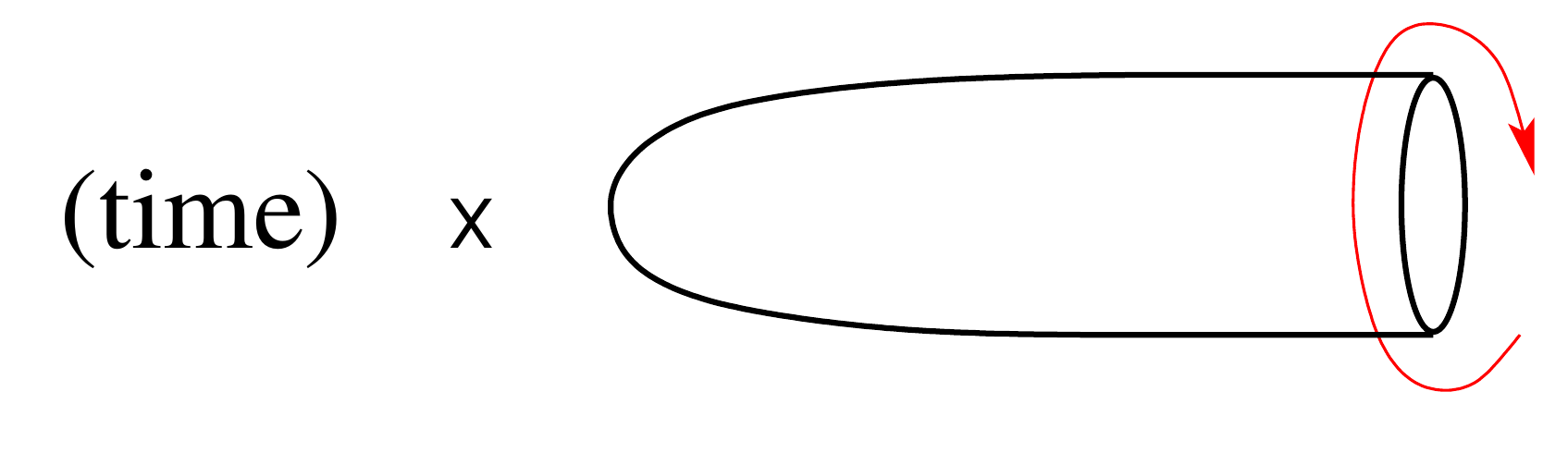}
\caption{Another representation of the background in Figure~1.}
\label{fig:cigar1}
\end{figure}

\subsubsection*{Physics behind the proposal}

From physics point of view, the homological invariants $\CH_a[M_3]$ can be realized by the following M-theory geometry,
\beq
\begin{matrix}
{\mbox {\textrm{$N$ fivebranes:}}}~~~\qquad & \R & \times & M_3& \times & D^2  \\ \qquad & &
&   \cap &  & \cap \\
{\mbox{\rm space-time:}}~~~\qquad & \R&  \times  & T^*M_3 & \times & TN   \\ \qquad &
&  &  \circlearrowright &  & \circlearrowright   \\
{\mbox{\rm symmetries:}}~~~\qquad & &  & \text{``$U(1)_N$''} &   &  U(1)_q\times U(1)_R .
\end{matrix}
\label{CatGeom1}
\eeq
or, equivalently, any of its dual descriptions (some of which will be discussed below).
Here, the first two lines summarize the geometry of the fivebranes and their ambient space,
whereas the last line describes their symmetries.
The reason ``$U(1)_N$'' appears in quotes is that it is a symmetry of our physical system only when $M_3$
is a Seifert 3-manifold, unlike the ``universal'' symmetry group $U(1)_q\times U(1)_R$.

In order to preserve supersymmetry for a general metric on $M_3$, it has to be embedded in the geometry
of ambient space-time as a supersymmetric (special Lagrangian) cycle $M_3\subset CY_3$ which,
according to McLean's theorem, always looks like $M_3\subset T^* M_3$ near the zero section.
Equivalently, the geometry $T^* M_3$ represents a partial topological twist on the fivebrane world-volume,
upon which three of the scalar fields on the world-volume become sections of the cotangent bundle of $M_3$.
As a result, one can first reduce the 6d $(2,0)$ theory --- the world-volume theory of M5-branes --- on $M_3$ to obtain a 3d $\CN=2$ SCFT usually denoted as $T[M_3;G]$, where $G=U(N)$ or $SU(N)$, $N$ being the number of M5-branes. All SUSY-protected objects like partition functions, index and BPS spectra of the resulting theory $T[M_3;G]$
do not depend\footnote{A ``folk theorem'' states that any continuous deformation of the metric on $M_3$ results in a $Q$-exact term of the supergravity background. However, there may be dependence on discrete data such as the Atiyah 2-framing \cite{atiyah1990framings}.} on the metric of $M_3$, and give rise to numerical as well as homological invariants of $M_3$.

Similarly, in order to preserve supersymmetry of the brane system \eqref{CatGeom1} along the other world-volume
directions of the fivebranes, one needs to introduce a SUSY-preserving background along $D^2$.
Moreover, it needs to be done in a way that preserves the rotation symmetry $U(1)_q\times U(1)_R$
and allows to keep track of the corresponding quantum numbers (spins) of BPS states, as required for categorification.
The suitable background can be described in a number of equivalent ways: as the Omega-background
along $TN \cong \R^4_{q,t}$ in which $D^2 \cong \R^2$ is embedded as a linear subspace,
or as a $U(1)_q\times U(1)_R$ invariant Lagrangian submanifold (the ``cigar'') $D^2$ in the Taub-NUT space $TN$
where one keeps track of the spin with respect to the rotation symmetry, {\it cf.} Figure~\ref{fig:cigar1}.
To emphasize that one keeps track of both spins under $U(1)_q\times U(1)_R$ symmetry,
the adjective {\it refined} is often added to the invariant, BPS state, or other object under consideration.

When $M_3$ is a Seifert manifold, the brane system \eqref{CatGeom1} enjoys an extra symmetry $U(1)_N$ that appears in a degeneration limit of the metric on $M_3$ and can be used to redefine the R-symmetry of the SCFT $T[M_3]$. When $M_3$ is $\Sigma \times S^1$, the symmetry $U(1)_N\times U(1)_R$ is further enhanced to the $SU(2)_N\times SU(2)_R$ R-symmetry of the 3d $\CN=4$ theory $T[\Sigma\times S^1]$; when $M_3$ is a generic Seifert manifold, one combination gives the R-symmetry of the 3d $\CN=2$ theory $T[M_3]$ which we denote as $U(1)_R$, while another is a flavor symmetry $U(1)_{\beta}$, see \cite[sec.~3.4]{Gukov:2016gkn} for details.

After reduction on $M_3$, the system \eqref{CatGeom1} gives a theory $T[M_3;G]$ in space-time $D^2\times \R$, illustrated in Figure~\ref{fig:impurity}, and we can consider its Hilbert space with a certain boundary condition at $\partial D^2=S^1$. For $N=2$ and $G=SU(2)$ --- the case that we will be mostly considering in this paper --- these boundary conditions turn out to be labeled by $a\in \Tor H_1(M_3,\Z)/\Z_2$. Then, we arrive at a set of doubly-graded homological invariants of $M_3$ labeled by $a$,
\begin{equation}
	\CH_a[M_3] \; = \; \CH_{T[M_3]}(D^2;a) \; = \; \bigoplus_{i\in \Z+\Delta_a,\atop j\in\Z}\CH^{i,j}_a,
\end{equation}
given by the BPS sector of the Hilbert space of $T[M_3;SU(2)]$. This is the subspace annihilated by two of the four supercharges of the 3d $\CN=2$ supersymmetry (different choices are related by automorphisms of the superconformal algebra, resulting in isomorphic BPS spaces). The grading $i$ counts the charge under the $U(1)_{q}$ rotation of $D^2$ and the ``homological'' grading $j$ corresponds to R-charge of the $U(1)_R$ R-symmetry. When $M_3$ is Seifert, the $U(1)_{\beta}$ symmetry will give rise to the third grading\footnote{Not to be confused with the extra ``HOMFLY grading'' {\it e.g.} on the right-hand side of \eqref{conifoldphases}.} on $\CH_a[M_3]$.

One can understand $\CH_a[M_3] = \CH_{T[M_3]}(D^2;a)$ as the massless multi-particle BPS spectrum of $T[M_3]$ with a label $a\in \Tor H_1(M_3,\Z)/\Z_2$ being a discrete charge. From M-theory point of view, the BPS particles of $T[M_3]$ arise from M2-branes ending on the pair of M5-branes that realize the $A_1$ 6d $\CN=(2,0)$ theory. The boundaries of M2-branes wrap 1-cycles $(\tilde{a},-\tilde{a})$ so that $[\tilde{a}]=a\in \Tor H_1(M_3,\Z)/\Z_2$. This is similar to the counting of BPS states in \cite{Ooguri:1999bv}. Note, however, that BPS particles that arise from M2-branes ending on a non-torsion 1-cycles of $M_3$ have mass and do not enter into the IR BPS spectrum. Therefore, it is elements in $\Tor H_1(M_3,\Z)/\Z_2$ that give rise to physical boundary conditions which specifies a superselection sector labeled by this brane charge, resulting in a physical BPS Hilbert space $\CH_a$ with integrality property. In contrast, a flat connection, given by an element of $(\Tor H_1(M_3,\Z))^*/\Z_2$ doesn't correspond to any physical boundary conditions. Instead, it is a linear combination of physical boundary conditions leading to a mixture of different charge sectors.

From this point of view, the S-transform in Conjecture \ref{Conjecture1} carries out the change of basis between \emph{charges} (valued in $\Tor H_1(M_3,\Z)/\Z_2$) and \emph{holonomies} (valued in $(\Tor H_1(M_3,\Z))^*/\Z_2$) using the natural pairing between them via the ``Aharonov-Bohm phase.'' More precisely, the M2-branes ending on M5-branes produce particles in the effective 3d theory carrying electric-magnetic charge $b\in \Tor H_1(M_3,\Z)/\Z_2$.\footnote{Note that in the case of $N$ M5-branes wrapping $M_3$, the M2-branes produce states charged under the magnetic $U(1)^N/S_N$ (not necessarily $U(N)$) symmetry, as explained in \cite{Ooguri:1999bv}.} On the other hand, $a\in(\Tor H_1(M_3,\Z))^*/\Z_2$ specified the holomony, and, by viewing $(\Tor H_1(M_3,\Z))^*$ as the group of characters of $\Tor H_1(M_3,\Z)$, we have
\be
S_{ab}\propto\sum_{ \text{$\Z_2$ orbit of $a$} }\chi_a(b)=\sum_{ \text{$\Z_2$ orbit of $a$} }e^{4i\pi \lk(a,b)}.
\ee
Geometrically, $S_{ab}$ is the trace of the holonomy of the flat connection labeled by $a$ along the 1-cycle representing homology class $b$.

Alternatively, one can understand the boundary conditions in the type IIB duality frame of the brane system \eqref{CatGeom1}, where S-transform can be interpreted
as the S-duality of type IIB string theory. Indeed, a quotient of the eleven-dimensional space-time by a circle action $U(1)_q$
lands us in type IIA string theory, which can be further T-dualized along the ``time'' circle $S^1$.
(Equivalently, these two dualities can be combined into one step, which is the standard M-theory / type IIB duality.)
The resulting system involves D3-branes ending on a D5-brane, and S-duality of type IIB string theory maps it
into a stack of D3-branes ending on an NS5-brane \cite{Witten:2011zz}.
Note, that the natural choice of boundary conditions at infinity for a system of D3-branes ending on an NS5-brane is an arbitrary (not necessarily abelian) $SL(2,\C)$ flat connection on $M_3$. Such choice of a flat connection corresponds to considering analytically continued $SU(2)$ Chern-Simons theory on the Lefschetz thimble associated to that flat connection.
However, in general, the corresponding partition function is not continuous in a disk $|q|<1$ due to Stokes phenomena.
Instead, in (\ref{WRT-ab-contrib}) we consider quantities $Z_a(q)$ which are labeled by {\it abelian} flat connections only,
and analytic inside $|q|<1$. As explained in \cite{Gukov:2016njj}, for a given value of $\text{arg}\,k$ one can express $Z_a(q)$ as a linear combination of the Feynman path integral on Lefschetz thimbles. If one were to write the S-transform in the basis corresponding to Lefschetz thimbles, instead of $Z_a(q)$, the S-matrix would be $k$-dependent.

By definition, each homological block $\hat{Z}_a(q)$ is the graded Euler characteristics (\ref{BlockEuler}) of $\CH_a$
that can be computed as an supersymmetric partition function of $T[M_3]$ on $D^2\times_q S^1$ with an $\CN=(0,2)$ supersymmetric
boundary condition $a$ and metric corresponding to rotation of the disk $D^2$ by $\mathrm{arg}\,q$ when we go around the $S^1$,
\begin{equation}
	\hat{Z}_a(q)=Z_{T[M_3]}(D^2\times_q S^1;a).
	\label{BlockD2S1}
\end{equation}
If one knows the Lagrangian description of $T[M_3]$, this partition function can be computed using localization, see \eg~\cite{Yoshida:2014ssa}.

Note, if the Lagrangian description of $T[M_3]$ contains chiral multiplets charged under the gauge symmetry, carrying zero R-charge (at the unitarity bound) and with Neumann boundary conditions, then the integral computing the $D^2\times_q S^1$ partition function (\ie~the half-index) will be singular in general. For example, for $g$ adjoint chiral multiplets the partition function has the following form:
\begin{equation}
	\frac{1}{2}\int_{|z|=1}\frac{dz}{2\pi i z}\,\frac{f(z,q)}{(1-z^2)^{g-1}(1-z^{-2})^{g-1}}\,,\qquad f(z,q)=f(z^{-1},q)\in \Z[z,z^{-1}][[q]].
\end{equation}
A natural way to regularize it is to take the principle value prescription:
\begin{multline}
	\frac{1}{2}\,\text{v.p.}\int_{|z|=1}\frac{dz}{2\pi i z}\,\frac{f(z,q)}{(1-z^2)^{g-1}(1-z^{-2})^{g-1}}\,\equiv\\
	\frac{1}{4}\left(\int_{|z|=1+\epsilon}+\int_{|z|=1-\epsilon}\right)\frac{dz}{2\pi i z}\,\frac{f(z,q)}{(1-z^2)^{g-1}(1-z^{-2})^{g-1}}\,.
\end{multline}
As we will see in many examples, this regularization prescription is in agreement with the relation between $\hat{Z}_a(q)=Z_{T[M_3]}(D^2\times_q S^1;a)$ and the WRT invariant. This is also the source of the $2^{-c}$ factor in (\ref{Block-qSeries}).\footnote{Such factor can appear even without singularity, when there is only $g=1$ adjoint chiral with R-charge 0. It cancels the contribution of the vector multiplet, leaving
\begin{equation}
	\frac{1}{2}\int \frac{dz}{2\pi i z}\,f(z,q)=\frac{1}{2}[z^0\text{ coefficient of }f(z,q)],
\end{equation}
where one can take $f(z,q)=1$, as an example.
}

\subsubsection*{Resurgence in 3d $\CN=2$ theories}

The phenomenon of non-abelian contributions being distributed among the abelian ones can be translated entirely into the language of 3d $\CN=2$ field theories. In fact, this phenomenon is not limited to theories $T[M_3]$ associated with 3-manifolds and should be regarded more broadly, as a decomposition of BPS quantities in gapless (conformal) theories into BPS spectra of its massive deformations.

Thanks to supersymmetric localization, partition functions of 3d $\CN=2$ theories on $D^2\times_q S^1$ can be reduced
to finite-dimensional integrals, which in the limit $q = e^{\hbar} \to 1$ have the form
\be
Z (D^2\times_q S^1) \; = \; \int dz \, \Delta (z) \, \exp \left( \frac{1}{\hbar} \tilde \CW (z) + \ldots \right)
\label{generalZvortex}.
\ee
The multi-variable $z = (z_1 , \ldots, z_r)$ typically comes from gauge factors, and $\tilde \CW (z)$ can be interpreted
as a twisted superpotential of the 3d theory on a circle. Its critical points are the vacua of the 2d $\CN=(2,2)$ effective
field theory, which for a finite-size circle come in infinite towers. Indeed, when the 3d theory is compactified on a circle of
finite radius, the variables $z_i$ are $\C^*$-valued and the equations for vacua also have the exponentiated form:
\be
\exp \left( \frac{\partial \tilde \CW}{\partial \log z_i} \right) \; = \; 1.
\ee
For this reason, each vacuum comes in an infinite family, much like saddle points of Chern-Simons functional.
Also, as in Chern-Simons theory, the critical points of $\tilde \CW$ have the leading order behavior
\be
\hbar^{n/2} e^{\frac{1}{\hbar} \tilde \CW (z_*)} \cdot (\text{power series in}~\hbar)
\ee
where the critical exponent $n$ is the degree of degeneracy of the critical point $z = z_*$;
the larger the value of $n$ the more singular the critical point is (see \cite{appetizer}
for a careful treatment of such factors).

\begin{figure}[ht]
\centering
\includegraphics[width=3.0in]{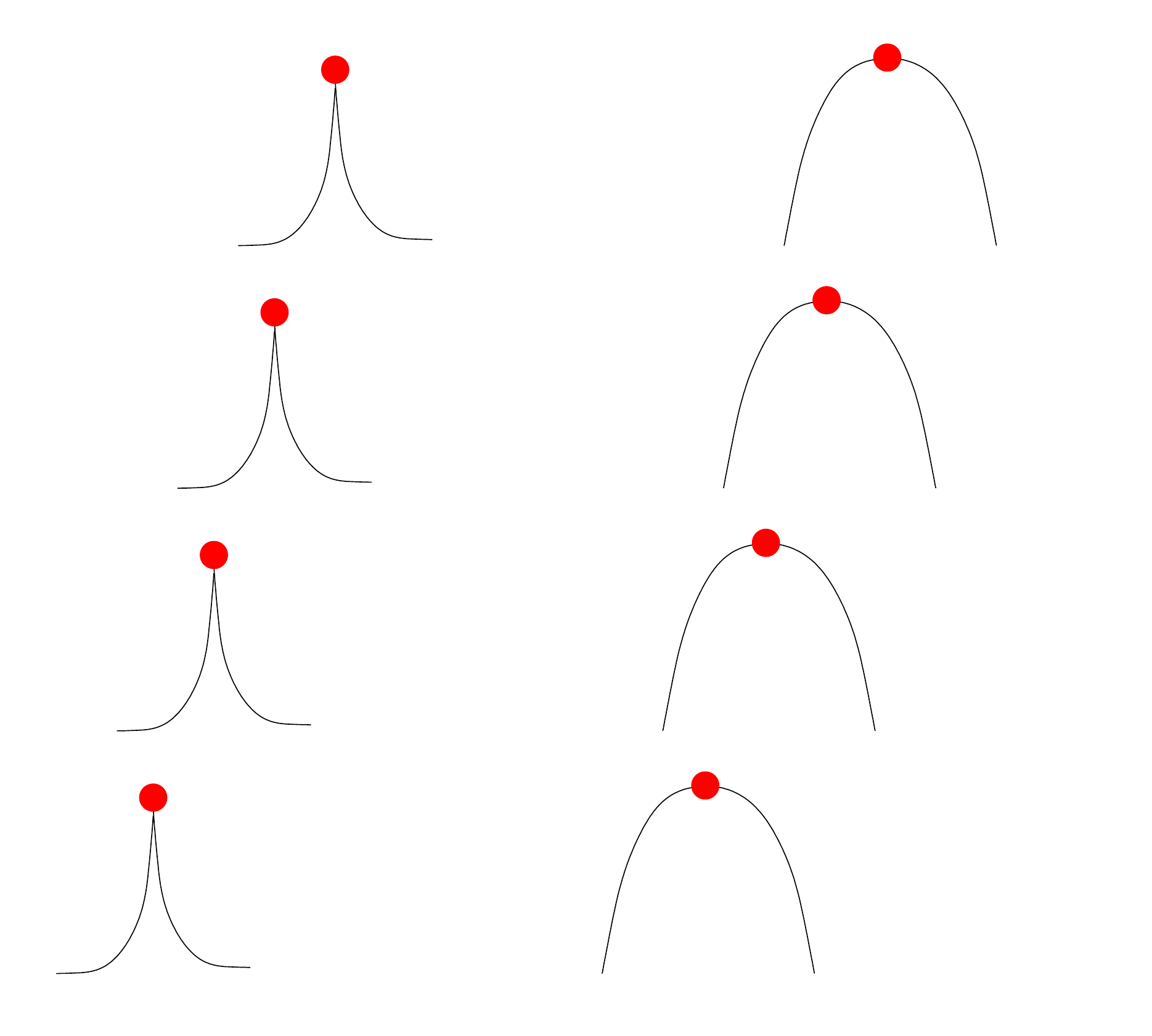}
\caption{Critical points of $\tilde \CW$ in 3d $\CN=2$ theories come in infinite towers because, as in complex
Chern-Simons theory, the space of fields is not simply-connected. Moreover, a generic theory has degenerate,
non-Morse critical points, which arise, {\it e.g.}, when the fields at the critical point have a large stabilizer.
In localization integrals, such critical points ``absorb'' the contributions of regular, Morse critical points as trans-series.}
\label{fig:critpts}
\end{figure}

Therefore, in the context of completely general 3d $\CN=2$ theories we find two key elements of the resurgent analysis \cite{Gukov:2016njj}:
the critical points come in infinite $\Z$-towers and they may have different degree of degeneracy signalled by powers of $\hbar = \log q$.
Hence, applying the resurgent analysis of \cite{Gukov:2016njj} to the integral \eqref{generalZvortex},
we conclude that contributions of critical points with smaller values of the exponent $n$ will be ``distributed''
among critical points with larger values of $n$.

In order to see this phenomenon in physically-motivated 3d $\CN=2$ theories, such as 3d SQED or SQCD,
we simply need to look at examples with non-trivial range of values of $n$.

\subsection{Realization in terms of counting solutions of differential equations}

In the previous section we defined homological invariants of 3-manifolds $\CH_a[M_3]$ as Hilbert spaces of 6d (2,0) theory on $M_3\times D^2\times \R$ where $\R$ is the time direction. It is possible to give another definition that, in a sense, is closer to the definition of instanton \cite{MR956166} or monopole (Seiberg-Witten) Floer homology of 3-manifolds \cite{kronheimer2007monopoles} familiar to many topologists. We will mostly follow \cite{Witten:2011zz}, where the detailed discussion can be found.

Let us represent $D^2$ as a cigar fibered over $\R_+$ with fibers being the orbits of the $U(1)$ action rotating the disc (see Figure~\ref{fig:cigar1}).\footnote{Exchanging the circle fiber of $D^2$ with $S^1$ in \eqref{BlockD2S1} has the effect of replacing the group $G$ with its GNO/Langlands dual $^LG$.} The tip of the cigar is a degenerate fiber projected to the origin of $\R_+$. The 6d (2,0) theory corresponding to the Lie algebra $^L\mathfrak{g}:=\text{Lie}(^LG)$ on $M_3\times D^2\times \R$ then can be effectively described as 5d maximally supersymmetric Yang-Mills theory with gauge group $^LG$ on $M_3\times \R_+\times \R$, topologically twisted along $M_3$. One can further reduce it to the 2d $\CN = (2,2)$ LG theory on  $\R_+\times \R$ with the target space being the space of $^LG_\C$ flat connections modulo gauge transformations and the superpotential being the holomorphic Chern-Simons functional $\text{CS}(\CA)$  where $\CA=(A_\mu+i\phi_\mu)dx^\mu$ is the $^LG_\C$ connection 1-form \cite{Dimofte:2010tz}. The imaginary part $\phi_\mu$ originates from three (out of five) adjoint scalar fields in 5d theory that transform as a vector under $SO(3)_\text{R}\subset SO(5)_\text{R}$ R-symmetry group and become a one-form on $M_3$ after topological twisting~\cite{Blau:1996bx,Blau:1997pp}. The boundary condition at the origin of $\R_+$ (the image of the cigar under projection) is a Nahm-pole type boundary condition that requires a certain singular behavior of $\phi_\mu$, which in local coordinates takes the following form:
\begin{equation}
	\phi_\mu = \frac{\sum_{i=1}^3 e^i_\mu t_i}{y}+\ldots,\qquad y\rightarrow 0
\end{equation}
where $e^i_\mu$ is the vierbein for some choice of metric on $M_3$, $t_i$ are the images of the standard generators of $\mathfrak{su}(2)$ under a principal embedding $\mathfrak{su}(2)\rightarrow {}^L\mathfrak{g}$,\footnote{For $^LG=SU(N)$ (or $U(N)$) the principal embedding is such that fundamental representation of $SU(N)$ restricts to an irreducible $N$-dimensional representation of $SU(2)$.} and $y$ is the coordinate along $\R_+$. The real part of the $^LG_\C$ connection obeys the Dirichlet boundary condition that fixes it at $y=0$ to be equal to the spin-connection on $M_3$ via the principal embedding $\mathfrak{su}(2)\rightarrow {}^L\mathfrak{g}$.

There is also a choice of boundary condition at the infinite end of $\R_+$ that corresponds to the choice of the boundary condition at the boundary of the disk. As in any 2d $\CN = (2,2)$ Landau-Ginzburg model, there is a natural family of boundary conditions labeled by connected components of the critical points of the superpotential \cite{Hori:2000ck}. They are given by Lagrangian branes supported on Lefschetz thimbles associated to those connected components.

The BPS Hilbert space $\CH_a[M_3]$ labeled by the connected component
\begin{equation}
a\in \pi_0\CM_\text{flat}(M_3, ^LG_\C)
\label{aHW}
\end{equation}
then can be constructed as a Floer homology as follows. As usual, classically the BPS states of the 2d $\CN = (2,2)$ Landau-Ginzburg model on $\R_+\times \R$, where $\R$ is the time direction, are given by solutions of time-independent BPS equations, that is gradient flows with respect to CS functional on $\R_+$ with specified boundary conditions. On the quantum level the BPS Hilbert space is given by the cohomology of Morse-Witten complex of such classical BPS states graded by the Morse index. The differential is given in terms of counting solutions of BPS equations interpolating between two time-independent solutions at $\pm\infty$ in the time direction. Such differential equations, when written in terms of 5d theory on $M_3\times \R\times \R_+$ are often referred as Haydys-Witten equations\cite{Haydys:2010dv,Witten:2011zz}. The Hilbert space $\CH_a[M_3]$ has two gradings. One is the usual Morse-index/R-charge/homological grading. The other grading, corresponding to the generating variable $q$ in the index $\hat{Z}_a(q)=\Tr_{\CH_a[M_3]} (-1)^R q^{L_0}$, is given by the instanton number
\begin{equation}
L_0=\frac{1}{32\pi^2}\int_{\R_+\times M_3}\Tr \,F \wedge F - \text{CS}_\text{grav}\;\;\;\in\;\;\; \Z+\text{Re}\,\text{CS}(a).
\end{equation}
Here $\text{CS}_\text{grav}$ is the action of the gravitational CS on $M_3$, which, taking into the account the boundary condition at the origin of $\R_+$, makes the instanton number well defined. Note that this, at least naively, defines $\CH_a[M_3]$ for connected components of \textit{all} flat connections.\footnote{On a closer look, though, there is a significant difference between solutions with abelian and non-abelian flat connections at $y=+\infty$. When the flat connection is irreducible, the $y$-component of the adjoint-valued Higgs field $\phi_y$ vanishes and remains zero for all values of $y$, while it is not the case when the flat connection at $y = + \infty$ is reducible.} However, in the previous sections, when discussing categorification of WRT invariant, we used only a subset of spaces $\CH_a[M_3]$ labeled by connected components of \textit{abelian} flat connections. The deep reasoning for this phenomenon is yet to be understood, but resurgence theory provides some explanation on the decategorified level at least for some class of 3-manifolds \cite{Gukov:2016njj}.

The definition of $\CH_a[M_3]$ given in this section, although quite explicit, is still on the physical level of rigor, and in order to make it mathematically accurate one still has to overcome various problems, such as singularities and compactness of the corresponding moduli spaces. When moduli spaces are not compact, one needs either to find a suitable compactification (which, in the present context, does not spoil topological invariance) or supplement the standard Morse-Floer theory with more powerful tools. Indeed, even at the decategorified level, the ``counting'' of solutions requires integration over moduli spaces and, even if all problems are solved, may not produce integers. A familiar example is the calculation of Gromov-Witten invariants via integration over the moduli spaces of stable maps, where maps with non-trivial stabilizers lead to rational rather than integer invariants.

The present problem, however, is much more challenging than compactness issues that one encounters in Gromov-Witten or Donaldson-Witten theory. One standard tool that replaces Morse theory in the non-compact setting is the Conley index. While its application to the current problem will be discussed elsewhere, here we can make a few brief comments that resonate with the rest of the present paper. The moduli spaces in question, $\CM (M_3, \CB)$, depend on the choice of a 3-manifold $M_3$ and the boundary condition $\CB$ at $y = + \infty$. A generic choice of the boundary condition, of course, will not result in integer invariants (even at the decategorified level) and only particular choices of boundary conditions will lead to moduli spaces where one might hope to address compactness issue without destroying integrality and topological invariance. Our present study suggests that boundary conditions labeled by abelian flat connections are precisely the ones which have this property.

Another issue is that counting the number of solutions of such differential equations, even if properly defined, is computationally very hard. Therefore one should explore other possibilities to mathematically define $\CH_a[M_3]$.

\subsection{Surgeries, triangulations, and Atiyah-Floer}

In order to give a mathematical formulation of the homological invariants
that can be useful for practical calculations on an arbitrary 3-manifold,
the starting point must be a construction of the 3-manifold itself.
Then, different ways to present the same 3-manifold may lead to different but
equivalent definitions of $\CH_a[M_3]$ which, in physics, correspond to different ways
of looking at the fivebrane system \eqref{CatGeom1}.

For example, in addition to its original combinatorial definition \cite{khovanov1999categorification},
the Khovanov homology of knots by now has definitions based on matrix factorizations \cite{khovanov2008matrix},
algebraic geometry \cite{MR2411561}, symplectic geometry \cite{MR2254624,MR2313538,AbouzaidSmith}, {\it etc.}
As expected, they are all equivalent \cite{Cautis,MackaayWebster} and offer different useful perspectives.
Similarly, we expect that our homological invariants $\CH_a[M_3]$ of 3-manifolds admit several mathematical definitions,
thus providing new opportunities and bridges among different areas of mathematics.

A standard way to produce a symplectic definition is based on the Atiyah-Floer conjectures, which relate gauge theory version of
Floer homology and its symplectic version (see {\it e.g.} \cite{Gukov:2007ck} for a review and an application in a closely related context).
The idea is to represent a 3-manifold $M_3$ as a Heegaard decomposition
\be
M_3 \; = \; M_3^{-} \cup_{\Sigma} M_3^+
\ee
of two handlebodies $M_3^{\pm}$ joined by a long ``neck'' $\cong \R \times \Sigma$. Then, topological reduction on $\Sigma$
gives a 2d sigma-model with target space $\CM (G, \Sigma)$ and space-time (``world-sheet'') $\R \times I$,
with the boundary conditions $\CB^{\pm}$ at the end-points of the interval $I$ determined by $M_3^{\pm}$.
Then, the Floer homology of the higher-dimensional theory can be equivalently formulated as a symplectic Floer homology
in the topological A-model with target space $\CM (G, \Sigma)$ and two Lagrangian submanifolds $\CB^+$ and $\CB^-$:
\be
\CH_a[M_3] \; = \; HF^*_{\text{symp}} (\CB^+, \CB^-).
\ee
As with other candidates for the mathematical definition of $\CH_a[M_3]$, the main challenge is to
describe precisely the moduli spaces  $\CM (G, \Sigma)$ and $\CB^{\pm}$.
For example, to obtain $\CM (G, \Sigma)$ it is convenient to first reduce 6d $(0,2)$ theory on $\Sigma$ times
the circle fiber of $D^2$ to obtain a sigma-model with target space $\CM_{\text{flat}} (G_{\C}, \Sigma) \cong \CM_H (G,\Sigma)$,
the Hitchin moduli space (in which $M_3^+$ and $M_3^-$ cut out holomorphic Lagrangian submanifolds).

Another way to build 3-manifolds is to start with a simpler 3-manifold, say 3-sphere, remove a tubular neighborhood of
some knot (or link) and then glue it back in a different way. This operation is called surgery and, in fact, every 3-manifold
can be realized as a sequence of surgery operations on a 3-sphere $S^3$. In view of this, it is natural to ask if
3-manifold homology $\CH_a[M_3]$ of the resulting manifold $M_3$ is related to homology of knots and links.

As is well known in the theory of quantum group invariants, at the decategorified level the answer is ``yes''
and here we argue that this is also true at the homological level. Indeed, a quantum group invariant (WRT invariant)
of a 3-manifold $M_3$ obtained by surgery on $K$ is completely determined by polynomial invariants of $K$,
with a proviso that one needs to know all colored invariants of $K$. In the case of $G=SU(2)$ these are colored
Jones polynomials and to obtain the WRT invariant of $M_3$ one basically need to sum over color (which is typical for any state sum type models).

Since at present all colored polynomial invariants of $K$ have rigorous mathematical definition one could try
to write a surgery formula where colored Jones polynomials of $K$ are replaced by the corresponding homologies
(or their Poincar\'e polynomials). Surprisingly, in a few examples that we checked this indeed gives
the 3-manifold homology $\CH_a[M_3]$, and it is natural to conjecture that this is true in general.
If so, one extremely concrete, mathematically well-defined and computable definition of $\CH_a[M_3]$
could be via surgery formula and sum over colored knot (link) homologies.

As a concrete example, consider a Lens space $M_3 = L(p,1)$ constructed as a surgery on the unknot.
In fact, this also gives an example of genus-$1$ Heegaard splitting since the unknot complement is a solid torus.
Let $P_n (q,t)$ be the Poincar\'e polynomial of $n$-colored $sl(2)$ homology categorifying the $n$-colored
Jones polynomial $J_n (q)$. As usual, it is convenient to package all $P_n (q,t)$ in a single function $F(x;q,t)$
such that
\be
P_n (q,t) \; = \; F(x;q,t) \vert_{x = q^n}.
\ee
Put differently, if $\hat A (\hat x, \hat y; q,t)$ is the operator that generates recursion relation among $P_n (q,t)$,
such that $\hat x P_n (q,t) = q^n P_n (q,t)$ and $\hat y P_n (q,t) = P_{n+1} (q,t)$, then \cite{Fuji:2012pm,Gukov:2015gmm}:
\be
\hat A \, F (x;q,t) \; = \; 0.
\ee

In terms of the function $F(x;q,t)$ that packages Poincar\'e polynomials of colored unknot homology,
the Poincar\'e polynomial of $\CH_a[M_3]$, where $M_3 = L(p,1)$,
is given by the ``Laplace transform'' \cite{Gukov:2016njj,BBLlaplace,MR2457479} of $F(x;q,t)$,
\be
P(\CH_a[M_3];q,t) \; = \; \oint \frac{dx}{2\pi i x} \, F(x;q,t) \, \theta_p^{(a)} (x;q)
\ee
{\it i.e.}~by a convolution with the theta-function $\theta_p^{(a)} (x;q)$.
Note, all the information about homological grading (that is, $t$-dependence) comes entirely from $F(x;q,t)$,
{\it i.e.}~from the colored unknot homology, and not from the Laplace transform that produces 3-manifold invariant
out of the knot invariant. In other words, categorification of the surgery formula affects only the Jones polynomial
part (by replacing it with the Poincar\'e polynomial).

Another way to build 3-manifolds and, therefore, another candidate for constructing $\CH_a[M_3]$ is based on triangulations.
This approach is especially attractive due to its combinatorial nature and has been extensively used as a platform
for mathematical definitions of various supersymmetric partition functions of 3d $\CN=2$ theory $T[M_3]$,
see {\it e.g.} \cite{Andersen:2011bt,Andersen:2013rxa,Andersen:2015tma,Garoufalidis:2013rca,MR3522084,Garoufalidis:2013axa,Garoufalidis:2016ckn}.
Reducible flat connections, again, are special, for the same reason as before. However, in partition functions,
the non-trivial stabilizer of reducible flat connections works against them and formally makes their contribution
vanish in a Chern-Simons theory with non-compact gauge group. This has to be contrasted with homological invariants,
where redicible solutions produce infinite contributions ({\it e.g.}~$\CT_+$ in Heegaard Floer homology)
compared to finite contributions of irreducible solutions. This suggests that in partition functions too,
both reducible and irreducible contributions can be seen together if we consider framed gauge transformations\footnote{A framed gauged transformation acts
trivially at a chosen point on $M_3$.} and then instead of further dividing by $G$ work equivariantly.
To the best of our knowledge, this has not been done and presents a nice and tractable problem for future work.\footnote{A closely
related $U(1)$-equivariant rather than $G$-equivariant regularization of complex Chern-Simons partition function
was used \cite{equivariant} for 3-manifolds of the form $M_3 = S^1 \times \Sigma$.
As expected, this equivariant localization does help to
see contributions of both abelian and non-abelian flat connections in the same partition function.}

\subsection{Superconformal index of ${T[M_3]}$ and its factorization}

Although the new invariants $\hat{Z}_a(q)$ and their categorification $\CH_a[M_3]$ have a direct connection to the WRT invariant via (\ref{WRT-decomposition}), from the viewpoint of 3d $\CN=2$ theory $T[M_3;SU(2)]$, it is not the most natural or simplest object to consider. The main reason is that, in principle, there are infinitely many possible boundary conditions that could be considered.\footnote{If we understand the boundary condition as a coupling of the 3d theory to a 2d theory living on the boundary, then the infinite number of possibilities can be seen, for example, from the fact that we can always introduce a 2d theory decoupled from the bulk.} Identifying the correct subset that appears in (\ref{BlockD2S1}) may be subtle, yet possible, as we shall see in concrete examples.

The more natural object is the superconformal index of $T[M_3]$ or, equivalently, the partition function on $S^2\times S^1$ with a certain supersymmetry preserving background:
\begin{equation}
	\CI(q) \; \equiv \; \Tr_{\CH_{S^2}}(-1)^Fq^{R/2+J_3} \; = \; Z_{T[M_3]}(S^2\times_q S^1)
\end{equation}
where $\CH_{S^2}$ is the space of BPS states of 3d $\CN=2$ SCFT $T[M_3]$ or, equivalently, $Q$-cohomology of all physical local operators, $F$ is the fermion number, $R$ is the generator of the $U(1)_R$ R-symmetry and $J_3$ is the Cartan generator of the $SO(3)$ isometry of $S^2$. By construction, this index has the desired integrality property\footnote{For reasons similar to the ones described at the end of the previous section, in general a negative power of 2 can appear as an overall factor. We omit it in some generic formulas to avoid clutter and instead focus attention on the conceptual structure. As mentioned earlier and as we shall see in examples, the effect of such fractions on categorification can typically be traced to the existence of bosonic modes with zero $q$-grading.} and can be categorified,
\begin{equation}
	\CH_{S^2}=\bigoplus_{i,j}\CH^{i,j}_{S^2} \,,\qquad \CI(q)=\sum_{i,j\in\Z}q^{i}(-1)^j\dim\CH_{S^2}^{i,j}\;\in \Z[[q]] \,.
	\label{IndexEuler}
\end{equation}
Another advantage of $\CH_{S^2}$ compared to $\CH_{D^2}$ is that it has a natural ring structure which is given by multiplication of BPS operators. One of the statements of 3d/3d correspondence is that the partition function $Z_{T[M_3]}(S^2\times_q S^1)$ computes the partition function of complex Chern-Simons on $M_3$ with real part of the ``level'' being $0$ and analytically continued imaginary part $\propto \tau\equiv \log q/(2\pi i)$.
Motivated by the topological/anti-topological fusion \cite{Cecotti:1991me}
and its recent 3d incarnation \cite{Pasquetti:2011fj,Beem:2012mb,Cecotti:2013mba,Blau:2016vfc}, we would like to make the following conjecture:
\begin{conj}
For $\hat{Z}_a(q)$ as in Conjecture \ref{Conjecture1}, the following holds
	\begin{equation}
		\CI(q)=\sum_{a\,\in\, \Tor H_1(M_3,\Z)/\Z_2}\, |\CW_a|\hat{Z}_a(q)\hat{Z}_a(q^{-1})
		\qquad \in\Z[[q]]
		\label{IndexFactorization}
	\end{equation}
where $\hat{Z}_a(1/q)$ is an apropriate extension\footnote{For a generic 3-manifold $M_3$, the analytic continuation of the series $\hat{Z}_a(q)\in \Z[[q]]$ convergent in $|q|<1$ may not exist outside $|q|=1$, at least in the standard way. However, one possible way to define it for general $M_3$ is
\begin{equation}
	\hat{Z}_a(q^{-1}) \; \equiv \; \left.\hat{Z}_a(q)\right|_{M_3\rightarrow \bar{M}_3},
\end{equation}
where $\bar{M}_3$ denotes $M_3$ with the reversed orientation. Note that therefore $\CI(q)$, unlike $\hat{Z}_a(q)$, is insensetive to the orientation.
} of $\hat{Z}_a(q)$ to the region $|q|>1$.
\label{Conjecture2}
\end{conj}

\subsection{Further refinement}

As we discussed above, for Seifert 3-manifolds the theory $T[M_3]$ has an extra $U(1)_{\beta}$ flavor symmetry. In particular, this is the case for $M_3=L(p,1)$, which will serve as an important example to us later. This flavor symmetry in $T[M_3]$ results in the presence of an extra $\Z$-grading in homological invariants $\CH_{D^2}$ and $\CH_{S^2}$ considered in the previous sections and a possibility to consider the corresponding refined indices (equivariant Euler characteristics):
\begin{equation}
	\hat{Z}_a(q;t)=\sum_{i\in \Z+\Delta_a,\atop j,\ell \in\Z}q^{i}t^{\ell}(-1)^j\dim\CH^{i,j;\,\ell}_a,
	\label{BlockEuler}
\end{equation}
and
\begin{equation}
\CI(q;t)=\sum_{i,j,\ell\in\Z}q^{i}t^{\ell}(-1)^j\dim\CH_{S^2}^{i,j;\,\ell}.
\end{equation}
where, from the physics point of view, $\ell$ is the $U(1)_{\beta}$ charge.
The refined indices obviously provide more information about the underlying vector spaces and, as we will see in examples, can be used sometimes to compute (conjecturally) the full homological invariants via the ``homological-flavor locking'' phenomenon that we will explain later.

The refined version of the factorization formula \eqref{IndexFactorization} reads
\begin{equation}
		\CI(q;t)=\sum_{a\,\in\, \Tor H_1(M_3,\Z)/\Z_2}\, |\CW_a|\hat{Z}_a(q;t)\hat{Z}_a(q^{-1};t^{-1}).
		\label{IndexRefDecomp}
\end{equation}

\subsection{Topologically twisted index of ${T[M_3]}$}

Another interesting invariant of $M_3$ which can be realized as an observable of $T[M_3]$ and has a categorification by construction is the topologically twisted index on $S^2\times S^1$ \cite{Benini:2015noa}. Namely, one can consider the 3d $\CN=2$ theory $T[M_3]$ with a background value of the $U(1)_R$-symmetry connection equal to the spin connection  on $S^2$. In terms of the effective 2d $\CN=(2,2)$ theory obtained by compactifying $T[M_3]$ on $S^1$, this is the familiar A-twist along the $S^2$.

The topologically twisted index and the underlying homological invariant of $M_3$ have the same fugacities/gradings as the superconformal index:
\begin{equation}
\CI_\top(q;t) \; = \; \sum_{i,j,\ell \in\Z}q^{i}t^{\ell}(-1)^j\dim\CH_{S^2_\top}^{i,j;\,\ell}.
\end{equation}
As in the superconformal index $\CI(q;t)$, here the parameter $q$ plays the role of the Omega-background parameter corresponding to rotating $S^2$ along one of the axes. However, in general the topologically twisted index $\CI_\top(q;t)$ has a much simpler structure (as will be explained later) compared to the superconformal index $\CI(q)$. Namely, $\CI_\top(q;t)$ is a rational function of $q,t$ (\ie~it has a form of the index of quantum mechanics with two supercharges), whereas $\CI(q,t)$ can be as transcendental as, say, a quantum dilogarithm or (Jacobi) mock modular form. Nevertheless, the topologically twisted index is expected to have a similar factorization into homological blocks:
\begin{equation}
		\CI_\top(q,t)=\sum_{a\,\in\, \Tor H_1(M_3,\Z)/\Z_2}\, |\CW_a|\hat{Z}_a(q;t)\hat{Z}_a(q^{-1};t).
\label{Itop-decomp}
\end{equation}
The difference from (\ref{IndexRefDecomp}) is due to the fact that the supersymmetric backround chosen for superconformal $S^2\times S^1$ index can be interpreted as doing topological A-twist along one of the $D^2\times S^1$ halves and anti-A-twist along the other  $D^2\times S^1$ half,
\be
\CI (q;t)
\; =
{\,\raisebox{-.5cm}{\includegraphics[width=6.0cm]{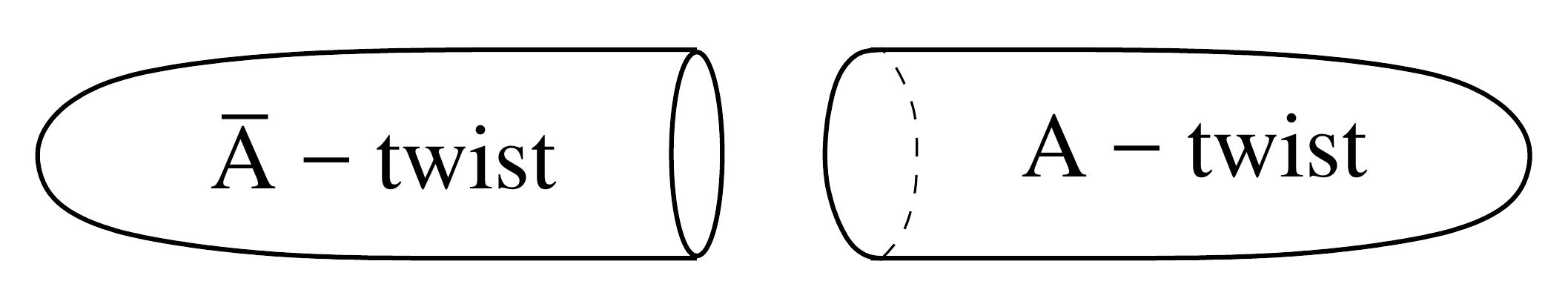}}\,}
\ee
whereas the background for the topologically twisted $S^2\times S^1$ index is such that the same A-twist is performed along both halves:
\be
\CI_\top(q;t)
\; =
{\,\raisebox{-.5cm}{\includegraphics[width=6.0cm]{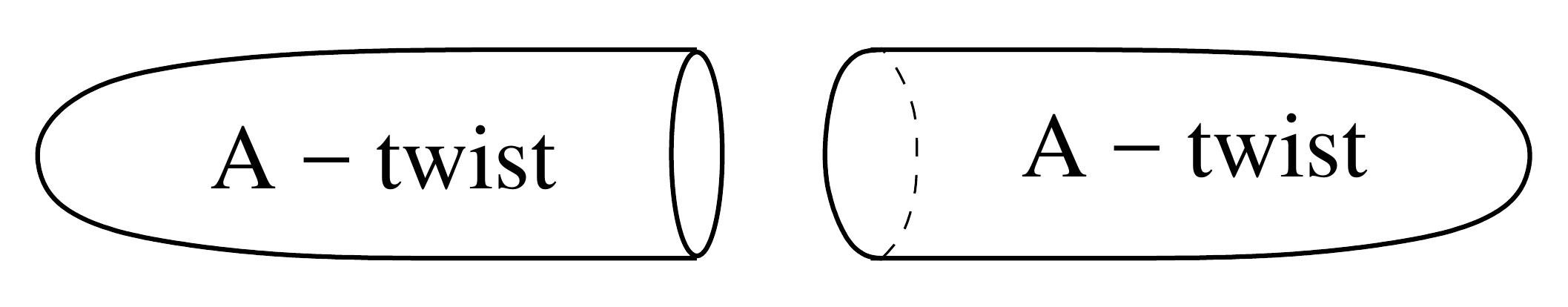}}\,}.
\ee

\subsection{Generalization to ${U(N)}$}
\label{sec:higherrk}

For simplicity, in this paper we mostly consider Chern-Simons theory on $M_3$ with gauge group being $G=SU(2)$ and the 3d/3d dual theory $T[M_3;SU(2)]$. However, in principle, everything we say above and below for $SU(2)$ can be easily generalized to the $SU(N)$ or $U(N)$ case; there are no obstructions for that. In particular, the formulae appearing in Conjecture \ref{Conjecture1} generalize as follows for $G = U(N)$:
\begin{equation}
	Z_{U(N)_k}[M_3]=(i\sqrt{2k})^{N(b_1(M_3)-1)/2}\sum_{a\;\in \;(\Tor H_1(M_3,\Z))^{N}/S_N}e^{\pi ik\,\sum_{i=1}^N\lk(a_i,a_i)}\,Z_a(q),
\label{WRT-UN-decomp}
\end{equation}
\begin{equation}
Z_a(q)=\sum_{b\;\in \;(\Tor H_1(M_3,\Z))^{N}/S_N}e^{\pi ik\,\sum_{i=1}^N\lk(a_i,a_i)}\,S_{ab}
	\, \left.\hat{Z}_b(q)\right|_{q\rightarrow e^{\frac{2\pi i }{k}}},
\end{equation}
\begin{equation}
	S_{ab}=\frac{\sum_{\sigma\in S_N}e^{2\pi i \sum_{i=1}^N\lk(a_i,b_{\sigma(i)})}}{|\mathrm{Stab}_{S_N}(a)|\cdot|\Tor H_1(M_3,\Z)|^{N/2}},
\end{equation}
\begin{equation}
 \hat{Z}_a(q)\in (N!)^{-c}\,q^{\Delta_a}\Z[[q]].
\end{equation}
As before, using the linking pairing one can identify $(\Tor H_1(M_3))^N/S_N$ with the set of connected components of abelian flat connections:
\begin{multline}
	(\Tor H_1(M_3))^N/S_N\stackrel{\lk}{\cong}\, \mathrm{Hom}\left(\Tor H_1(M_3),U(1)^N\right)/S_N
 \cong \pi_0\CM_\text{flat}^\text{ab}(M_3,U(N))
\label{UN-ab-set}
\end{multline}
where $S_N$ is the permutation group of a set with $N$ elements.
The factorization formulae (\ref{IndexRefDecomp}) and (\ref{Itop-decomp}) generalize as follows
\begin{equation}
		\CI(q,t)\quad = \sum_{a\,\in\, (\Tor H_1(M_3,\Z))^N/S_N}\, |\CW_a|\hat{Z}_a(q,t)\hat{Z}_a(q^{-1},t^{-1}),
\label{I-UN-decomp}
\end{equation}
\begin{equation}
		\CI_\top(q,t) \quad =\sum_{a\,\in\, (\Tor H_1(M_3,\Z))^N/S_N}\, |\CW_a|\hat{Z}_a(q,t)\hat{Z}_a(q^{-1},t),
\label{Itop-UN-decomp}
\end{equation}
with
\begin{equation}
	\CW_a\equiv \text{Stab}_{S_N}(a).
\end{equation}

Physically, there are several ways to understand the special role of abelian flat connections or, more generally,
reducible flat connections for $G$ of higher rank. As we already mentioned in section \ref{sec:summaryConj}, from
the viewpoint of quantum field theory (on fivebrane world-volume or its various reductions and limits)
this follows directly from the resurgent analysis. On the other hand,
by looking at the same system \eqref{CatGeom1} from the vantage point of the Calabi-Yau 3-fold,
the set \eqref{UN-ab-set} which labels the new invariants $\hat{Z}_a(q)$ and their categorification $\CH_a[M_3]$
can be understood as charges of the BPS states or enumerative invariants of the Calabi-Yau 3-fold.
We elaborate on this perspective in the following two subsections, expanding the web of dualities and interpretations.

\subsection{Relation to open GW invariants on ${T^*M_3}$}

Consider $U(N)$ level-$k$ Chern-Simon theory on a rational homology sphere $M_3$,
\begin{equation}
	b_1(M_3) \; = \; 0 \,,
\end{equation}
\ie~$H_1(M_3,\Z)=\Tor H_1(M_3,\Z)$ is a finite abelian group. (What follows can be considered as a generalization from the case $H_1(M_3,\Z)\cong \Z$ considered in \cite{Labastida:2000yw}).

Similar to  the $SU(2)$ case, $\sum_{i=1}^N\lk(a_i,a_i)=\mathrm{CS}(a)$ is the Chern-Simons invariant of the abelian flat connection $a$, and $Z_a(q)$ is a Borel resummation \cite{Gukov:2016njj} of the asymptotic expansion around $a$. Correspondingly, $\hat{Z}_b(q)$ is a Borel resummation of the series of the following type:
\begin{equation}
	\hat{Z}_b(q)=\sum_{m\geq 0}n^b_m\,q^{\Delta_b+m}\;\;\stackrel{\text{resum}}{=\joinrel=\joinrel=}\;\; \hat{Z}^\text{pert}_b(k)
=\sum_{m\geq 1} N^b_m\left(\frac{2\pi i}{k}\right)^m
	\qquad \in\Q[[2\pi i/k]].
	\label{resum}
\end{equation}
Consider
\begin{equation}
	e^{2\pi i\,a}\;\;\in \;\;\mathrm{Hom}\left(H_1(M_3),U(1)^N\right)/S_N
\end{equation}
as a formal variable in a generating series. Then $S_{ab}$, for each $b$, up to a factor, can be viewed as a basis element of the symmetric part of the group ring\footnote{For example, when $N=2$ and $H_1(M_3,\Z)=\Z_2$ we have:
\begin{equation}
	\Z\left[\left(H_1(M_3,\Z)\right)^N\right]^{S_N}\cong
	\Z[x_1,x_2]^{S_2}/\{x_1^2=1,x_2^2=1\}
\end{equation}
and
\begin{equation}
	\begin{array}{rl}
		S_{a,(0,0)} & \propto 1 \\
		S_{a,(1,0)} & \propto x_1+x_2 \\
		S_{a,(1,1)} & \propto x_1x_2 \\
	\end{array}
\end{equation}
where $x_i=e^{\pi i a}$.
}
 of $(H_1(M_3,\Z))^N$:
\begin{equation}
	|H_1(M_3)|^{N/2}|\text{Stab}_{S_N}(b)|^{-1}\,S_{ab}\in
	\Z\left[\left(H_1(M_3,\Z)\right)^N\right]^{S_N},\qquad b\in (H_1(M_3))^N/S_N,
	\label{group-ring-basis}
\end{equation}
if we identify
\begin{equation}
	\prod_{i=1}^Ne^{2\pi i\lk(a_i,c_i)}\equiv 1\cdot c
	\label{ring-id}
\end{equation}
in the group ring of $(H_1(M_3))^N$.

Then
\begin{equation}
	\left(2k|H_1(M_3)|\right)^{N/2}\left|\text{Stab}_{S_N}(a)\right|\,Z_a(q)\;\;\in\;\;
	\Z[[q^{1/|H_1|}]]\;\otimes_\Z\;
	\Z\left[\left(H_1(M_3,\Z)\right)^N\right]^{S_N}
	\label{Zq}
\end{equation}
while
\begin{equation}
	\sum_{b\in (H_1(M_3))^N/S_N}S_{ab}\hat{Z}^\text{pert}_b(k)\;\;
	\in\;\;
		\Q[[2\pi i/k]]\;\otimes_\Z\;
		\Z\left[\left(H_1(M_3,\Z)\right)^N\right]^{S_N}.
		\label{Zk}
\end{equation}

On the other hand, one can consider open topological strings (A-model) on $T^*M_3$ with a Lagrangian brane along $M_3$ equipped with a rank-$N$ bundle and an abelian flat connection (local system) $a$. The free energy is given by
\begin{equation}
	F^\text{top}_a[T^*M_3](g_s)=
	\sum_{g,\, h}g_s^{2g+h-2}\sum_{c\in (H_1(M_3,\Z))^h}
	N_{g;c_1,\ldots,c_h}
	\prod_{j=1}^h\,e^{2\pi i\sum_{i=1}^N \lk (a_i,c_j)}
	\label{Ftop}
\end{equation}
where the open GW invariants
\begin{equation}
	N_{g;c_1,\ldots,c_h}\;\;\in \;\; \Q
\end{equation}
count pseudoholomorphic maps from a genus-$g$ Riemann surface with $h$ holes to $T^*M_3$ such that the image of the $j$-th boundary component lands in a homology class $c_j\in H_1(M_3,\Z)$. The factor
\begin{equation}
	e^{2\pi i\sum_{i=1}^N \lk (a_i,c_j)}\;=\;\text{Hol}_a(c_j)
\end{equation}
is the holonomy of the abelian flat connection $a$ along $c_j$. Note that
\begin{equation}
	N_{g;c_1,\ldots,c_h}=0\qquad \text{unless}\qquad c_1+\ldots+c_h= 0\;\;\in H_1(M_3,\Z)
\end{equation}
because there are no non-trivial 2-cycles in $T^*M_3$.
When the flat connection $a$ is trivial, the generating function (\ref{Ftop}) takes a familiar form
\begin{equation}
	F^\text{top}_0[T^*M_3](g_s)=
	\sum_{g,\, h}g_s^{2g+h-2}\,
	N_{g;h}\,N^{h}
\end{equation}
where the averaged version of open GW invariants
\begin{equation}
	N_{g;h}=\sum_{c\in (H_1(M_3,\Z))^h}\,N_{g;c_1,\ldots,c_h}
\end{equation}
``forgets'' where the boundary components map to.

Again, using the identification (\ref{ring-id}), the generating function (\ref{Ftop}) can be regarded as
\begin{equation}
	F^\text{top}_a[T^*M_3](g_s)\;\; \in \;\;
	\Q[[g_s]] \otimes_\Z
	\Z\left[\left(H_1(M_3,\Z)\right)^N\right]^{S_N}.
	\label{Fk}
\end{equation}
Then, the relation between CS and open topological strings can be formulated as follows:\footnote{Normalizations of $Z_{U(N)_k}$, as well as the first few terms in $F^\text{top}$, are subject to ambiguity, and we won't attempt to fix them here.}
\begin{equation}
	\left.\exp\,F^\text{top}_a[T^*M_3](g_s)\right|_{g_s=\frac{2\pi i}{k}}\;\; \propto \;\;
	\sum_{b}S_{ab}\hat{Z}_b^\text{pert}(k)
	\label{top-CS}
\end{equation}
where the exponent is taken using multiplication rules in the ring in (\ref{Fk}), the same as the ring\footnote{Note, that any element of this ring has the form as in the right-hand side of (\ref{top-CS}) since (\ref{group-ring-basis}) are basis elements.} in (\ref{Zk}). The integer numbers $n^b_m$ in (\ref{resum}) are the ``massless variants'' of open DT/Ooguri-Vafa invariants of $T^*M_3$.

\subsection{Langlands duality and flat connections}
\label{sec:Langlands}

Motivated by the results of this paper, combined with \cite{Gukov:2016njj}, we can make another conjecture:

\begin{conj}
Let $G$ be a compact Lie group (not necessarily of ADE type) and $^LG$ its Langlands dual.
Then, under S-duality, the boundary condition $\CB_a$ corresponding to a connected component in $\CM_{\text{flat}}^{\text{ab}} (M_3, G_{\C})$
of $G_{\C}$ flat connections on a 3-manifold $M_3$ maps to
\be
\CB_a \quad \mapsto \quad
{}^L\CB_a \; = \; \sum_{{b \atop \text{abelian}}} S^{ab} \; (\, \CB_b + \sum_{{c \atop \text{non-abelian}}} n_{bc} \; \CB_c \, )
\label{Langlands}
\ee
where $S_{ab}$ is the S-matrix as in Conjecture \ref{Conjecture1} and $n_{bc}$ are the transseries coefficients from \cite{Gukov:2016njj}.
\label{conjLanglands}
\end{conj}

In formulating this Conjecture, we used $\CB_a$ (instead of $a$) to emphasize its role as a boundary condition
on $\partial (D^2\times_q S^1) = T^2$. The mapping class group $SL(2,\Z)$ of the 2-torus acts on the boundary
conditions and we claim that, at least for boundary conditions associated with abelian flat connections,
the S-transformation is given by \eqref{Langlands}.
Indeed, at the ``far end'' of the cigar, the geometry of the fivebrane world-volume looks like $\R \times M_3 \times T^2$,
and we can first reduce on $T^2$ to obtain 4d $\CN=4$ super-Yang-Mills on $\R \times M_3$.
The mapping class group of $T^2$ then becomes the S-duality group of 4d $\CN=4$ SYM,
in particular, exchanging $G$ and $^LG$ under the electric-magnetic duality.

When $G$ is of Cartan type A or D, the left-hand side of \eqref{Langlands} is the choice of boundary condition we
used in \eqref{aHW}, whereas the right-hand side describes the boundary condition that gives $\CH_a [M_3]$
and is used in most of the paper. Then, Conjecture \ref{conjLanglands} is basically a statement that both
sets of boundary conditions lead to integrality and are compatible.

\subsection{3-manifolds and the ``bottom row''}

Using a relation between the physical realizations of 3-manifold homology and a HOMFLY-PT homology,
respectively, we can produce a purely {\it mathematical} relation between the familiar knot homologies
and less familiar 3-manifold homologies of this paper.

Namely, the familiar setup for homology of a knot $K\subset S^3$ involves $Q$-cohomology (space of BPS states) of the following system:
\be
\begin{array}{rcl}
\multicolumn{3}{c}{~~\text{doubly-graded}~} \\[.1cm]
\hline
{\mbox{\rm space-time:}} && ~~  \R \times T^* S^3 \times TN \\
{\mbox{\rm $N$ M5-branes:}} && ~~  \R \times ~S^3~ \times ~D^2 \\
{\mbox{\rm M5$'$-branes:}} && ~~  \R \times ~L_K~ \times  ~D^2
\end{array}
\quad \xleftarrow[~\text{transition}]{~\text{phase}} \!\!\! \to \quad
\begin{array}{rcl}
\multicolumn{3}{c}{~~~\text{triply-graded}~~} \\[.1cm]
\hline
&& \R \times X \times  ~TN \\
{\mbox{\rm M5$'$:}} && \R \times L_K \times  D^2
\end{array}
\label{conifoldphases}
\ee
where $L_K$ is the conormal bundle of the knot $K$,
and $X$ is the resolved conifold, {\it i.e.} the total space of $\CO (-1) \oplus \CO (-1)$ bundle over $\cp^1$.

Note, on the triply-graded (``resolved'') side the original M5-branes disappear
and we are only left with a system of M5$'$-branes in a non-trivial Calabi-Yau background $X$.
This is very similar to a physical realization of 3-manifold homology in \eqref{CatGeom0} or \eqref{CatGeom1}
where, compared to the right-hand side of \eqref{conifoldphases}, the Lagrangian 3-manifold $M_3$
plays the role of $L_K$.
Indeed, according to the McLean's theorem, the neighborhood of the Lagrangian submanifold $L_K \subset X$
in \eqref{conifoldphases} can be identified with the total space of the cotangent bundle,
\be
N (L_K) \; \cong \; T^* L_K
\ee
which is precisely the setup of \eqref{CatGeom1}.
There is an important difference, however.

While in \eqref{CatGeom1} the Calabi-Yau space is simply the cotangent bundle to the Lagrangian 3-manifold $M_3$,
in \eqref{conifoldphases} it only looks like $T^* L_K$ in the neighborhood of $L_K$.
Globally, the topology of $X$ is different from $T^* L_K$ and, in particular, has non-trivial relative homology group
\be
H_2 (X,L_K) \cong \Z.
\ee
It plays an important role in the physical realization of the colored HOMFLY-PT homology of the knot $K$;
namely, the conserved charge captured by this relative homology is the so-called $a$-grading of the HOMFLY-PT homology of $K$.

As in \cite{Gorsky:2013jxa}, we can bridge the gap between the two systems \eqref{CatGeom1} and \eqref{conifoldphases}
by taking the limit\footnote{Note, in relating HOMFLY-PT homology to $sl(N)$ knot homology one sets $a = q^N$, so that $\log (a) \sim N$.}
\be
\log (a) = \text{Vol} (\cp^1) \to \infty.
\label{alimit}
\ee
This limit has a simple interpretation in almost every duality frame. For example, from the vantage point of the Calabi-Yau 3-fold $X$
it corresponds to the limit
\be
X \quad \leadsto \quad \C^3
\label{Xlimit}
\ee
that, on the toric diagram of the resolved conifold $X$, corresponds to moving two trivalent vertices far away from each other.
Keeping only one vertex in sight, we end up with $\C^3$ (whose enumerative invariants are counted by the refined topological vertex).
After making this replacement in \eqref{conifoldphases}, we obtain a simpler fivebrane system, whose BPS spectrum ($Q$-cohomology)
categorifies only the bottom (resp.~top) row of the HOMFLY-PT polynomial that contains the terms with minimal (resp.~maximal)
$a$-degree \cite{Gorsky:2013jxa}.

\begin{table}[htb]
\centering
\renewcommand{\arraystretch}{1.3}
\begin{tabular}{|@{\quad}c@{\quad}|@{\quad}c@{\quad}| }
\hline Knot $K$ & $\CH_{\text{bottom}} (K)$
\\
\hline
\hline ${\bf 3_1}$ & $q^{-2} + q^2 t^2$ \\
\hline ${\bf 4_1}$ & $t^{-2}$ \\
\hline ${\bf 5_1}$ & $q^{-4} + t^2 + q^4 t^4$ \\
\hline ${\bf 5_2}$ & $q^{-2} + t + q^2 t^2$ \\
\hline ${\bf 6_1}$ & $t^{-2} + q^{-2} t^{-1}$ \\
\hline
\end{tabular}
\caption{The reduced Poincar\'e polynomial of the bottom row of the HOMFLY-PT homology for simple knots colored by $\lambda = \Box$. The unreduced HOMFLY-PT homology is infinite-dimensional and its Poincar\'e ``polynomial'' is, in fact, a power series in $q$ and $t$.}
\label{tab:examples}
\end{table}

Now, after taking the limit \eqref{alimit}--\eqref{Xlimit}, the second cohomology of $X$ is trivial and there is no winding
around $\cp^1$ that we had in the original system \eqref{conifoldphases}. This, of course, corresponds to the fact that, by taking the limit,
we lost the $a$-grading focusing only on terms with the lowest $a$-degree.
The $q$-degree and $t$-degree, however, are still present and come from $U(1)_q \times U(1)_R$ symmetry acting on $D^2 \subset TN$.

To summarize, by taking the limit \eqref{alimit}--\eqref{Xlimit} in the physical system \eqref{conifoldphases},
we obtain a relation between the bottom row of the HOMFLY-PT homology of $K$ and the homology of the 3-manifold $M_3 = L_K$:
\be
\CH_{\text{bottom}} (K) \; \cong \; \CH (M_3).
\label{knotsvsmflds}
\ee
In fact, the geometry and topology of $M_3 = L_K$ is closely related to that of the knot complement $S^3 \setminus K$,
see \cite{Aganagic:2013jpa}.
Namely, for a knot (= a link with one component), both $M_3 = L_K$ and $S^3 \setminus K$ have $H_1 \cong \Z$.
In particular, in both cases the abelian flat connections are labeled by $\C^*$-valued holonomies $x_1, \ldots, x_N$.

Note, \eqref{knotsvsmflds} is a purely mathematical relation whose left-hand side is knot homology
and whose right-hand side is a 3-manifold homology.
One can explore it and try to make it more concrete, in particular using the physical setup \eqref{CatGeom1} and \eqref{conifoldphases}.
First, we need to develop a more precise dictionary between the RHS and LHS of \eqref{knotsvsmflds}.
Following the above derivation, we see that the M5$'$-branes in \eqref{conifoldphases}
correspond to M5-branes in \eqref{CatGeom1}.
Therefore, the $sl(N)$ homology of $M_3 = L_K$ gets related to the bottom row of $\lambda$-colored HOMFLY-PT homology of $K$,
where $\lambda$ is a partition with $N$ rows (or $N$ columns).
Moreover, as in \cite{Ooguri:1999bv,Labastida:2000yw},
we need to relate the ``partition basis'' (natural on the LHS of \eqref{knotsvsmflds})
to the ``holonomy basis'' (natural on the RHS of \eqref{knotsvsmflds}).
And, finally, it is important to keep in mind that in relating the physical systems \eqref{CatGeom1} and \eqref{conifoldphases}
we obtain a relation \eqref{knotsvsmflds} that involves unreduced knot homology\footnote{The reduced HOMFLY-PT homology
categorifies normalized HOMFLY-PT polynomial; it is finite-dimensional for $\lambda = \Box$ and links with one component
({\it i.e.} for knots). The unreduced version, on the other hand, categorifies unnormalized HOMFLY-PT polynomial and
is infinite-dimensional for any color $\lambda$. The former has no obvious analogue for 3-manifolds.} in the sense of \cite{khovanov2008matrix}.

As a warm-up, let us take a closer look at the simple case of $N=1$ and $\lambda = \Box$.
Recall, that $L_K$ is a special Lagrangian submanifold in $\R^6 \cong \C^3$.
The moduli space $\CM (L_K)$ of its Lagrangian deformations, together with a local system that it carries,
is the space of SUSY vacua of 3d $\CN=2$ theory $T[L_K;U(1)]$ on a circle,
\be
\CM (L_K) \; \cong \; \CM_{\text{SUSY}} (T[L_K;U(1)]) \; \cong \; \CM_{\text{flat}} (L_K, U(1)_{\C}),
\label{MLK}
\ee
where in the last relation we used the McLean theorem.
The cohomology of this moduli space is precisely the bottom row of the HOMFLY-PT homology
of $K$ colored by $\lambda = \Box$ \cite{Shende:2014ira}:
\be
\CH_{\text{bottom}} (K) \; \cong \; H^* (\CM (L_K)).
\ee
In \cite{Shende:2015mzx}, the moduli space $\CM (L_K)$ is described as a cluster variety.

Another simple example is $K = \text{unknot}$ and $\lambda = S^r$ (or $\lambda = \Lambda^r$),
a single-row Young tableaux. The corresponding superpolynomial is
\be
\CP^{S^r}_{\text{bottom}} (\text{unknot}) \; = \; \prod_{i=1}^r \frac{1}{1-q^{2i} t^{2(i-1)}} \,.
\label{SrHOMFLY}
\ee
These expressions can be found as $x^r$-coefficients of $Z_{\text{vortex}} (x;q,t)$ for a single 3d $\CN=2$ chiral multiplet.
More generally, for $K = \text{unknot}$ colored by Young diagrams with $n$ rows (or columns)
we end up with the $S^1 \times D^2$ vortex partition function of $n$ 3d $\CN=2$ chiral multiplets.
This has to be compared with $gl(n)$ homology of $M_3 = S^3 \setminus (\text{unknot})$.

It would be useful to develop the relation \eqref{knotsvsmflds} between the familiar HOMFLY-PT homology and the less
familiar 3-manifold homology further. We leave this to future work and now turn to detailed analysis of the latter.

\section{Examples}

In this section, we present many examples, illustrating the general proposal outlined in the previous section.
In particular, our goal is two-fold:
first, we wish to use the proposed physics definition of the new homological invariants $\CH_a[M_3]$
to compute them in many concrete examples, to the extent that one can start exploring the structure
of the results and explicitly test the Conjectures \ref{Conjecture1} and \ref{Conjecture2}, which is our second goal.
We start with the simplest three-manifold and gradually move to the study of more complex ones.

\subsection{${M_3=S^3}$}\label{sec:S3}

For $M_3=S^3$ and $G=U(N)$, the 3d $\CN=2$ theory $T[S^3]$ is a 3d $\CN=2$ Chern-Simons theory with gauge group $U(N)$ at level $1$ and an adjoint chiral multiplet $\phi$, whose R-charge is equal to 2 \cite{equivariant}. This theory is dual (the duality is usually referred to as the ``duality appetizer'' \cite{Jafferis:2011ns,Kapustin:2011vz,appetizer}) to a system of $N$ free chirals, making it simple to analyze. The R-charges of the free chirals are given by $R=2,4,\ldots, 2N$, and they have charges $1, 2,\ldots, N$ under the $U(1)_{\beta}$ flavor symmetry that rotates the original adjoint chiral $\phi$ by a phase.

As a result, the superconformal index of the theory $T[S^3]$ can be expressed a simple product, 	
\begin{equation}
	{\CI}_{U(N)}(q,t)=\prod_{i=1}^N\frac{(t^{-i}q^{1-i};q)_\infty}{(t^iq^i;q)_\infty},
	\label{Index-UN-free}
\end{equation}
where, as usual, the $q$-Pochhammer symbol is defined as
\begin{equation}
 (z; q)_n = \prod_{j=0}^{n-1} (1-z q^j).
\end{equation}
Since $H_1(S^3)=0$, there is only one homological block $\hat{Z}_0(q,t)$, realized as the $S^1\times D^2$ partition function of the $N$ free chirals with Neumann boundary conditions,
\begin{equation}
	\hat{Z}_0(q,t)=\prod_{i=1}^N\frac{1}{(t^iq^i;q)_\infty}.
\end{equation}
In the terminology of \cite{Gukov:2016gkn} this is an \textit{unreduced} homological block.\footnote{Although the terminology
``reduced'' and ``unreduced'' here is very similar to the one used in knot homology, there is no direct connection.}
In order to relate it to the WRT invariant of $S^3$ as in Conjecture \ref{Conjecture1}, before taking the unrefined limit $t\rightarrow 1$, one has to divide by the contribution of the Cartan components of the adjoint chiral $\phi$. The result looks like
\begin{equation}
 \hat{Z}_0(q)=\left.{(tq;q)^N_\infty}\,\hat{Z}_0(q,t)\right|_{t\rightarrow 1}=
\prod_{j=1}^{N-1}(1-q^j)^{N-j}.
\end{equation}
In the case of $G=SU(2)$, the dual theory consists of just one free chiral multiplet with R-charge $4$ and $U(1)_{\beta}$ charge $2$, whose index is
\begin{equation}
	\CI_{SU(2)}(q,t)=\frac{(t^{-2}q^{-1};q)_\infty}{(t^2q^2;q)_\infty}.
\end{equation}
and we have
\begin{equation}
	\hat{Z}_0(q,t)=\frac{1}{(t^2q^2;q)_\infty},\qquad \hat{Z}_0(q)=(1-q^2).
\end{equation}
Using the standard prescription for extending the quantum dilogarithm outside $|q|<1$,\footnote{Note that this is not an ordinary analytic continuation. The latter actually does not exists because $|q|=1$ is a natural boundary.}
\begin{equation}
	(x;q^{-1})_\infty\propto \frac{1}{(xq;q)_\infty},
	\label{poch-cont}
\end{equation}
we have, for $G=U(N)$,\footnote{Note that there is an ambiguous overall constant in (\ref{poch-cont}). We fix it in (\ref{S3-anti-block}) by requiring that the unrefined quantities are related by the ordinary analytic continuation. Namely,
\begin{equation}
 \left.\hat Z_0(1/q,1/t)\cdot(t^{-1}q^{-1};q^{-1})^N_\infty\right|_{t\rightarrow 1}=
\left.\hat Z_0(1/q,1/t)\cdot(1/t;q)^{-N}_\infty\right|_{t\rightarrow 1}=
\hat{Z}_0(1/q)=\prod_{j=1}^{N-1}(1-q^{-j})^{N-j}.
\end{equation}
}
\begin{equation}
	\hat Z_0(1/q,1/t)=\prod_{i=1}^N\frac{1}{(t^{-i}q^{-i};1/q)_\infty}=
	\frac{1}{N!}\prod_{i=1}^N{(t^{-i}q^{1-i};q)_\infty}.
	\label{S3-anti-block}
\end{equation}
We see that the superconformal index \eqref{Index-UN-free} indeed admits a factorization {\it \`a la} (\ref{IndexRefDecomp}),
\begin{equation}
 \CI(q,t)=|\CW_0|\,\hat{Z}_0(q,t)\hat{Z}_0(q^{-1},t^{-1}),
\end{equation}
with only one homological block in this case.

\subsubsection{Categorifying the index}

As $T[S^3,U(N)]$ is dual to a system of free chirals, its space of BPS states on $S^2$ --- which gives the homological invariant associated with $M_3 = S^3$ --- factorizes as product
\be
\CH (S^3) \; = \; \CH_2 \otimes \CH_4 \otimes \ldots \otimes \CH_{2N},
\ee
where $\CH_{2i}$ is the BPS Hilbert space of a single 3d $\CN=2$ chiral multiplet with R-charge $2i$ and $U(1)_{\beta}$ flavor charge $i$.
Thus, we now turn to the problem of categorifying the index of a free chiral multiplet.

Recall that for a 3d $\CN=2$ chiral multiplet $\Phi$ of R-charge $r$ and flavor charge $f$, the superconformal index --- equal to the equivariant Euler characteristic of the BPS Hilbert space $\CH_{r,f}$ --- is given by
\be
\CI_r (q,t) = \chi_{q,t}(\CH_{r,f})= \sum_{i,j,\ell \in\Z} q^i t^\ell (-1)^j \dim \CH_{r,f}^{i,j;\ell}= \prod_{j=0}^{\infty} \frac{1 - t^{-f}q^{1-r/2+j}}{1-t^f q^{r/2+j}}.
\label{CI}
\ee
Each term in this product corresponds to a generator of the BPS Hilbert space $\CH_{r,f}$. One can identify the denominator of \eqref{CI} as the contribution of the bosonic modes $\partial^j \phi$, and the numerator as the contribution of the fermionic modes $\partial^j \bar \psi_+$. Then $\CH_{r,f}$ is freely generated by these generators as a supercommutative algebra,
\be
\CH_{r,f} \; = \; \C [x_i, \xi_i] \; \cong \; \Omega^{\bullet} (\text{Sym}^{\infty} (\C))
\label{chiralhomology}
\ee
with an infinite set of even generators $x_0, \, x_1, \, x_2, \, \ldots$
and odd generators $\xi_0, \, \xi_1, \, \xi_2, \, \ldots$ coming from
\bea
x_i & \quad \longleftrightarrow \quad & \partial^j \phi, \\
\xi_i & \quad \longleftrightarrow \quad & \partial^j \bar \psi_+. \nonumber
\eea
In fact, this result has already appeared in math and physics literature
(see {\it e.g.}~\cite{Gorsky:2013jxa,Gukov:2016gkn} and references therein), and it is isomorphic to the colored HOMFLY-PT homology of the unknot.

The charges of the generators are summarized below
\begin{center}
  \begin{tabular}{ c | c |c | c | c}
	 & $R$ & $F$ & $j_3$ & $\frac{R}{2}+j_3$\\
    \hline
    $\phi$ & $r$ & $1$ & 0 & $\frac{r}{2}$ \\ \hline
    $\bar{\psi}_+$ & $1-r$ & $-1$ & $\frac{1}{2}$ & $1-\frac{r}{2}$ \\ \hline
    $\partial_+$ &0 & 0 & 1 & 1 \\
    \hline \hline
		$\partial^j \phi$ & $r$ & 1 & $j$ &$\frac{r}{2} +j$
		\\ \hline
		$\partial^j \bar \psi_+$ & $1-r$ & $-1$ & $\frac{1}{2}+j$ & $1-\frac{r}{2} +j $ \\
  \end{tabular}\ .
\end{center}
Here $R$ and $F$ are generators of the R-symmetry $U(1)_R$ and the flavor symmetry $U(1)_{\beta}$, and $j_3$ generates a $U(1)$ subgroup of the $SU(2)$ isometry of the $S^2$. Using this data, one can also obtain the Poincar\'e polynomial of $\CH^{i,j;\ell}_{r,f}$
\be
\CP_r (q,z,t)= \sum_{i,j,\ell\in\Z}q^i z^j t^\ell \dim \CH_{r,f}^{i,j;\ell}=\prod_{j=0}^{\infty} \frac{1 + z^{1-r}t^{-f}q^{1-r/2+j}}{1-z^r t^f q^{r/2+j}}.
\ee

One interesting observation of \cite{Gukov:2016gkn} is that, for some simple 3-manifolds $M_3$, it is often the case that one can trade the $\ell$-grading for the homological $j$-grading, making the Poincar\'e polynomial effectively computable. This phenomenon can be seen at the level of $\CH_{r,f}$ for a single chiral. Indeed, the $(i,j,\ell)$-degrees of the generators are not independent, and under the substitution of variables $z^r\rightarrow t^{f}$ and $z\rightarrow -1$, the Poincar\'e polynomial for the first two gradings becomes \eqref{CI},
$$
\CP_r (q,z,1)= \prod_{j=0}^{\infty} \frac{1 + z^{1-r}q^{1-r/2+j}}{1-z^r q^{r/2+j}} \quad \leadsto \quad
\CP_r (q,-1,t) =\CI_r(q,t)= \prod_{j=0}^{\infty} \frac{1 - t^{-f}q^{1-r/2+j}}{1-t^f q^{r/2+j}} \,.
$$

The homology \eqref{chiralhomology} obtained from the index has the same form as the homology on $D^2\times_q S^1$ found in \cite{Gukov:2016gkn}. This could be justified by the following argument. The geometry $S^1\times S^2$ is conformally flat, and the SUSY variation is obtain from that on flat space. The latter, in turn, is equivalent to partially A-twisted 3d $\CN=2$ theory, which is precisely the setting of \cite{Gukov:2016gkn}.

\subsubsection{The $t=q^{\beta}$ reduction from BPS spectral sequence}\label{sec:Spec}

One interesting property of the invariants \eqref{Index-UN-free} and \eqref{CI} is their behavior under the specialization $t=q^{\beta}$ with $\beta\in \Z$. First, they all become polynomial after taking this limit. For example,
\be
\CI_{r,f=1} (q,t) \quad \overset{t=q^\beta}{\longrightarrow}\quad \prod_{j=0}^{\infty} \frac{1 - q^{1-r/2-\beta+j}}{1- q^{r/2+\beta+j}}=(q^{1-r/2-\beta};q)_{r+2\beta-1},
\ee
is simply a polynomial (as opposed to a power series).
Second, they exhibit ``positivity'' in the sense that if one adds a minus sign setting $t=-q^{\beta}$, one finds
a polynomial with only positive coefficients:
\be
\CI_{r,f=1}(q,t=-q^\beta)=\prod_{j=0}^{r+2\beta-2} \left(1+q^{1-r/2-\beta+j}\right).
\ee
Here, these phenomena originate from the fact that, upon setting $t=q^\beta$, there is a lot of cancellation between contributions from bosonic and fermionic generators, with only $\bar{\psi}_+$,$\partial \bar{\psi}_+$,\ldots, $\partial_+^{2\beta}\bar{\psi}_+$ remaining. However, as we will see in later part of this paper, these phenomena are actually very universal and also appear in a wide variety of 3d $\CN=2$ theories $T[M_3]$ with the symmetry $U(1)_{\beta}$. Therefore, it is interesting to understand the $t=q^\beta$ reduction at the categorified level.

The natural uplift of the reduction of superconformal indices
\be
\CI_{T[M_3]}(q,t) \quad \leadsto \quad \CI_{T[M_3]}^{(\beta)}(q)=\CI(q,q^\beta)
\ee
is a BPS spectral sequence \cite{Gukov:2015gmm}:
\be\label{BPSSpec}
\CH_{Q}^{\text{BPS}} \quad \leadsto \quad \CH_{Q+Q'}^{\text{BPS}},
\ee
which starts with the space of BPS states of $T[M_3]$ on $S^2$ and converges to the space of $Q+Q'$-cohomology with a deformed supercharge. For concreteness, we will focus on the example of a single chiral $\Phi=(\phi,\psi)$ with R-charge $r=2$, which can be understood as the theory $T[S^3, U(1)]$. The first step is to explicitly identify the space of BPS states $\CH_{S^2}$ found in the previous section as the $Q$-cohomology for a supercharge in the following way.

All four supercharges of a 3d $\CN=2$ theory can be preserved on $S^2\times \R$. We pick a supercharge $Q=\bar{Q}_-$ parametrized by a Killing spinor $\bar{\zeta}_+$ with
\be
j_3(\bar{\zeta}_+)=\frac{1}{2},\quad R(\bar{\zeta}_+)=-1.
\ee
The supergravity background is given by
\be
V_\mu=A_\mu^{(R)}=-i\delta_{\mu 3},
\ee
with the other components of the background multiplet set to zero.\footnote{We have chosen the round metric and the veilbein
\be
e_1=d\theta,\quad e_2=\sin \theta d\varphi, \quad e_3=dx_3.
\ee
} Then, the on-shell SUSY variation given by $\bar{Q}_-$ is
\beq\label{QTrans}
\delta\phi&=0\\
\delta\bar{\psi}&=0\\
\delta\bar{\phi}&=\sqrt{2} \bar{\zeta}_+\bar{\psi}_-,\\
\delta\psi&=-\sqrt{2} i\gamma^\mu \bar{\zeta}_+ D_\mu \phi+\sqrt{2} i \sigma \phi \bar{\zeta}_+,
\eeq
where we have included a background $U(1)_{\beta}$ vector multiplet $(A_\mu,\lambda,\bar{\lambda},\sigma,D)$, and $D_\mu$ is a covariant derivative that involves the Levi-Civita connection, $A_\mu$ and $A_\mu^{(R)}$. Then the $Q$-closed states are $\phi$, $\bar{\psi}_\pm$ and their derivatives. Among them $\bar{\psi}_-$, $D_3\phi$ and $\partial_-\phi$ are $Q$-exact, while $D_3\bar{\psi}_+$ and $\partial_-\bar{\psi}_+$ are eliminated by the equation of motion. So we have found that the space of BPS states is exactly the cohomology of $Q=\bar{Q}_-$.

One might expect a spectral sequence to arise when $Q$ is deformed into $Q+Q'$, whose second page is given by
\be
E_2^{\bullet,\bullet}=H^{\bullet}(\CH^{\bullet}_Q,Q').
\ee
The simplest scenario is to have only non-trivial differentials on the second page, enabling us to read off $\CH^{\bullet}_{Q+Q'}$ from the third page, like most examples studied in \cite{Gukov:2015gmm}.

When we take $t=q^\beta$, we are effectively shifting the $U(1)_R$ charge by $2\beta$ multiple of the $U(1)_{\beta}$ charge. As a result, the operators $\phi$ and $\partial_+^{2\beta+1}\bar{\psi}_+$ have the same $R/2+j_3$ quantum number and their contribution to the index cancels. Then, naively, we expect to have the following action of $Q'$ on $\phi$:
\be
[Q',\phi]\sim\partial_+^{2\beta+1}\bar{\psi}_+.
\ee
However, no supercharges can achieve this, even if we turn on more general supergravity backgrounds, as $\phi$ and $\bar{\psi}$ live in different representations of the supersymmetry algebra and don't mix.

Another possibility is to have $\phi$ and $\partial_+^{2\beta+1}\bar{\psi}_+$ to be eliminated separately. This can be achieved by turning on $-2\beta+1$ units of $U(1)_{\beta}$ flux along $S^2=\cp^1$.\footnote{In order for the background to be supersymmetric, one also needs to turn on a constant $\sigma$ for the background gauge multiplet, which won't affect the analysis in this section.} Now $\phi$ and $\bar{\psi}_+$ are holomorphic sections of
\be
\phi\in H^0(\cp^1,O(-2\beta+1)),\quad \bar{\psi}_+\in H^0(\cp^1,O(2\beta)).
\ee
The former has no sections, while the latter has $2\beta+1$ sections given by $\bar{\psi}_+$, $\partial_+\bar{\psi}_+,\ldots, \partial_+^{2\beta}\bar{\psi}_+$.

At the level of SUSY transformations, we now have a term in $\delta \psi$ of the following form
\be
(\delta \psi)_- \sim \gamma^-\bar{\zeta}_+ D_-\phi,
\ee
making all modes of $\phi$ exact under $Q+Q'$. On the other hand, the equation of motion for $\bar{\psi}_+$ will impose
\be
D_-\bar{\psi}_+=0,
\ee
but there are solutions to this equation, because $\bar{\psi}$ have negative charge under $U(1)_{\beta}$.
It would be interesting to pursue this analysis further in non-trivial examples and also make contact with \cite{Beem:2013sza} and \cite{Dedushenko:2016jxl},
where similar phenomena were studied in theories with larger supersymmetry.

\subsection{${M_3=L(p,1)}$}\label{sec:Lp1}

\subsubsection{Refined superconformal index}

We now move to the case of $M_3=L(p,1)\cong S^3/\Z_p$. The theory $T[L(p,1),U(N)]$ is an $\CN=2$ Chern-Simons theory at level $p$, with a chiral multiplet $\Phi$ in the adjoint representation with R-charge $R(\Phi)=2$ \cite{Gadde:2013sca,Chung:2014qpa, equivariant}. Its superconformal index is given by (see \eg~\cite{Imamura:2011su})
\begin{multline}
	\label{IndexInt}
{\CI}_{U(N)}(q,t) = \sum_{{m_1 \geqslant \cdots \geqslant m_N} \in \mathbb{Z}}  \frac{1}{\left| {\cal W}_m \right|} \int\limits_{|z_i|=1} \prod_j \frac{dz_j}{2\pi i z_j}  \prod_{i}^{N} \left( z_i \right)^{{ p}m_i} \prod_{i \neq j}^{N} t^{- \left| m_i -m_j \right|/2} q^{-R \left| m_i -m_j \right|/4}\;\times \\
\left( 1-q^{ \left| m_i -m_j \right|/2} \frac{z_i}{z_j} \right)
\prod_{i \neq j}^{N} \frac{\left( \frac{z_j}{z_i} t^{-1} q^{ \left| m_i -m_j \right|/2+1-R/2}; q \right)_{\infty}}{\left( \frac{z_i}{z_j} t q^{ \left| m_i -m_j \right|/2+R/2}; q \right)_{\infty}} \;\times  \left[\frac{(t^{-1}q^{1-R/2}; q)_{\infty}}{(t q^{R/2}; q)_{\infty}}\right]^N.
\end{multline}
Here $R$ stands for the R-charge of the adjoint chiral multiplet $\Phi$ and the fugacity $t$ is associated to the $U(1)_{\beta}$ flavor symmetry which acts on the adjoint chiral multiplet via $\Phi\mapsto e^{i\theta}\Phi$. Using some computer algebra (\eg~Mathematica) one can calculate explicitly $\CI(q,t)$ as a series in $q$ up to a relatively high order. The coefficients are polynomials in $t$, that is
\begin{equation}
  \CI(q,t)\;\in\;\Z[t][[q]].
\end{equation}

For gauge group $G=SU(2)$, the expression for the index simplifies to
\begin{multline}
	\CI_{SU(2)}(q,t)=\frac{1}{2}\sum_{m\in \Z}\int \frac{dz}{2\pi i z}\,
	z^{2pm}\,t^{-2|m|}\,q^{-2|m|}\,(1-z^{\pm 2}q^{|m|})
\\
	 \times
	\frac{(1/t;q)_\infty(z^{\pm 2}q^{|m|}/t;q)_\infty}{(qt;q)_\infty(z^{\pm 2}q^{|m|+1}t;q)_\infty}
	\label{I-Lp1-SU2}
\end{multline}
where we use the standard notation
\begin{equation}
 f(z^{\pm 2})\equiv f(z^2)f(z^{-2}).
\end{equation}

The \textit{unreduced} homological blocks can be calculated using the following formula \cite{Gukov:2016gkn}:
\begin{multline}
	 \hat Z_a(q,t)=Z_{T[M_3]}(D^2\times_q S^1;a)=\\
\frac{1}{|\CW_a|}\,\frac{1}{(tq;q)_\infty^N}
	\int\limits_{|z_i|=1} \prod_{i=1}^N \frac{dz_i}{2\pi i z_i}
	\prod_{i\neq j}\frac{(z_i/z_j;q)_\infty}{(z_i/z_jtq;q)_\infty}
	\,\Theta_{a}^{\mathbb{Z}^N,p}(z;q),
	\label{TM3mod}
\end{multline}
where $\Theta_{a}^{\mathbb{Z}^N;p}(z;q)$ is the theta function of the rank-$N$ lattice $\Z^N$ with quadratic form $p\cdot\mathrm{Id}$:
\be
\Theta_{a}^{\Z^N;p}(z, q)=\sum_{n\in p\Z^N+a}q^{\sum_{i=1}^N{n_i^2}/2p}\prod_{i=1}^Nz_i^{n_i}.
\ee

We would like to check that the following relation holds:
\begin{equation}
	\CI_{U(N)}(q,t)=\sum_{a\in \Z_p^N/S_N} |\CW_a|\hat{Z}_a(q,t)\hat{Z}_a(1/q,1/t).
	\label{Index-factorization}
\end{equation}
Compared to the case $p=1$ ($M_3=S^3$) considered in section \ref{sec:S3},  there are now multiple homological blocks. Another technical complication is that the formula (\ref{TM3mod}) only defines $\hat Z_a(q,t)$ for $|q|<1$, since the theta function is only given in terms of series convergent in $|q|<1$, with no canonical analytic continuation outside of the unit disk.

This problem will be resolved in section \ref{sec:cyclotomic}. For now, let us note that in the unrefined case ($t=1$) such problem does not appear because
\begin{equation}
	\hat{Z}_a(q)=\left.\frac{\hat{Z}_a(q,t)}{(tq;q)_\infty^N}\right|_{t\rightarrow 1}
=
\frac{1}{|\CW_a|}\int\limits_{|z_i|=1} \prod_{i=1}^N \frac{dz_i}{2\pi i z_i}
	\prod_{i\neq j}(1-z_i/z_j)
	\,\Theta_{a}^{\mathbb{Z}^N,p}(z;q)\qquad \in \Z[q]
\end{equation}
is just a polynomial and is obviously well-defined for any $|q|<\infty$. Note, that the factorization of the superconformal index in the unrefined case was essentially checked in \cite{appetizer}. There it was shown that
\begin{equation}
 \CI_{U(N)}(q)=\sum_{a\in \Z_p^N/S_N} |\CW_a|{Z}_a(q){Z}_a(1/q)
\label{ILens-decomp-flat}
\end{equation}
where\footnote{The full index $\CI_{U(N)}(q,t)$ of $T[L(p,1)]$ vanishes in the unrefined limit $t\rightarrow 1$,
\be
\CI_{U(N)}(q,t)\sim (1-t^{-1})^N \CI_{U(N)}(q),
\ee
due to the contribution of Cartan components of the adjoint chiral multiplet with R-charge 2, which saturate the unitarity bound. The rescaling in (\ref{I-UN-unref}) removes their contribution and makes the limit finite. Another method of regularization is to take the same limit but in the form $t=q^\epsilon,\,\epsilon\rightarrow 0$, which correponds to taking R-charge to be $2-\epsilon$, and then just remove an appropriate power of $\epsilon$:
\begin{equation}
 \CI'_{U(N)}(q)=\lim_{\epsilon\rightarrow 0} \CI_{U(N)}(q,t)\cdot \epsilon^{-N}=(\log q)^N \CI_{U(N)}(q).
\end{equation}
This is the approach that was used in \cite{appetizer}. Then the $(\log q)^N\propto k^{-N}$ factor agrees with the factorization if we include the overall factor $k^{-N/2}$ in (\ref{WRT-UN-decomp}) into the definition of $Z_a(q)$.
 }
\begin{equation}
 \CI_{U(N)}(q)=\left. \CI_{U(N)}(q,t)\cdot \frac{(tq;q)_\infty^N}{(1/t;q)_\infty^N} \right|_{t\rightarrow 1}
\label{I-UN-unref}
\end{equation}
and $Z_a(q)=\sum_b S_{ab}\hat{Z}_b(q)$ is as in (\ref{WRT-UN-decomp}), the contributions of flat connection $a$ to WRT invariant/CS partition function on $M_3=L(p,1)$. Then
\begin{equation}
  \CI_{U(N)}(q)=\sum_{a\in \Z_p^N/S_N} |\CW_a|\hat{Z}_a(q)\hat{Z}_a(1/q)
\end{equation}
follows from (\ref{ILens-decomp-flat}) and $S^2=\text{Id}$ (\ref{S-squared}) along with the symmetry
\be
|W_a|S_{ab}=|W_b|S_{ba}.
\ee

\subsubsection{Topologically twisted index of ${T[L(p,1)]}$}
\label{sec:TopInd}

For group $G=SU(2)$, the topologically twisted index (refined by angular momentum) of $T[L(p,1)]$ reads \cite{Benini:2015noa}:
\begin{equation}
	\CI_\text{top}(q,t)=\frac{1}{2}\sum_{m\in \Z}\int\limits_\text{JK} \frac{dz}{2\pi i z}
	\frac{z^{2(p+2)m}\,q^{-m}\,(1-z^2q^{m})(1-z^{-2}q^{m})}{(z^2tq^{1-m};q)_{2m-1} (z^{-2}tq^{1+m};q)_{-2m-1}(tq;q)_{-1}},
	\label{Itop-Lp1-SU2}
\end{equation}
where, as usual, the Pochhammer symbol with negative integer in the subscript is defined via the following identity:
\begin{equation}
	(x;q)_n=\frac{1}{(xq^n;q)_{-n}}.
\end{equation}
The contour in (\ref{Itop-Lp1-SU2}) is chosen according to the Jeffrey-Kirwan residue prescription. Namely, we either choose poles at $z=0$, $z=\pm \sqrt{t}\,q^{\cdots}$ or at $z=\infty$, $z=\pm 1/\sqrt{t}\,q^{\cdots}$. The result is strikingly simple:
\begin{equation}
	\CI_\text{top}(q,t)=
	\left\{
	\begin{array}{cr}
		\frac{1}{(t^2q^2;q)_{-3}}\equiv (t^2/q;q)_3,& p=1,\\
		1-t^4,& p=2,\\
		1-t^3,& p\geq 3.
	\end{array}
	\right.
\end{equation}
As expected, the result for $p=1$ is in agreement with the dual description by a free chiral multiplet with R-charge $4$. Note that for $p>1$ the result turns out to be $q$-independent. In fact, for large $p$, the twisted index of $T[L(p,1);G]$ with a general Lie group $G$ becomes
\be
\CI_{\text{top}}~\stackrel{p\gg 1}{\longrightarrow}~ P_{-t}(G),
\ee
where $P_{-t}(G)$ denotes the Poincar\'e polynomial of $G$ in variable $-t$. For $G=SU(2)$,
\be
P_{-t}(G) \; = \; 1-t^3
\ee
and, for $SU(3)$, one has \cite{EVFCBI}:
\be
\CI_{\text{top}}~\stackrel{p\gg 1}{\longrightarrow}~ 1-t^3-t^5+t^8=P_{-t}(SU(3)).
\ee
Additionally, the BPS spectrum can be identified with the cohomology of $G$,
\be
\CH_{\text{tw-BPS}}^\bullet=H^{\bullet}(G).
\ee
An argument for this relation mentioned above is essentially given in section 5 of \cite{Andersen:2016hoj} after Proposition 3, as summarized in the following.

The twisted index of $T[L(p,1)]$ computes the equivariant index of certain K-theory class over $\frak{M}_H(\cp^1;G)$, the Hitchin moduli stack over $\cp^1$,\footnote{In this equation, $L$ is the determinant line bundle and $\CF_q$ is an object in the derived category of coherent sheaves on $\frak{M}_H$ with dependence on $q$. In general, as explain in \cite{equivariant, EVFCBI}, the twisted partition function of $T[L(k,1)]$ on $S^1\times \Sigma$ gives the ``equivariant Verlinde formula,'' which in turn can be written as a K-theory index over $\frak{M}_H(\Sigma)$. This fact, combined with the projection map $\frak{M}_H\rightarrow \frak{M}$, is the starting point for the proof of the equivariant Verlinde formula \cite{Andersen:2016hoj, 2016arXiv160801754H}.}
\be
\CI_{\text{top}}(q,t)=\mathrm{Ind}_t(\frak{M}_H,\CF_q\otimes L^p).
\ee
Here, $\frak{M}_H$ is a complicated derived stack, but using the projection map $\frak{M}_H(\cp^1)\rightarrow \frak{M}(\cp^1)$ to the moduli stack of $G_\C$-bundles over $\cp^1$, the above index can be written as\footnote{Here $S_tT_{\frak{M}}$ denotes the total symmetric power of the tangent complex of $\frak{M}$. }
\be\label{ITop=Ind}
\CI_{\text{top}}(q,t)=\mathrm{Ind}(\frak{M},S_tT_{\frak{M}}\otimes\CF_q\otimes L^p),
\ee
and can be further decomposed into a summation over different strata of $\frak{M}(\cp^1)$, given by Grothendieck's classification theorem.

For sufficiently large $p$, the contribution from the unstable strata all vanish. And the semi-stable stratum is the classifying stack $BG_\C$ (it could be viewed as a single point --- the moduli space --- with $G_\C$ as the stabilizer). As both $L^p$ and $\CF_q$ are trivial over $BG_\C$, both $q$ and $p$ dependence disappears, and the problem only depends on the choice of $G$. In this simple case, one can actually directly compute the cohomology groups (BPS states) that contribute to the index in \eqref{ITop=Ind}
\be
\CH_{\text{tw-BPS}}^\bullet=H^\bullet(\frak{M},S_tT_{\frak{M}}\otimes\CF_q\otimes L^p).
\ee
The tangent space of $BG_\C$ is given by the two-step complex $\frak{g}\rightarrow 0$ with $\frak{g}$ placed at degree $-1$. So in the end, one has
\be
\CH_{\text{tw-BPS}}^j=H^{-j}_{G}(\Lambda^{j}\frak{g}[j])=(\Lambda^{j}\frak{g})^G=H^{j}(G),
\ee
with the BPS states represented by the elements in the cohomolgy of the group $G$, or equivalently, elements in $(\Lambda^\bullet\frak{g})^{G}$. And, in this concrete example, one again finds that $t$-degree agrees with cohomological degree. The ``(co-)homological-flavor locking'' in this case is a result of the tangent complex $\frak{g}\rightarrow 0$ being concentrated in degree $-1$.

In fact, the above explanation using the stack language can be readily ``translated'' into a physics argument as follows. When $p$ is sufficiently large, in the topologically twisted index of $T[M_3]$ only
the zero-flux sector $m=0$ survives --- this is what mathematicians would call the ``semi-stable stratum.'' Then, the index essentially becomes the
index of the $\CN=2$ SQM (which can be considered as a reduction of a 2d $\CN=(0,2)$ theory) with gauge group $G$ and an adjoint Fermi multiplet
$\Psi$ of $U(1)_{\beta}$ charge 1. In particular, for $G=SU(2)$ we have
\begin{equation}
	\CI_\text{top}(q,t)=\frac{1}{2}\sum_{m\in \Z}\int\limits_\text{JK} \frac{dz}{2\pi i z}
	\,(1-z^{\pm2})(1-tz^{\pm 2})(1-t)=(1-t^3).
	\label{Itop-Lp1-SU2}
\end{equation}
In the IR, such theory should be effectively described by gauge-invariant combinations of Fermi-fields, corresponding precisely to generators of $G$-invariant part of the exterior algebra $(\Lambda\frak{g})^G$, which,
in turn, can be understood as generators of the cohomology of the Lie group.
For $G=SU(N)$ these correspond to $\Tr\Psi^3$, $\Tr\Psi^5$, ..., $\Tr\Psi^{2N-1}$ in the SQM.

Although the topological index looks much simpler compared to the superconformal index, it should also admit a factorization into the homological blocks via \eqref{Itop-decomp}. For $G=SU(2)$, we would like to check
\begin{equation}
		\CI_\text{top}(q,t)=\sum_{a\in \Z_p/\Z_2}
		|\CW_a|\,\hat{Z}_a(q,t)\hat{Z}_a(1/q,t).
		\label{Itop-factorization}
\end{equation}
It is easy to see how this works for $p=1$ ($M_3=S^3$):
\begin{equation}
	\hat{Z}_0(q,t)\cdot\hat{Z}_0(1/q,t)=
	\frac{1}{(t^2q^2;q)_\infty}\cdot\frac{1}{(t^2q^{-2};q^{-1})_\infty}=
	\frac{(t^2q^{-1};q)_\infty}{(t^2q^2;q)_\infty}=(t^2/q;q)_3.
\end{equation}
In the case $p>1$, again, the problem of defining $\hat{Z}_a(1/q,t)$ arises. It will be resolved for both superconformal and topologically twisted index in section \ref{sec:cyclotomic}.

\subsubsection{${t=q^\beta}$ reduction}
\label{sec:beta-reduction}

Consider the case of $G=SU(2)$. Instead of taking the unrefined limit $t\rightarrow 1$, one can consider more general\footnote{A similar reduction was considered in \cite{Aganagic:2011sg} for refined Chern-Simons theory as a way to circumvent certain technicalities.} limit $t\rightarrow q^\beta$, with $\beta\in\Z$. One can show that in such a limit the reduced (that is with removed contribution of the Cartan component of the adjoint chiral) the superconformal index given by formula (\ref{Itop-Lp1-SU2}) becomes a Laurent polynomial in $q$ or a rational function, depending on the sign of $\beta$:
\begin{equation}
	\CI^{(\beta)}(q)\equiv
	 \lim_{t\rightarrow q^\beta} \CI(q,t)\,\frac{(qt;q)_\infty}{(1/t;q)_\infty}\;\;\in
\left\{
\begin{array}{cl}
\Z[q,q^{-1}], & \beta\in \Z_+,\\
\Z(q), & \beta\in -\Z_+.
\end{array}
\right.
	 \label{ISU2-beta}
\end{equation}
Similarly, for topologically twisted index we have
\begin{equation}
	\CI^{(\beta)}_\top(q)\equiv
	 \lim_{t\rightarrow q^\beta} \CI_\top(q,t)\,(qt;q)_{-1}=
	  \lim_{t\rightarrow q^\beta} \frac{\CI_\top(q,t)}{1-t},
	 	 \label{ItopSU2-beta}
\end{equation}
while for the homological blocks we obtain
\begin{equation}
	\hat{Z}_a^{(\beta)}(q)\equiv
	 \left.{(qt;q)_\infty}\,\hat{Z}_a(q,t)\right|_{t=q^\beta}\;\;\in q^{\frac{a^2}{p}}
\cdot\,\left\{
\begin{array}{cl}
\Z[q], & \beta\in \Z_+,\\
\Z(q), & \beta\in -\Z_+.
\end{array}
\right.
	 \label{BlockBeta}
\end{equation}
Explicitly, we have
\begin{equation}
	\hat{Z}_a^{(\beta)}(q)=\frac{1}{|\CW_a|}\int\frac{dz}{2\pi i z}
	(z^2;q)_{\beta+1}(z^{-2};q)_{\beta+1}\,\sum_{n\in p\Z+a}q^{n^2/p}\,z^{2n}.
	\label{BlockBeta1}
\end{equation}
Taking into account that
\begin{equation}
	(q^{-1}t^{-1};q^{-1})_\infty=\frac{1}{(1/t;q)_\infty},
\end{equation}
the formula (\ref{Index-factorization}) then reduces to the following relation between \textit{Laurent polynomials} (for $\beta\in \Z_+$) or \textit{rational functions} (for $\beta\in-\Z_+$):
\begin{equation}
	\CI^{(\beta)}(q)=\sum_{a\in \Z_p/\Z_2}
	|\CW_a|\,\hat{Z}_a^{(\beta)}(q)\hat{Z}_a^{(\beta)}(1/q).
	\label{Ibeta-factorization}
\end{equation}
For the topologically twisted index we have instead
\begin{equation}
	\CI_\text{top}^{(\beta)}(q)=\sum_{a\in \Z_p/\Z_2}
	|\CW_a|\,\hat{Z}_a^{(\beta)}(q)\hat{Z}_a^{(-\beta)}(1/q).
	\label{Itopbeta-factorization}
\end{equation}
These relations are easy to check for various values of $p$ and $\beta$. For example, for $p=5$ we should check
\begin{equation}
	\CI^{(\beta)}(q)=2\hat{Z}_0^{(\beta)}(q)\hat{Z}_0^{(\beta)}(1/q)+
	\hat{Z}_1^{(\beta)}(q)\hat{Z}_1^{(\beta)}(1/q)+
	\hat{Z}_2^{(\beta)}(q)\hat{Z}_2^{(\beta)}(1/q)
\end{equation}
and
\begin{equation}
	\CI_\text{top}^{(\beta)}(q)=\hat{Z}_0^{(\beta)}(q)\cdot 2\hat{Z}_0^{(-\beta)}(1/q)+
	\hat{Z}_1^{(\beta)}(q)\hat{Z}_1^{(-\beta)}(1/q)+
	\hat{Z}_2^{(\beta)}(q)\hat{Z}_2^{(-\beta)}(1/q).
\end{equation}
The particular expressions for a few first values of $\beta$ are as follows.

For $\beta=0$,
\begin{equation}	
	\begin{array}{c}
	
	\hat{Z}_0^{(0)}(q)=1,\;
	\hat{Z}_1^{(0)}(q)=-q^{1/5},\;
	\hat{Z}_2^{(0)}(q)=0.\; \\
		\CI^{(0)}(0)=3 ,\\
	\CI_\text{top}^{(0)}(0)=3.
	\end{array}
\end{equation}

For $\beta=1$,\footnote{Notice that after removing a $(1-t)$ factor from $\CI^{\text{top}}(q,t)$, we are left with $(1-t^3)/(1-t)=1+t+t^2$.}
\begin{equation}		
	\begin{array}{c}

	\hat{Z}_0^{(1)}(q)=1+q+q^2,\;
	\hat{Z}_1^{(1)}(q)=-q^{1/5}(1+2q+q^2),\;
	\hat{Z}_2^{(1)}(q)=q^{9/5}.\;
	 \\
	2\hat{Z}_0^{(-1)}(q)=1,\;
	\hat{Z}_1^{(-1)}(q)=0,\;
	\hat{Z}_2^{(-1)}(q)=0.\;
	 \\
	 \CI^{(1)}(q)=\cfrac{3}{q^2}+\cfrac{8}{q}+ 13+8\,q+3\,q^2,
	 \\
\CI_\text{top}^{(1)}(q)=1+q+q^2 .
	\end{array}
\end{equation}

For $\beta=2$,
\begin{equation}
	\begin{array}{c}

	 	\hat{Z}_0^{(2)}(q)=1+q+2q^2+2q^3+2q^4+q^5+q^6,\\	
	 	\hat{Z}_1^{(2)}(q)=-q^{1/5}(1+2q+3q^2+3q^3+3q^4+2q^5+q^6),\\
	 	\hat{Z}_2^{(2)}(q)=q^{9/5}(1+q+2q+q^4).
	 	 \\
	2\hat{Z}_0^{(-2)}(q)=-q^2/(1-q^2),\\	
	\hat{Z}_1^{(-2)}(q)=-q^{6/5}/(1-q^2),\\
	\hat{Z}_2^{(-2)}(q)=-q^{4/5}/(1-q^2).
	 \\
	 \CI^{(2)}(q)=\cfrac{3}{q^6}+\cfrac{8}{q^5}+\cfrac{21}{q^4}
	 +\cfrac{35}{q^3}+\cfrac{55}{q^2}+\cfrac{65}{q}+76+65q+55q^2+35q^3+21q^4+8q^5+3q^6,
	\\
\CI^{(2)}_\text{top}(q)=1+q^2+q^4.
	\end{array}
\end{equation}

And	for $\beta=3$,
\begin{equation}
	\begin{array}{c}
	 	\hat{Z}_0^{(3)}(q)=q^{12}+q^{11}+2 q^{10}+3 q^9+4 q^8+4 q^7+5 q^6+4 q^5+4 q^4+3 q^3+2 q^2+q+1,\\	
	 	\hat{Z}_1^{(3)}(q)=-q^{1/5} \left(q^{12}+2 q^{11}+3 q^{10}+4 q^9+6 q^8+7 q^7+8 q^6+7 q^5+6 q^4+5 q^3+3 q^2+2 q+1\right),\\
	 	\hat{Z}_2^{(3)}(q)=q^{9/5}(q^{10}+2 q^8+2 q^7+2 q^6+3 q^5+3 q^4+2 q^3+3 q^2+q+1).
	 	 \\
	2\hat{Z}_0^{(-3)}(q)=(q^7+q^5+q^4+q^3+q^2)/\left((1-q^3)(1-q^4)\right),\\	
	\hat{Z}_1^{(-3)}(q)=q^{11/5} \left(q^4+q^3+q^2+q+1\right)/\left((1-q^3)(1-q^4)\right),\\
	\hat{Z}_2^{(-3)}(q)=q^{14/5} \left(q^3+q^2+2 q+1\right)/\left((1-q^3)(1-q^4)\right).
	 \\
\CI^{(3)}_\text{top}(q)=1+q^3+q^6
	\end{array}
\end{equation}
\begin{equation*}
	\cdots
\end{equation*}
Notice that the coefficients of the $q$-expansions are always \emph{positive} (up to an overall sign), and this reflects that the BPS space $\CH^{(\beta)}_a$ obtained from the spectral sequence \eqref{BPSSpec}, now doubly-graded by the eigenvalues under the $U(1)_q\times U(1)_R$ action,
\be
\CH^{(\beta)}_a=\bigoplus_{i,j\in \Z}\CH^{(\beta)i,j}_a,
\ee
enjoys some highly non-trivial properties --- subspaces with odd $j$-degree are either all empty, or have their contributions canceled when we take the equivariant Euler characteristic. Moreover, the spectral sequence \eqref{BPSSpec} is consistent with the following property, which one can check explicitly order by order in $q$:
\begin{equation}
(-qt;q)_\infty \hat{Z}_a(q,-t) = \hat{Z}_a^{(\beta)}(q)+(1+q^\beta t^{-1})P_+(q,t)
\label{beta-Poincare-seq}
\end{equation}
where all the coefficient are positive (up to an overall common sign). Namely,
\begin{equation}
(-qt;q)_\infty \hat{Z}_a(q,-t),\,P_+(q,t) \in \pm\Z_+[t][[q]],\qquad \hat{Z}_a(q)\in \pm\Z_+[q]
\end{equation}
The second term in (\ref{beta-Poincare-seq}) represent the pairs of states that go away in the cohomology w.r.t. the deformed supercharge.

Also, the complexity of the blocks $\hat{Z}^{(\beta)}_a$ and $\hat{Z}^{(-\beta)}_a$ will grow with $\beta$, this is expected from the discussion in section~\ref{sec:Spec} via spectral sequence, where we have seen more cancellations for smaller $\beta$ and fewer ones for larger $\beta$.

\subsubsection{Cyclotomic expansion}
\label{sec:cyclotomic}

In this secition we show how to obtain expression for $\hat{Z}_a(1/q,t)$ with $|q|<1$ using what is usually called the \textit{cyclotomic expansion}.

Let us first consider the general construction. Let $F(q,t)$ be some function of $q$ and $t$,
and suppose we know its restrictions at $t=q^\beta$, with $\beta\in\Z$,
\begin{equation}
	F^{(\beta)}(q)\equiv F(q,q^\beta).
\end{equation}
Then, one can write the following formal cyclotomic-like expansion for $F(q,t)$:
\begin{equation}
	F(q,t)=\sum_{m=0}^\infty\tilde{F}^{(m)}(q)\cdot\frac{t^m(1/t;q)_m}{(q;q)_m}.
	\label{cyclotomic}
\end{equation}
The coefficient of the cyclotomic expansion, $\tilde{F}^{(m)}(q)$, are  related to ${F}^{(m)}(q)$ by a ``triangular'' linear transform:
\begin{equation}
	\begin{array}{rcl}
		\tilde{F}^{(0)}(q) & = & F^{(0)}(q)\\
		\tilde{F}^{(1)}(q) & = & F^{(0)}(q)-F^{(1)}(q)\\
		\tilde{F}^{(2)}(q) & = & F^{(0)}(q)-(1+q^{-1})F^{(1)}(q)+q^{-1}F^{(2)}(q)\\
		\tilde{F}^{(3)}(q) & = & F^{(0)}(q)-(1+q^{-1}+q^{-2})F^{(1)}(q)
		+(q^{-1}+q^{-2}+q^{-3})F^{(2)}(q)-q^{-4}F^{(3)}(q)\\
		\ldots
	\end{array}
	\label{Cyc-transform}
\end{equation}
Note, that the denominators $(q;q)_m$ in (\ref{cyclotomic}) are introduced for convenience, so that
\begin{equation}
	F^{(\beta)}(q)\;\in\Z[q,q^{-1}]\qquad\Longleftrightarrow \qquad\tilde{F}^{(m)}(q) \in\Z[q,q^{-1}]
\end{equation}
but in principle they can be absorbed into the definition of cyclotomic coefficients $\tilde{F}^{(m)}(q)$.

If the minimal degree of $q$ in $\tilde{F}^{(m)}(q)$ grows with $m$, one can produce a $q$-expansion from the cyclotomic expansion (\ref{cyclotomic}). However, such procedure is quite formal and in general might not actually give the actual $q$-expansion that converges to $F(q,t)$ when $|q|<1$. But it can be done for homological blocks. In particular, from (\ref{BlockBeta}) one obtains the following cyclotomic-like expansion:
\begin{equation}
	\hat{Z}_a(q,t)=\frac{1}{(qt;q)_\infty}\sum_{m=0}^\infty\tilde{\hat{Z}}_a^{(m)}(q)\cdot\frac{t^m(1/t;q)_m}{(q;q)_m}
	\label{Zcyclotomic}
\end{equation}
where the coefficients
\begin{equation}
	\tilde{\hat{Z}}_a^{(m)}(q)\;\;\in\Z[q]
\end{equation}
 turn out to be polynomials in $q$ (non-Laurent) with minimal power growing linearly in $m$. Expanding each term in $q$ then gives us a $q$-series of the following form:
\begin{equation}
	\hat{Z}_a(q,t)=\sum_{n= 0}^\infty P_n(t)q^n,\qquad P_n(t)\in \Z[t],
	\label{Block-qexp}
\end{equation}
which coincides with the $q$-series that can be produced directly from (\ref{TM3mod}).
On the other hand, since $P_m(t)$ contain only positive powers of $t$, one can produce $\hat{Z}^{(\beta)}(q)$ --- or, equivalently, the coefficients of the cyclotomic expansion $\tilde{\hat{Z}}^{(\beta)}(q)$ --- from (\ref{TM3mod}) just by making the substitution $t=q^\beta$ and observing that the resulting $q$-series truncates.

Therefore, we assume that for anti-blocks one can write similarly:
\begin{equation}
	\hat{Z}_a(1/q,t)=(1/t;q)_\infty\sum_{m=0}^\infty\tilde{\hat{Z}}_a^{(-m)}(1/q)\cdot\frac{t^m(1/t;q)_m}{(q;q)_m}
	\label{AZcyclotomic}
\end{equation}
where $\tilde{\hat{Z}}_a^{(-m)}(1/q)$ are related to ${\hat{Z}}_a^{(-\beta)}(1/q)\equiv (q^{-1}t;q^{-1})_\infty\hat{Z}_a(q^{-1},t)|_{t=q^\beta}$ in the same way as $\tilde{F}^{(m)}(q)$ are related to $F^{(\beta)}(q)$ in (\ref{Cyc-transform}).

For example, in the case $p=5$, using expressions for $\hat{Z}_a^{(-\beta)}(q)$ calculated in Section \ref{sec:beta-reduction} we have
\begin{equation}
	\begin{array}{rl}
		2\hat{Z}_0(1/q,t)=&\left(-t^3+t^2-t+1\right)+\left(t^4-3 t^3+5 t^2-3 t\right)
   q+\left(-t^5+6 t^4-11 t^3+11 t^2-6 t+1\right) q^2\\ &\qquad+\left(t^6-6
   t^5+17 t^4-28 t^3+26 t^2-12 t+2\right)
   q^3+O\left(q^4\right),\vspace{1ex}
   \\
		\hat{Z}_1(1/q,t)=&q^{-1/5}\Big(\left(t^2-t\right)+\left(t^4-3 t^3+3
   t^2-2 t+1\right) q+\left(-t^5+4 t^4-9 t^3+10 t^2-5 t+1\right)
   q^2 \\& \qquad+\left(-4 t^5+15 t^4-24 t^3+21 t^2-10 t+2\right)
   q^3+O\left(q^{4}\right)\Big),\vspace{1ex}
   \\
		\hat{Z}_2(1/q,t)=&q^{1/5}\Big(\left(-t^3+2 t^2-t\right)+\left(2
   t^4-5 t^3+5 t^2-3 t+1\right) q+ \\& \qquad+\left(-2 t^5+7 t^4-13 t^3+14
   t^2-7 t+1\right) q^2+O\left(q^3\right)\Big).
	\end{array}
\end{equation}
Now we can check (\ref{Index-factorization}) and (\ref{Itop-factorization}) directly. From (\ref{I-Lp1-SU2}), or equivalently from (\ref{Zcyclotomic}), we have
\begin{equation}
	\begin{array}{rl}
		\hat{Z}_0(q,t)&=1+q+(2-t+t^2)q^2+(3-2t+2t^2)q^3+O(q^4),\vspace{1ex}\\
		\hat{Z}_1(q,t)&=q^{1/5}\left(-1+(-2+t)q+(-3+2t-2t^2)q^2+(-5+4t-3t^2+t^3)q^3+O(q^4)\right),\vspace{1ex}\\
		\hat{Z}_2(q,t)&=q^{4/5}\left((1-t)q+(1-2t+t^2)q^2+O(q^3)\right).
	\end{array}	
\end{equation}
And from (\ref{I-Lp1-SU2}) we have:
\begin{multline}
	\CI(q,t)=\left(1-\frac{1}{t^3}\right)+\left(-1+\frac{1}{t}+\frac{1}{t^2}-\frac{1}{t^3}\right)
   q+\left(t^2-2+\frac{1}{t}-\frac{1}{t^3}+\frac{1}{t^4}\right)
   q^2+\\+\left(\frac{1}{t}-\frac{1}{t^2}-\frac{1}{t^5}+\frac{1}{t^6}
   \right) q^3+O\left(q^4\right).
\end{multline}
Then, one can check explicitly that indeed
\begin{equation}
	\CI(q,t) \; = \; \hat{Z}_0(q,t)\cdot 2\hat{Z}_0(1/q,1/t)+
	\hat{Z}_1(q,t)\hat{Z}_1(1/q,1/t)+
	\hat{Z}_2(q,t)\hat{Z}_2(1/q,1/t) \,.
\end{equation}
Similarly, the topologically twisted index,
\begin{equation}
	\CI_\text{top}(q,t) \; = \; 1-t^3 \,,
\end{equation}
can indeed be decomposed as
\begin{equation}
	\CI_\text{top}(q,t) \; = \; \hat{Z}_0(q,t)\cdot 2\hat{Z}_0(1/q,t)+
	\hat{Z}_1(q,t)\hat{Z}_1(1/q,t)+
	\hat{Z}_2(q,t)\hat{Z}_2(1/q,t) \,.
\end{equation}

\subsubsection{Positivity of coefficients}

In \cite{Gukov:2016gkn}, it was shown that, up to an overall $\pm$ sign,
the refined homological blocks (with or without contribution from the Cartan component of the adjoint chiral) have the following positivity property:
\beq\label{block-pos}
\pm \hat{Z}_a(q,-t) \quad & \in \quad q^{\Delta_a} \Z_+[t][[q]] \\
\pm (-qt;q)_\infty\hat{Z}_a(q,-t) \quad & \in \quad q^{\Delta_a} \Z_+[t][[q]]
\eeq
so that it is naturally to conjecture that $\pm \hat{Z}_a(q,-t)$ actually coincides with the Poincar\'e polynomial of the underlying doubly graded homology (after a shift in overall degrees),
\begin{equation}
\CH_a[M_3] \; = \; \bigoplus_{i\in \Z+\Delta_a\atop j\in \Z} \CH_a^{i,j}.
\end{equation}
This is equivalent to the statement that the triply graded homology (refined by the flavor charge) is supported only on the ``diagonal''
\begin{equation}
\CH_a^{i,j;\ell}=\delta_{j,\ell}\CH_a^{i,j}.
\end{equation}
Since we have advertised that the superconformal/twisted index and their categorifications are ``better-behaved,'' one may ask whether they have a similar positivity property. The answer is affirmative.

\paragraph{Superconformal index.}

In contrast to the homological blocks, the coefficients of $\CI(q,t)$ are not all positive/negative and there does not seem to be an easy change of variables which achieves that (contrary to what happend for $\hat{Z}_a(q,t)$, see \cite{Gukov:2016gkn}). However, there seem to be an easy factor which makes them positive. In particular, in the case of $G=SU(2)$ and $p>1$, one observes that
\begin{equation}
	\CI_{T[L(p,1),SU(2)]}(q,t) \; = \; P(q,t) \cdot (t^{-1};q)_\infty\,\cdot (1-qt)
\end{equation}
where
\begin{equation}
	P(q,t) \; \in \; \Z_+[t][[q]] \,
\end{equation}
is a positive power series. So, naively one can expect that the spectrum of BPS operators of $T[L(p,1),SU(2)]$ is given by cohomology
\begin{equation}\label{PositiveFactor}
	\CH_\text{BPS} \; = \; H^*( \CH_{P} \otimes \CH_\text{Fermi} , d)
\end{equation}
for some differential $d$, where $\CH_P$ categorifies $P(q,t)$ and $\CH_\text{Fermi}$ is the space
generated by a tower of fermions with the index $(t^{-1};q)_\infty$ and one single fermions with index $(1-tq)$. And one would expect the tower is related to $\bar{\psi}^3_+$, the Cartan component of $\bar{\psi}_+$ in the (anti-)chiral multiplet, and its derivatives.

\paragraph{Twisted index.}

The twisted index also shares this positivity property, which can be understood in greater depth. For $T[L(p,1);G]$ with large $p$, we know from the discussion in section~\ref{sec:TopInd} that
\be
\CH_\text{tw-BPS}^\bullet=H^\bullet(G)=(\Lambda^\bullet \frak{g})^G.
\ee
Then the equivariant Euler characteristics will factorizes into a positive part times a simple factor,
\be\label{TopIndPos}
\CI_{\text{top}}=\prod_i^{\mathrm{rk}\,G}(1-t^{2m_i+1})=(1-t)^{\mathrm{rk}\,G}\cdot \prod_i^{\mathrm{rk}\,G}(1+t+\ldots +t^{2m_i}),
\ee
where $m_i$ are exponents of the group $G$. At the level of homological invariants, one may be tempted to factor out the exterior algebra of the Cartan subalgebra $\frak{t}\subset\frak{g}$
\be
(\Lambda \frak{g})^G\sim\Lambda \frak{t} \cdot (\Lambda \frak{g})^G/\Lambda \frak{t}.
\ee
However, the last quotient doesn't make sense as $\Lambda \frak{t}$ is not a submodule of $(\Lambda \frak{g})^G$ --- this parallels the discussion for the superconformal index where the gauge-non-invariant operator $\bar{\psi}^3_+$ plays the role of $\frak{t}$ here. Instead, one has to consider the Leray spectral sequence
\be
E_2^{\bullet,\bullet}=H^{\bullet}(G/T,H^\bullet(T)) \Rightarrow H^\bullet(G)
\ee
associated with the fibration
\be
T\rightarrow G\rightarrow G/T.
\ee
The $E_2$ page factorizes as
\be
H^\bullet(T)\otimes H^{\bullet}(G/T)
\ee
where $H^\bullet(T)$ is fermionic (generated by elements in $H^1(T)$) while $H^{\bullet}(G/T)$ is bosonic (with only even-degree cohomology groups). But the differentials in the spectral sequence are no longer trivial. And this is the reason that we expect \eqref{PositiveFactor} as opposed to a direct factorization of $\CH_{\text{BPS}}$.

\paragraph{A summary.}

Up to now, we have encountered several positivity-related phenomena, similar but each with its own flavor, and it may be worthwhile to summarize and compare.
\begin{enumerate}
	
	\item Positivity of homological blocks $\hat{Z}_a$. Besides overall signs, all homological blocks are positive in variables $q$ and $-t$. This is likely due to ``homological-flavor locking'' --- the homological degree is not independent from the flavor degree and one completely determines the other, {\it cf.} \cite[sec.3.4]{Gukov:2016gkn}. We have already illustrated how this happens for $T[S^3]$ in section~\ref{sec:S3}, and expect this to be a general phenomena. In fact, such behavior was already observed for homological invariants of knot \cite{Aganagic:2011sg,Gorsky:2013jxa,Cherednik:2011nr}.

	\item Positivity of the superconformal index $\CI$ and twisted index $\CI_{\text{top}}$. After factorizing out a ``fermionic factor,'' all coefficients appearing in the two indices will be positive. This may seems to be qualitatively different from 1, where we have positivity in $(q,-t)$-variable as opposed to $(q,t)$. However, the example of $\CI_\text{top}(T[L(p,1)])$ with large $p$ suggests that 1 and 2 are intimately related. Namely, from \eqref{TopIndPos}, the twisted index in this case is
\be
\CI_{\text{top}}=P_{-t}(G)=\prod_i^{\mathrm{rk}\,G}(1-t^{2m_i+1})=(1-t)^{\mathrm{rk}\,G}\cdot (\text{positive polynomial}),
\ee
where one sees that the third quantity is positive in $-t$ via ``homological-flavor locking'', while removal of a factor will achieve strict positivity in the fourth quantity.

	\item Positivity of ``$\beta$-reduced'' homological blocks $\hat{Z}^{(\beta)}_a$ and indices $\CI^{(\beta)}$, $\CI^{(\beta)}_{\text{top}}$. For $\CI^{(\beta)}$ and $\CI^{(\beta)}_{\text{top}}$, their positivity directly follows from 2 by setting $t=q^\beta$. On the other hand, positivity for $\hat{Z}^{(\beta)}_a$ is not \textit{a priori} obvious and require non-trivial cancellations that also involve the $q$-degree. This hints at the non-trivial role played by the $q$-grading in the ``homological-flavor locking.''
\end{enumerate}

\subsubsection{Comparison with refined CS}
\label{sec:refCS}

Let $R_n$ denote by a representation of $SU(2)$ of dimension $n$. The Hilbert space of the refined
Chern-Simons theory \cite{Aganagic:2011sg} on $T^2$ is the same as in the unrefined case.
Namely, it is generated by integral representations of the affine Kac-Moody algebra at ``bare'' level $k-2$,
which in our notations\footnote{Our notations slightly differ from those in \cite{Aganagic:2011sg}:
\begin{equation}
	\begin{array}{c}
		k_\text{here} \; = \; k_\text{AS}+2, \\
		\beta_\text{here} \; = \; \beta_\text{AS}-1, \\
		t_\text{here} \; = \; t_\text{AS}/q.
	\end{array}		
\end{equation}
} have $n = 1,\ldots,k-1$:
\begin{equation}
	\CH_{T^2}=\bigoplus_{n=1}^{k-1} \C |R_n\rangle.
\end{equation}
Let
\begin{equation}
	q=e^{\frac{2\pi i}{k+2\beta}},\quad\text{and} \quad t=q^\beta.
\end{equation}
The $T$ and $S$ matrices read (up to simple $q,t$-independent phase factors)
\begin{equation}
	T_{nm}\equiv \langle R_n |T|R_m\rangle=
	\delta_{nm}\cdot q^{\frac{n^2}{4}}t^{\frac{n}{2}+\frac{\beta}{4}}
\end{equation}
\begin{equation}
	S_{nm}\equiv \langle R_n |S|R_m\rangle=d_{R_n}(t)M_{R_m}(tq^n)
\end{equation}
where the refined quantum dimension $d_{R_n}$ is given by
\begin{equation}
	d_{R_n}(tq)=\frac{1}{(2k+4\beta)^{1/2}}
		\prod_{m=0}^{\beta} (q^{-(m+1)/2}t^{-1/2}-q^{(m+1)/2}t^{1/2})M_{R_n}(tq)
\end{equation}
and the $SU(2)$ Macdonald polynomials are given by\footnote{As usual, $q$-numbers are defined as \begin{equation}[n]_q \equiv \frac{q^{n/2}-q^{-n/2}}{q^{1/2}-q^{-1/2}}. \end{equation}}
\begin{equation}
	M_{R_n}(x)=\sum_{j=0}^{n-1}x^{j-n/2+1/2}
	\prod_{i=1}^j\frac{[n-i]_q[i+\beta]_q}{[n-i+\beta]_q[i]_q},
\end{equation}
which at the special value of the argument simplifies to
\begin{equation}
	M_{R_n}(tq)=\prod_{m=0}^\beta
	\frac{q^{-(m+n)/2}t^{-1/2}-q^{(m+n)/2}t^{1/2}}{q^{-(m+1)/2}t^{-1/2}-q^{(m+1)/2}t^{1/2}}.
\end{equation}
The scalar product on $\CH_{T^2}$ is given by
\begin{equation}
	\langle R_n|R_m\rangle = g_n \delta_{nm},
\end{equation}
\begin{equation}
	g_n=\prod_{m=0}^\beta\frac{q^{-(n+m)/2}t^{-1/2}-q^{(n+m)/2}t^{1/2}}{q^{-(n-m)/2}t^{-1/2}-q^{(n-m)/2}t^{1/2}}.
\end{equation}
The refined CS partition function on $M_3=L(p,1)$ is then given by
\begin{multline}
	Z_{SU(2)}^\text{ref. CS}[L(p,1)]=\sum_{n=1}^{k-1} g^{-1}_n\,(S_{1n})^2(T_{nn})^{-p}\\
	=(2k+4\beta)^{-1}
	\sum_{n=1}^{k-1}q^{-\frac{pn^2}{4}-\frac{pn\beta}{2}-\frac{p\beta^2}{4}}
	\prod_{j=0}^\beta (q^{-\frac{n+j+\beta}{2}}-q^{\frac{n+j+\beta}{2}})
	(q^{-\frac{n-j+\beta}{2}}-q^{\frac{n-j+\beta}{2}}).
\end{multline}
Changing the summation variable from $n$ to $m=n+\beta$ gives
\begin{multline}
	(2k+4\beta)\,Z_{SU(2)}^\text{ref. CS}[L(p,1)]=
	\sum_{m=\beta+1}^{k+\beta-1}q^{-\frac{pm^2}{4}}
	\prod_{j=0}^\beta (q^{-\frac{m+j}{2}}-q^{\frac{m+j}{2}})
	(q^{-\frac{m-j}{2}}-q^{\frac{m-j}{2}})\\
	=\frac{(-1)^\beta q^{-\frac{\beta(\beta+1)}{2}} }{2}
	\sum_{m\in \Z_{2(k+2\beta)}}q^{-\frac{pm^2}{4}}\prod_{j=0}^\beta (1-q^{j+m})(1-q^{j-m}),
	\label{ref-Lp1-1}
\end{multline}
where we have extended the range of summation using the parity and vanishing properties of the summand.
Using the Gauss sum reciprocity formula (which is a particular case of (\ref{reciprocity}))
\begin{equation}
	\sum_{m\in \Z_{2(k+2\beta)}}q^{-\frac{pm^2}{4}}\cdot q^{\frac{n\ell}{2}}=
	\frac{e^{-\frac{\pi i}{4}}(2k+4\beta)^{1/2}}{|p|^{1/2}}
	\sum_{a\in \Z_p} e^{2\pi i (k+2\beta)\frac{a^2}{p}}\cdot
	e^{2\pi i \frac{a\ell}{p}}\cdot
	q^{\frac{\ell^2}{4p}},
\end{equation}
we can rewrite (\ref{ref-Lp1-1}) as follows:
\begin{multline}
	(2k+4\beta)^{1/2}\,Z_{SU(2)}^\text{ref. CS}[L(p,1)]=\\
	=t^{-\frac{\beta+1}{2}}\sum_{a,b\in \Z_p /\Z_2} e^{2\pi i (k+2\beta)\text{CS}(a)}\cdot
	S_{ab}\cdot \frac{1}{|\CW_b|}\,\CL_{b}^{(p)}\left[\prod_{j=0}^\beta (1-q^{j}z^2)(1-q^{j}z^{-2})\right],
\end{multline}
where, again, we drop the overall phase factor
and $\text{CS}(a)$ is the CS invariant of a flat connection $a$.
As before, the $S$-transform is given by\footnote{Notice, that this $S$-transform matrix is different from the $S$-matrix in refined Chern-Simons theory!}
\begin{equation}
	S_{ab}=\frac{e^{2\pi i\frac{ab}{p}}+e^{-2\pi i\frac{ab}{p}}}{|\CW_a|}
	\label{SZp}
\end{equation}
 and $\CL_{b}^{(p)}$ is the ``Laplace transform'':
\begin{equation}
	\begin{array}{cccc}
	\CL_{b}^{(p)}:& \Z[q]\otimes \Z[z^2,z^{-2}]&\longrightarrow & q^{\frac{b^2}{p}}\Z[q],
	\vspace{2ex}
	\\
	& z^{2\ell}&\longmapsto &
	\left\{\begin{array}{cl} q^{\ell^2/p}, & \ell=b \mod p, \\ 0, & \text{otherwise.}  \end{array}\right.
	\end{array}
	\label{LaplaceZp}
\end{equation}
Equivalently, it can be realized as follows:
\begin{equation}
	\CL_{b}^{(p)}:\qquad f(z) \longmapsto
	\int\frac{dz}{2\pi i z}\,
	f(z^{-1})\,\sum_{n\in p\Z+b}q^{n^2/p}\,z^{2n}.
\end{equation}
Therefore we have
\begin{equation}
	(2k+4\beta)^{1/2}\,Z_{SU(2)}^\text{ref. CS}[L(p,1)]
	=t^{-\frac{\beta+1}{2}}\sum_{a,b\in \Z_p /\Z_2} e^{2\pi i (k+2\beta)\text{CS}(a)}\cdot
	S_{ab}\cdot \hat{Z}_b^{(\beta)}(q)
	\label{ref-WRT-decomp}
\end{equation}
where $\hat{Z}^{(\beta)}_b(q)$ are the same as homological blocks (\ref{BlockBeta1}) obtained as the $D^2\times_q S^1$ partition function of $T[L(p,1),SU(2)]$,
\begin{equation}
	\hat{Z}^{(\beta)}_b(q)=
	\frac{1}{|\CW_b|}\left.\int\frac{dz}{2\pi i z}\,
	\frac{(z^2;q)_\infty(z^{-2};q)_\infty}{(z^2tq;q)_\infty(z^{-2}tq;q)_\infty}\,\sum_{n\in p\Z+b}q^{n^2/p}\,z^{2n}\right|_{t=q^\beta}.
\end{equation}
Therefore, we just showed that (\ref{ref-WRT-decomp}) can be understood as a refined version of the decomposition (\ref{WRT-decomposition}) of the WRT invariant into homological blocks $\hat{Z}_a(q)$.

\subsection{${M_3=O(-p)\rightarrow \Sigma_g}$}

Now, we consider a class of three-manifolds $M_3$ that could be viewed as a natural generalization of $L(p,1)$ --- the total space of a degree-$(-p)$ circle bundle over a genus-$g$ Riemann surface. We expect the 3d $\CN=2$ theory $T[M_3,SU(2)]$ to have the following Lagrangian description:
\begin{equation}
	T[M_3] \quad = \quad
	\boxed{\begin{array}{c}
	\CN=2 \text{ level-$p$ Chern-Simons term }\\
	+~\text{adjoint chiral $\Sigma$ with R-charge 2 }\\
	+~\text{$2g$ adjoint chirals $(\Phi_i,\tilde{\Phi}_i)_{i=1}^g$ with R-charge 0}.
	\end{array}}
	\label{TCircFib}
\end{equation}
When $p=0$, the theory has $\CN=4$ supersymmetry and coincides with the mirror description of the 3d reduction of $T[\Sigma_g]$ \cite{Benini:2010uu}. Having $p\neq 0$ results in $p$ units of RR 2-form flux after reducing the M5-branes first on the circle fiber, which then give rise to a level-$p$ Chern-Simons term via D4-brane world-volume couplings to RR fields. The assignment of R-symmetry charges is consistent with the cubic superpotential coupling $\Tr\,\tilde{\Phi_i}\Sigma\Phi_i$ and reproduces the the correct R-charge assignment for the Lens space theory $T[L(p,1)]$ when $g=0$.

When $p=0$, the theory $T[M_3]$, considered as an $\CN=2$ theory has $U(1)_{\beta}\times USp(2g)$ symmetry \cite{Benini:2010uu}, where $USp(2g)$ rotates $2g$ adjoint chirals and $U(1)_{\beta}$ is the anti-diagonal of $SU(2)\times SU(2)$ R-symmetry, with respect to which the fields $\Phi_i$, $\tilde{\Phi}_i$ and $\Sigma$ have charges $-1$, 0 and 1 respectively.\footnote{This is because $\Phi_i$ and $\tilde{\Phi}_i$ are combined into an $\CN=4$ \emph{twisted} hypermultiplet \cite{Benini:2010uu} and transform under the $SU(2)\times SU(2)$ in the same way as a vector multiplet. As a consequence, $U(1)_{\beta}$ only acts on $\Phi_i$, not $\tilde{\Phi}_i$.} When $p\neq 0$, the CS interaction does not break these symmetries, but the $U(1)_{t'}$ R-symmetry of the $\CN=4$ theory with $p=0$ will become the flavor symmetry $U(1)_\beta$ of the $\CN=2$ theory with $p\neq 0$. For our purposes, it is more convenient to consider instead $U(1)_{\beta}\times U(g)$ maximal rank subgroup with the following charge assignments:
\begin{equation}
	\begin{array}{c|cc}
		 & U(1)_{\beta} & U(g) \\
		\hline
		\Sigma & 1 & \mathbf{1} \\
		\Phi_i & -1 & \mathbf{\bar g} \\
		\tilde\Phi_i & 0 & \mathbf{g}
	\end{array}\,
\end{equation}

\subsubsection{Homological blocks and WRT invariant}
Consider the partition function of $T[M_3]$ on $D^2\times S^1$ with the following 2d $\CN=(0,2)$ boundary conditions (which generalize the boundary conditions for $T[L(p,1)]$ that produce $\hat{Z}_a(q,t)$):
\begin{itemize}
	\item Neumann for $\CN=2$ vector multiplet,
	\item Neumann for adjoint chiral $\Sigma$,
	\item Dirichlet for $g$ adjoint chirals $\Phi_i$,
	\item Neumann for $g$ adjoint chirals $\tilde\Phi_i$,
	\item 3d vector multiplet is coupled to modules of $\Z_2$-orbifolded $\sqrt{2p}\Z$ lattice VOA by gauging its flavor symmetry. Note that when $p=0$ the boundary conditions preserve 2d $\CN=(2,2)$ supersymmetry. This cancels gauge anomaly inflow due to the level-$p$ CS term.
\end{itemize}
The $D^2\times_q S^1$ partition then reads \cite{Yoshida:2014ssa}
\begin{multline}
	\hat{Z}_a(q,t,u)\equiv Z_{T[M_3]}(D^2_a\times_q S^1)= \frac{1}{|\CW_a|}\int_\gamma \frac{dz}{2\pi i z}
	\frac{(z^{\pm2};q)_\infty(q;q)_\infty}{(z^{\pm2}tq;q)_\infty(tq;q)_\infty}
	\times\\
\prod_{i=1}^g	\frac{(z^{\pm2}tu_iq;q)_\infty(tu_iq;q)_\infty}{(z^{\pm2}u_i;q)_\infty(u_i;q)_\infty}
\times\,\frac{\sum_{n\in p\Z+a}q^{n^2/p}\,z^{2n}}{(q;q)_\infty}.	
	\label{TCircFibBlock}
\end{multline}
The choice of the contour $\gamma$, which corresponds to a precise choice of the boundary condition for the vector multiplet
(or, equivalently, Lagrangian submanifold in the complexified maximal torus of $SU(2)$), will be discussed below.

The WRT invariant of $M_3$ can be computed as follows in terms of $S$ and $T$ matrices\footnote{We omit simple overall factors below.}:
\begin{equation}
	Z_{SU(2)_k}^\text{CS}[M_3]=\sum_{n=1}^{k-1} (S_{1n})^{2-2g}(T_{nn})^{-p}=
	{\sum}'_{n\in \Z_{2k}} (q^{n/2}-q^{-n/2})^{2-2g}q^{-\frac{pn^2}{4}}
\end{equation}
where we use notation such that matrix indices are ``colors'' (= dimensions of representations),
$q=e^{\frac{2\pi i}{k}}$ and a prime over the sum means that we skip singular terms.
After applying the Gauss reciprocity formula
\begin{equation}
	\sum_{m\in \Z_{2k}}q^{-\frac{pm^2}{4}}\cdot q^{\frac{n\ell}{2}} \; = \;
	\frac{e^{-\frac{\pi i}{4}}(2k)^{1/2}}{|p|^{1/2}}
	\sum_{a\in \Z_p} e^{2\pi i k\frac{a^2}{p}}\cdot
	e^{2\pi i \frac{a\ell}{p}}\cdot
	q^{\frac{\ell^2}{4p}},
\end{equation}
we arrive at
\begin{equation}
	Z_{SU(2)_k}^\text{CS}[M_3]
	\; = \; \sum_{a,b\in \Z_p /\Z_2} e^{2\pi ik\text{CS}(a)}\cdot
	S_{ab}\cdot \frac{1}{|\CW_b|}\,\left.\hat{Z}_b(q)\right|_{q\rightarrow e^{\frac{2\pi i}{k}}}
\end{equation}
where the $S$-matrix is given in (\ref{SZp}) and, \textit{formally},
\begin{equation}
	\hat{Z}_a(q) \; = \; \CL_a^{(p)}\left[(z-z^{-1})^{2-2g}\right]
\end{equation}
with Laplace transform given in (\ref{LaplaceZp}). To be precise, it can be computed using
\begin{equation}
	\hat{Z}_a(q)=\frac{1}{|\CW_a|}\cdot\text{v.p.}
	\int_{|z|=1}(z-z^{-1})^{2-2g}\,\sum_{n\in p\Z+a}q^{n^2/p}\,z^{2n}
	\label{WRTCircFibBlock}
\end{equation}
as a series in $q$, convergent in the unit disk $|q|<1$. One can see that (\ref{WRTCircFibBlock}) coincides with the unrefined partition function of $T[M_3]$ on $D^2\times S^1$, {\it cf.} (\ref{TCircFibBlock}), with removed contribution of Cartan components of the adjoint chiral:
\begin{equation}
	\hat{Z}_a(q)=\left.\hat{Z}_a(q,t,u)\,(qt;q)_\infty\,\prod_{i=1}^g\frac{(u_i;q)_\infty}{(tu_iq;q)_\infty}\right|_{t=1,u_i=1}.
\end{equation}
The principal value contour prescription in (\ref{WRTCircFibBlock}) suggests that the contour $\gamma$ in (\ref{TCircFibBlock}) should be chosen as shown in  Figure~\ref{fig:TCircFib-contour}. Note, that the labels $a$ of homological blocks now correspond to the \textit{connected components} of the space of abelian flat connections.

\subsubsection{Factorization of the superconformal index}

The superconformal index of 3d $\CN=2$ theory \eqref{TCircFib} reads
\begin{multline}
	\CI_{T[M_3]}(q,t)=\sum_{m\geq 0}\frac{1}{|\CW_m|}\int_{|z|=1}\frac{dz}{2\pi i z}\,z^{2pm}\,
(1-z^{\pm 2}q^m)\,q^{-2m(1-g)}\,t^{-2m(1-g)}
\times \\
\frac{(z^{\pm2}q^mt^{-1};q)_\infty(t^{-1};q)_\infty}{(z^{\pm 2}q^{m+1}t)_\infty(qt;q)_\infty}\times \\
\prod_{i=1}^g	
\frac{(z^{\pm2}u_i^{-1}q^{m+1};q)_\infty(u_i^{-1}q;q)_\infty}{(z^{\pm2}u_iq^{m};q)_\infty(u_i;q)_\infty}\cdot
\frac{(z^{\pm2}tu_iq^{m+1};q)_\infty(tu_iq;q)_\infty}{(z^{\pm2}t^{-1}u_i^{-1}q^m;q)_\infty(t^{-1}u_i^{-1};q)_\infty}.
\label{TCircFibIndex}
\end{multline}
Unlike in the case of Lens spaces ($g=0$), now there is some ambiguity in the choice of the contour,
because if $|t|=1$ there are poles at $|z|=1$. The natural
physical choice assumes that $|q|<1$ and $|u_i|=|t^{-1}u_i^{-1}|=q^{\epsilon}$ with $\epsilon>0$, which corresponds to choosing an infinitesimally small R-charge for adjoint hypers. However, such choice of the contour is different from the one in (\ref{TCircFibBlock}) and (\ref{WRTCircFibBlock}), illustrated in Figure~\ref{fig:TCircFib-contour}.

\FIGURE{
\centering
 \includegraphics[scale=0.8]{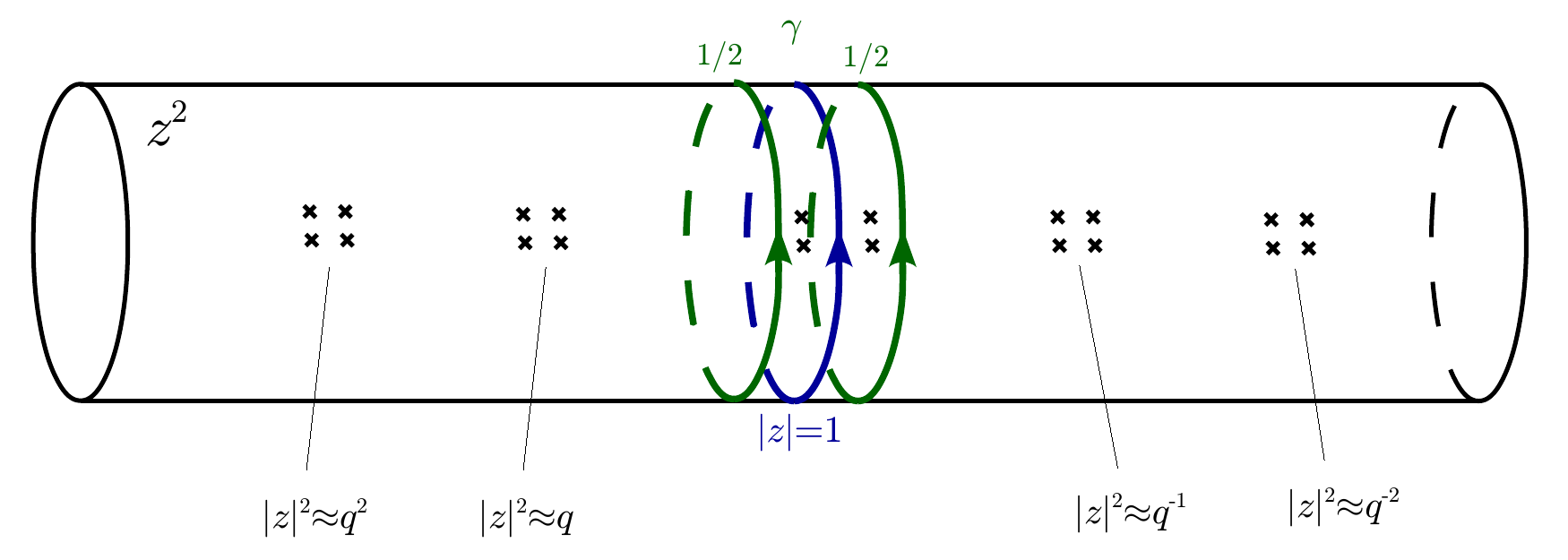}
\caption{Pole structure of the integrands in (\ref{TCircFibBlock}) and (\ref{TCircFibIndex}). The green contour corresponds to a choice of the contour $\gamma$ such that in the $t\rightarrow 1$ limit it agrees with the principal-value prescription in  (\ref{WRTCircFibBlock}) which reproduces the WRT invariant. The blue contour is the standard physical choice of contour for the superconformal index (\ref{TCircFibIndex}).}
\label{fig:TCircFib-contour}
}

Next, we would like to show that
\begin{equation}
	\CI_{T[M_3]}(q,t,u) \; = \; \sum_{a}|\CW_a|\hat{Z}_a(q,t,u)\hat{Z}_a(1/q,1/t,1/u).
	\label{FactorizationCircFib}
\end{equation}
As before, the problem with checking it directly is that we only have the expression for $\hat{Z}_a(q,t)$ which is valid inside a unit disc $|q|<1$, without any systematic way of analytically continuing it outside. Again, we can instead try to check (\ref{FactorizationCircFib}) in a particular limit of flavor fugacities such that $\hat{Z}_a(q,t)$ becomes a rational function and has canonical analytic continuation outside of $|q|<1$. However, one finds that there are several new technical complications compared to the case of Lens spaces ($g=0$). First, there is an ambiguity in the choice of the contours (\ie~whether we encircle the poles at $z^2=t^{\pm1/2}$ or not), illustrated in Figure~\ref{fig:TCircFib-contour}. Second,
for $g>1$ the limit $u_i\rightarrow 1$, $t\rightarrow q^{\beta}$ with $\beta\in \Z_{\geq 0}$ does not give a rational function, but rather a lacunary function that cannot be analytically continued outside $|q|<1$ due to the Fabry gap theorem.
In particular, one cannot repeat the trick with the cyclotomic expansion due to the absence of closed form expression for $t=1$.

However, one can avoid all these problems by taking the limit $u_i\rightarrow 1$, $t\rightarrow q^\beta$, with $\beta<0$.
In this limit, all poles at $|z|\approx 1$ disappear and the contour ambiguity becomes irrelevant.
Namely, for $\beta\in\Z_{-}$, let us define
\begin{equation} \hat{Z}_a^{(\beta)}(q) \; \equiv \; \left.\hat{Z}_a(q,t)\,(qt;q)_\infty\,\prod_{i=1}^g\frac{(u_i;q)_\infty}{(tu_iq;q)_\infty}\right|_{t=q^{\beta},u_i=1}
\end{equation}
and
\begin{equation}
	\CI^{(\beta)}(q) \; \equiv \; \left.\CI(q,t)\,\frac{(qt;q)_\infty}{(1/t;q)_\infty}
	\prod_{i=1}^g\frac{(u_i;q)_\infty(t^{-1}u_i^{-1};q)_\infty}{(u_i^{-1}q;q)_\infty(tu_iq;q)_\infty}
	\right|_{t=q^{\beta},u_i=1}.
\end{equation}
Then, one can check for various values of $g,p,\beta<0$ that indeed
\begin{equation}
	\CI^{(\beta)}(q) \; = \; \sum_a|\CW_a|\hat{Z}_a^{(\beta)}(q)\hat{Z}_a^{(\beta)}(1/q).
\end{equation}
For example, let us take $g=2$ and $p=5$. Then, we have the following.

$\beta=-1$:
\begin{equation}
	\begin{array}{c}
	2\hat{Z}_0^{(-1)}(q)=1 \,,\quad
	\hat{Z}_1^{(-1)}(q)=0 \,,\quad
	\hat{Z}_2^{(-1)}(q)=0.\; \\
		2\CI^{(-1)}(q)=1.
	\end{array}
\end{equation}

$\beta=-2$:
\begin{equation}
	\begin{array}{c}
	2\hat{Z}_0^{(-2)}(q)=q^{-2}+1 \,,\quad
	\hat{Z}_1^{(-2)}(q)=q^{-4/5} \,,\quad
	\hat{Z}_2^{(-2)}(q)=0.\; \\
		2\CI^{(-2)}(q)=4+q^2+\frac{1}{q^2}.
	\end{array}
\end{equation}

$\beta=-3$:
\begin{equation}
	\begin{array}{c}
	2\hat{Z}_0^{(-3)}(q)=\frac{1}{q^6}+\frac{1}{q^4}+\frac{2}{q^3}+\frac{1}{q^2}+1,\;
	\hat{Z}_1^{(-3)}(q)=-\frac{(q+1)^2
   \left(q^2-q+1\right)}{q^{24/5}},\;
	\hat{Z}_2^{(-3)}(q)=\frac{1}{q^{11/5}}.\; \\
		2\CI^{(-3)}(q)=q^6+\frac{1}{q^6}+4 q^4+\frac{4}{q^4}+8 q^3+\frac{8}{q^3}+5 q^2+\frac{5}{q^2}+8 q+\frac{8}{q}+18.
		\\
		\ldots
	\end{array}
\end{equation}
And indeed one can verify that in each case
\begin{equation}
	\CI^{(\beta)}(q)=2\hat{Z}_0^{(\beta)}(q)\hat{Z}_0^{(\beta)}(1/q)+
	\hat{Z}_1^{(\beta)}(q)\hat{Z}_1^{(\beta)}(1/q)+
	\hat{Z}_2^{(\beta)}(q)\hat{Z}_2^{(\beta)}(1/q).
\end{equation}
Note, when $g>1$ the fundamental group $\pi_1(M_3)$ is non-abelian and $\CM_\text{flat}(M_3,SL(2,\C))$ has irreducible flat connections.
We confirmed, however, that the WRT invariant and the superconformal index both have a natural decomposition into contributions labelled by connected components of \textit{abelian} flat connections.\footnote{It is important to keep in mind that the terms in this decomposition are not merely contributions of the corresponding connected components of abelian flat connections to the path integral, but rather contain contributions of irreducible flat connections as well \cite{Gukov:2016njj}. In other words, even though the sum in the decomposition runs only over abelian flat connections, all flat connections are accounted properly, reducible and irreducible; no one is left behind.}
This ``completeness'' of the homological blocks is a striking and highly non-trivial phenomenon.
Its physical origin was already discussed in section \ref{sec:summary} from the viewpoint of resurgent analysis
and charges which label BPS states / enumerative invariants in the physical system \eqref{CatGeom1}.
It would be interesting to explore it further in various duality frames and from other vantage points.

\subsubsection{Comparison with refined CS}

In the notations of section~\ref{sec:refCS},
the partition function of refined Chern-Simons theory on $M_3=O(-p)\rightarrow \Sigma_g$ reads \cite{Aganagic:2011sg}:
\begin{multline}
	Z_{SU(2)}^\text{ref.~CS}[M_3]=\sum_{n=1}^{k-1} g^{g-1}_n\,(S_{1n})^{2-2g}(T_{nn})^{-p}=\\
	=
	\sum_{n=1}^{k-1}q^{-\frac{pn^2}{4}-\frac{pn\beta}{2}-\frac{p\beta^2}{4}}
	\prod_{j=0}^\beta (q^{-\frac{n+j+\beta}{2}}-q^{\frac{n+j+\beta}{2}})^{1-g}
	(q^{-\frac{n-j+\beta}{2}}-q^{\frac{n-j+\beta}{2}})^{1-g}.
\end{multline}
By repeating manipulations in section~\ref{sec:refCS}, we arrive at (up to a simple overall factor)
\begin{equation}
	Z_{SU(2)}^\text{ref. CS}[M_3]
	=\sum_{a,b\in \Z_p /\Z_2} e^{2\pi i (k+2\beta)\text{CS}(a)}\cdot
	S_{ab}\cdot \,\hat{Z}^\text{(ref.CS;$\beta$)}_b(q)
\end{equation}
with
\begin{multline}
	|\CW_a|\hat{Z}^\text{(ref.CS;$\beta$)}_a(q)=\text{v.p.}\int_{|z|=1}\frac{dz}{2\pi iz}\,(z^{\pm2};q)_{\beta+1}^{1-g}\sum_{n\in p\Z+a}q^{n^2/p}\,z^{2n}=\\
	=\left.\text{v.p.}\int_{|z|=1}\frac{dz}{2\pi iz}\,\left(\frac{(z^{\pm2};q)_\infty}{(z^{\pm2}qt;q)_\infty}\right)^{1-g}\sum_{n\in p\Z+a}q^{n^2/p}\,z^{2n}\right|_{t=q^\beta}
\end{multline}
where ``v.p.'' stand for principal value integration.
This indeed agrees with the partition function of $T[M_3]$ on $D^2\times S^1$, {\it cf.} (\ref{TCircFibBlock}),
\begin{equation}	 \hat{Z}^\text{(ref.CS;$\beta$)}_a(q)=\left.\hat{Z}_a(q,t,u)\cdot(qt;q)_\infty\,\prod_{i=1}^g\frac{(u_i;q)_\infty}{(tu_iq;q)_\infty}\right|_{t=q^\beta,u_i=1}.
\end{equation}

\subsection{${M_3}=\text{plumbed}$}
\label{sec:plumbed}

\begin{figure}[ht]
\centering
 \includegraphics[scale=3]{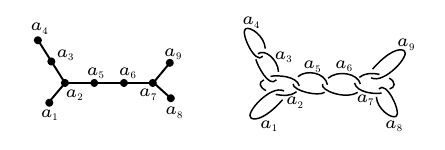}
\caption{An example of a plumbing graph $\Gamma$ (left) and the corresponding link $\CL(\Gamma)$ of framed unknots in $S^3$ (right). The associated 3-manifold $M_3(\Gamma)$ can be constructed by performing a Dehn surgery on $\CL(\Gamma)$.}
\label{fig:plumbing-example}
\end{figure}

Now we move to consider another very large class of 3-manifolds $M_3(\Gamma)$ associated to a plumbing graph $\Gamma$. There are various ways to define $M_3$ for a given $\Gamma$. One of them is to say that $M_3$ is obtained by a Dehn surgery on the corresponding link $\CL(\Gamma)$ of unknots (see Figure~\ref{fig:plumbing-example}). For simplicity, we also assume that $\Gamma$ is connected. Such class of 3-manifolds contains Seifert fibrations over $S^2$, which correspond to star-shaped plumbing graphs. Denote by $L$ the number of vertices of $\Gamma$. It is equal to the number of components of the link $\CL(\Gamma)$. We will also need $L\times L$ linking matrix of $\CL(\Gamma)$ which will be denoted by $M$:
\begin{equation}
 M_{v_1,v_2}=\left\{
\begin{array}{ll}
 1,& v_1,v_2\text{ connected}, \\
 a_v, & v_1=v_2=v, \\
0, & \text{otherwise}.
\end{array}
\right.\qquad v_i \in \text{Vertices of }\Gamma \;\cong\;\{1,\ldots,L\}.
\label{linking}
\end{equation}

\begin{figure}[ht]
\centering
\includegraphics[scale=1.7]{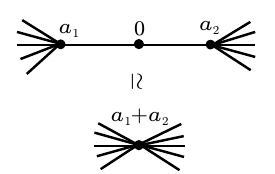}
\includegraphics[scale=1.7]{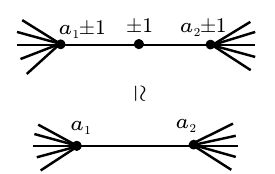}
\includegraphics[scale=1.7]{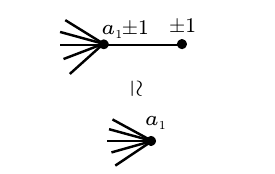}
\caption{3d Kirby moves that relate plumbing graphs which result in homeomorphic 3-manifolds.}
\label{fig:moves}
\end{figure}

The first homology group of $M_3(\Gamma)$ is given by the cokernel (over $\Z$) of the linking matrix:
\begin{equation}
H_1(M_3,\Z) \cong \Coker M = \Z^L/M\Z^L.
\end{equation}
In what follows we assume for simplicity that $M_3(\Gamma)$ is a rational homology sphere ($b_1(M_3)=0$), which implies that $\Gamma$ has no loops ({\it i.e.} it is a tree). The linking form on $H_1(M_3,\Z)$ is then given by
\begin{equation}
\lk(a,b) =(a,M^{-1}b)\;\mod \Z,\qquad a,b\in  \Z^L/M\Z^L.
\end{equation}
Different plumbing graphs can give homeomorphic 3-manifolds --- $M_3(\Gamma)\cong M_3(\Gamma')$ \textit{iff} $\Gamma$ and $\Gamma'$ can be connected by a sequence of 3d Kirby moves, as shown in Figure~\ref{fig:moves}\footnote{There is also a move which relates a connected graph to a disconected one. For the sake of brevity we don't discuss the case of disconnected plumbing graphs here.}.

The theory $T[M_3,SU(2)]$ can in principle be obtained by combining together the following 3d theories (see \eg~\cite{Gadde:2013sca}):
\begin{itemize}
	\item edge: $S$-duality wall in $\CN=4$ $SU(2)$ super-Yang-Mills ({\it i.e.} theory $T[SU(2)]$),
	\item 2-valent vertex: $T^a$-duality wall $~=~$ ``supesymmetric Chern-Simons term at level $a$,''
	\item $n$-valent vertex: $T[(\text{sphere with $n$ holes})\times S^1]$ =  $n$ copies of $T[SU(2)]$ gauged together.
\end{itemize}

\subsubsection{Homological blocks ${\hat{Z}_a}$}
\label{sec:plumbing-blocks}

The homological blocks can be calculated by the following formula (see appendix~\ref{app:Zq-from-WRT} for a derivation):
\begin{multline}
	\hat{Z}_b(q)=q^{-\frac{3L+\sum_v a_{v}}{4}}\cdot\text{v.p.}\int\limits_{|z_v|=1}
\prod_{v\;\in\; \text{Vertices}}
\frac{dz_v}{2\pi iz_v}\,
\left({z_v-1/z_v}\right)^{2-\text{deg}(v)}\cdot\Theta^{-M}_b(z)
	\\
	\equiv q^{-\frac{3L+\sum_v a_{v}}{4}}\cdot\text{v.p.}\int\limits_{|z_v|=1}
\prod_{v\;\in\; \text{Vertices}}
\frac{dz_v}{2\pi iz_v}\,
\left({z_v-1/z_v}\right)^{2}\times\\
\times\prod_{(v_1,v_2)\in \text{Edges}}\frac{1}{(z_{v_1}-1/z_{v_1})(z_{v_2}-1/z_{v_2})}
\cdot\Theta^{-M}_b(z)
	\label{Z-hat-def-0}
\end{multline}
where $\text{deg}(v)$ is the degree of a vertex $v$ and, as before, ``v.p.'' means taking principle value integral (\ie~take half-sum of contours ``dodging'' poles from both sides) and $\Theta^{-M}_b(x)$ is the theta function of the lattice corresponding to minus the linking form $M$ given by
\begin{equation}
	\Theta^{-M}_b(x)=\sum_{\ell \in 2M\Z^L+b}q^{-\frac{(\ell,M^{-1}\ell)}{4}}
	\prod_{i=1}^Lx_i^{\ell_i}.
\end{equation}

{\bf Remarks on} \eqref{Z-hat-def-0} {\bf :}
\begin{enumerate}

\item In the special case when the plumbing graph has just one vertex, the expression for the homological block reduces to (\ref{BlockBeta1}) with $\beta=0$.
	
\item Generically it is well-defined only if $M$ is negative definite, \ie~$M_3(\Gamma)$ is a link of a normal singularity, with $\Gamma$ being a resolution graph of the singularity.
	
\item The formula (\ref{Z-hat-def-0}) defines homological blocks $\hat{Z}_a(q)$ in terms of a plumbing graph $\Gamma$, but one can show that they are invariant under Kirby moves (Figure~\ref{fig:moves}) acting on $\Gamma$. Therefore, $\hat{Z}_a(q)$ actually depends only on the homeomorphism class of $M_3(\Gamma)$, and is indeed a topological invariant. A priori this is not obvious, since (\ref{Z-hat-def-0}) has been obtained (see Appendix  \ref{app:Zq-from-WRT}) as an analytic continuation of WRT invariant, which, in principle, does not have to be unique.
	
\item Even though the formula has the structure of the ``index of a quiver theory'', it is not clear how to properly realize it as a partition function of $T[M_3(\Gamma)]$ on $D^2_b\times S^1$ with some boundary condition.
				
\item Suppose $M<0$. Then $\hat{Z}_a(q)$ is a well-defined series in $q$ convergent in the unit disk $|q|<1$. Unless $M_3$ is a Lens space, $\hat{Z}_a(q)$ does not admit analytic continuation (in the usual sense) outside of $|q|<1$ due to Fabry gap theorem (because the powers of $q$ with non-zero coefficients grow quadratically).

\item We have
	\begin{equation}
	\hat Z_b(q)
	\;\in\; 2^{-c}q^{\Delta_b}\Z[[q]]
	\end{equation}
	where
	\begin{equation}
	 c\in \Z_+,\qquad c\leq L,
	\end{equation}
	\begin{equation}
	  b\in (2\text{Coker}\,M+\delta)\,/\Z_2
	\stackrel{\text{Set}}{\cong} H_1(M_3,\Z)\,/\Z_2,
	\end{equation}
	\begin{equation}
	    \delta\in \Z^L/2\Z^L,\qquad \delta_v = \text{deg}\,v \mod 2.
	\end{equation}
	\begin{equation}
	 \Delta_b=-\frac{3L+\sum_v a_{v}}{4}-\max_{\ell \in 2M\Z^L+b}\frac{(\ell,M^{-1}\ell)}{4}\,\in \Q
	\end{equation}
	where $\Z_2$ acts as
	\begin{equation}
	 b\mapsto -b
	\end{equation}
	and is a symmetry of (\ref{Z-hat-def-0}).
	
\end{enumerate}

\paragraph{Example 1:}
\begin{equation}
	\includegraphics{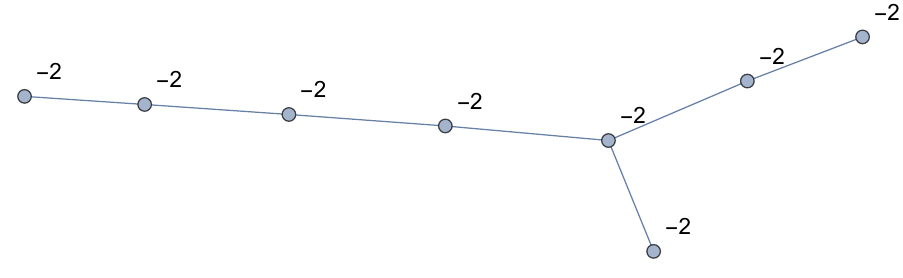}
\end{equation}
\begin{equation}
	H_1(M_3)=0
\end{equation}
\begin{equation}
	\hat{Z}_0=q^{-3/2}(1-q-q^3-q^7+q^8+q^{14}+q^{20}+q^{29}-q^{31}-q^{42}+\cdots)
\end{equation}

\paragraph{Example 2:}
\begin{equation}
{\,\raisebox{-2.0cm}{\includegraphics[width=8.0cm]{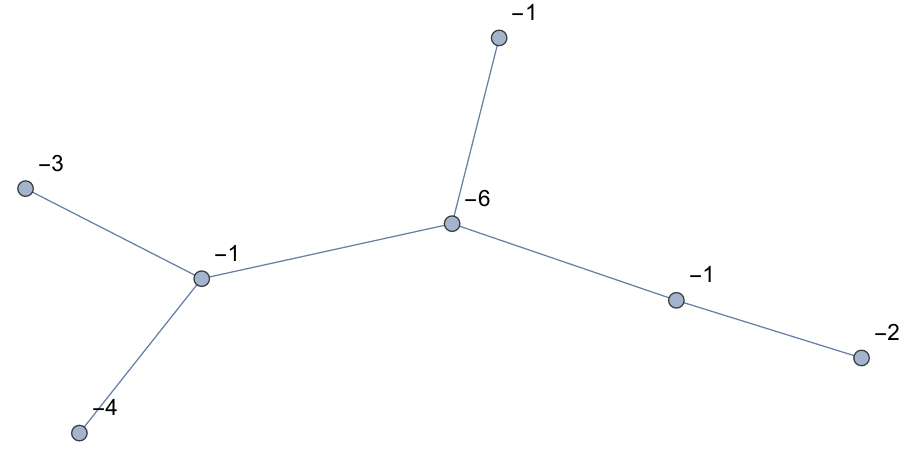}}\,}
\stackrel{\text{Kirby}}{\sim}
{\,\raisebox{-2.0cm}{\includegraphics[width=4.0cm]{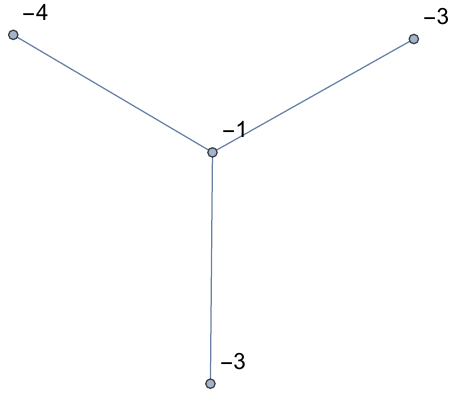}}\,}
\end{equation}
\begin{equation}
	H_1(M_3)=\Z_{3}
\end{equation}
\begin{equation}
	\hat{Z}=
	\left(
	\begin{array}{c}
	 1-q+q^6-q^{11}+q^{13}-q^{20}+q^{35}+O\left(q^{41}\right) \\
	 q^{5/3}\left(-1+q^3-q^{21}+q^{30}+O\left(q^{41}\right)\right) \\
	\end{array}
	\right)
\end{equation}

\paragraph{Example 3:} (non-Seifert)
\begin{equation}
	\includegraphics{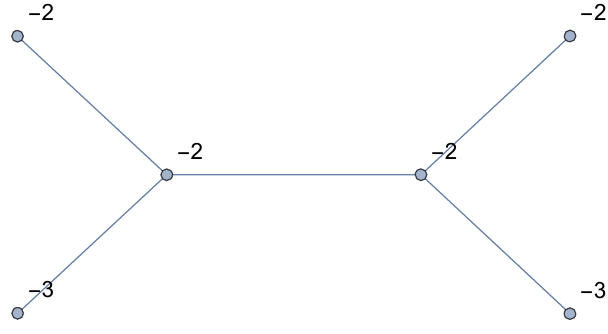}
\end{equation}
\begin{equation}
	H_1(M_3)=\Z_{13}
\end{equation}
\begin{equation}
	\hat{Z}=
	\frac{1}{4}\left(\tiny
	\begin{array}{c}
	q^{-1/2}(2-2q+2q^2-2q^4+4q^5+6q^{10}+8q^{11}-4q^{13}+2q^{14}+4q^{15}+O\left(q^{18}\right)) \\
	q^{5/26}(-3+2q+2q^2-4q^3-2q^7-q^8+2q^9-q^{10}+2q^{12}-4q^{13}+2q^{16}+O\left(q^{18}\right)) \\
	q^{7/26}(4+q+2q^3-2q^4-2q^6-3q^7-2q^8-2q^{10}-q^{11}-2q^{13}-4q^{14}-2q^{15}-4q^{16}+O\left(q^{18}\right)) \\
	q^{-7/26}(-3-3q^2+2q^4-2q^5+4q^7+2q^8+2q^9+2q^{10}+4q^{12}+4q^{13}+4q^{14}+2q^{15}+O\left(q^{18}\right)) \\
	q^{-11/26}(-1+2q-2q^2+4q^3-3q^6-2q^7+4q^8-2q^9+2q^{11}+2q^{13}+q^{14}-2q^{16}+O\left(q^{17}\right)) \\
	q^{-5/26}(2+2q^2-q^3-3q^5-2q^6+2q^7-4q^8-2q^{10}-2q^{11}-2q^{12}-2q^{13}-5q^{15}-2q^{16}+O\left(q^{17}\right))\\
	 q^{-15/26}(-1-2q+2q^2-q^4+2q^6-2q^7+2q^8+4q^{10}+2q^{12}+2q^{13}+2q^{15}+O\left(q^{17}\right)) \\
	\end{array}
	\right)
\end{equation}

\subsubsection{Problems on the ${T[M_3]}$ side}

One may wish to carry out the categorification program for $M_3(\Gamma)$ as well. However, it is not obvious how to write a correct expression for superconformal or topologically twisted index of the 3d $\CN=2$ theory $T[M_3]$ because the UV description with $\CN=2$ supersymmetry of its certain components (see the paragraph above Section~\ref{sec:plumbing-blocks}) is not known. In particular, there is no known Lagrangian description for half-BPS T-duality walls in $\CN=4$ super-Yang-Mills theory, and understanding the theory $T[M_3(\Gamma)]$ and its Hilbert space poses a challenge. We hope to study this problem in the future.

Let us note that a certain expression of the superconformal index of $T[M_3]$, with $M_3$ being an $SL(2,\Z)$-mapping class torus of punctured $T^2$ (which, without a puncture, would be homeomorphic to $M_3(\Gamma)$ with circular plumbing $\Gamma$) was considered in \cite{Gang:2012ff,Gang:2013sqa}. In particular, there it was checked that it was invariant under the first two of the Kirby moves in Figure~\ref{fig:moves} that correspond to relations in the $SL(2,\Z)$ group. However, as it was argued in \cite{Gang:2013sqa} it coincides with the result given by triangulation decomposition, as in \cite{Dimofte:2011ju,Dimofte:2011py}, and therefore, if interpreted as the partition function of the complex Chern-Simons theory, it misses contributions of the reducible flat connections \cite{Chung:2014qpa}. The latter as we have seen, play a major role for categorification of WRT invariant. Also, in principle one can extend their prescription for calculating superconformal indices associated with plumbing graphs with vertices of degree other than two, by requiring invariance under the third Kirby move in Figure~\ref{fig:moves}.


\section{Adding knots and links into ${M_3}$}

The goal of this section is to incorporate knots and links in a non-trivial 3-manifold $M_3$
and produce the corresponding homological invariants.
In particular, it will give us the very first examples where {\it both} knots and 3-manifolds are non-trivial.

The physical setup that we use to make concrete predictions for the new homological invariants $\CH_a[M_3;K,\CR]$
was already briefly mentioned in \eqref{conifoldphases}. As before, $K$ denotes a knot in $M_3$ while $\CR$ denotes a representation (color) of CS gauge group $G$. The case of links is similar and involves assigment of color to each component of the link. The space $\CH_a[M_3;K,\CR]$ can be explored from a number of vantage points
and in a variety of duality frames (all of which lead to the same result). In particular, from the vantage point
of the theory $T[M_3]$ on fivebrane world-volume, incorporating knots and links corresponds to introducing
a 1d defect (``impurity'') which preserves the same supersymmetry as the background along $D^2 \cong \R^2_q$
in our previous discussion, {\it cf.} Figure~\ref{fig:impurity}.

Alternatively, starting from the brane system \eqref{CatGeom1}, there are many possibilities \cite{Gukov:2015gmm}
to accommodate knots inside $M_3$. We focus on those which give rise to line operators in $T[M_3]$.\footnote{The other options lead to a 3d ``space-time filling defect,'' deforming the theory to a new one $T[M_3,K]$.} They correspond to codimension-four defects in the 6d (2,0) theory, engineered by either M2- or M5-branes (see \eg~\cite{Gang:2015bwa} for a discussion in the context of 3d-3d correspondence). For example, one can insert a stack of M2-branes in the following way:
\beq
\begin{matrix}
{\mbox {\textrm{$N$ fivebranes:}}}~~~\qquad & \R & \times & M_3& \times & D^2  \\ \qquad & &
&   \cap &  & \cap \\
{\mbox{\rm space-time:}}~~~\qquad & \R&  \times  & T^*M_3 & \times & TN   \\ \qquad & &
&   \cup &  & $\rotatebox[origin=c]{90}{$\in$}$ \\
{\mbox{\rm M2-branes:}}~~~\qquad &\R & \times & T^*K & \times  & O   \\
\end{matrix}.
\label{CatGeom2}
\eeq
We have chosen the line defect in $T[M_3]$ to sit at the tip $O\in D^2$ (also the Taub-NUT center) to preserve the $U(1)_q$ rotational symmetry.
Naively, the way the stack of M2-branes ends on $N$ fivebranes is determined by a partition of the number of M2 branes into $N$. Such partitions are in one-to-one correspondence with Young tableaux with at most $N$ rows, which are in turn can be identified with irreducible representations of $U(N)$. In the case of $SU(N)$ we have to mod out configurations where M2 branes are equally distributed among $N$ fivebranes. This leads to Young tableaux with at most $N-1$ rows that label irreducible representations of $SU(N)$.
In reality, mapping line operators in 3d/3d correspondence and in the fivebrane setup is a delicate business and should be dealt with care.

If one wishes to use M5-branes to represent knots, then besides $\R_{\text{time}}\times K$ they would also occupy
the two-dimensional co-normal bundle $N^*K$ of $K$ inside $T^*M_3$ and the cotangent space $T^*|_O$ to $D^2$ at $O$ \cite{Ooguri:1999bv}.
Both constructions, based on M2- and M5-branes, lead to Wilson loops --- in fact,
they are related by the Hanany-Witten type effect in M-theory or, in TQFT on $M_3$, by the Fourier transform
from representation basis to holonomy basis.

Using this physical setup, we compute the new homological invariants $\CH_a[M_3;K,\CR]$ for some simple examples
of non-trivial knots and 3-manifolds. In particular, we verify that suitable variants of the Conjectures~\ref{Conjecture1} and \ref{Conjecture2}
hold in the presence of knots and links.

\subsection{Links in ${M_3}$ and line operators in ${T[M_3]}$}
\label{sec:knots}

Consider a knot $K$ in $M_3$ colored by a representation $\CR$ of $G$.
As described above, it corresponds to a certain half-BPS line operator $\Gamma_{K,\CR}$ in $T[M_3;G]$:
\begin{equation}
	(K,\CR)\;\longmapsto \; \Gamma_{K,\CR}\;\in\;\CC.
	\label{Knots-LineOps}
\end{equation}
where $\CC$ is the category of BPS line operators in $\CN=2$ SCFT $T[M_3,G]$.
The homological invariant for a combined system of a knot and a 3-manifold
is then given by the BPS Hilbert space of $T[M_3]$ on the disk $D^2$ with $\Gamma_{K,\CR}$ inserted at $O$,
the center of the disk (see Figure~\ref{fig:impurity}):
\begin{equation}
	\CH_a[M_3;K,\CR]\equiv \CH_{T[M_3]}(D^2,\Gamma_{K,\CR};a),\qquad a\in (\text{Tor}\,H_1(M_3,\Z))^N/S_N
\end{equation}
where $a$ denotes the choice of boundary condition, the same as in the case without a knot.
It has the same gradings as before, namely the $q$-grading $i$ and the homological (R-charge) grading $j$:
\begin{equation}
	\CH_a[M_3;K,\CR]=\bigoplus_{i\in \Delta_a+\Z,\,j\in\Z}\CH_a^{i,j}[M_3;K,\CR]
\end{equation}
And, as before, the graded Euler characteristic gives the partition function on $D^2\times S^1$:
\begin{equation}
	\hat{Z}_a(q;\Gamma_{K,\CR})\equiv Z_{T[M_3]}(D^2\times S^1,\Gamma_{K,\CR};a)=
	\sum_{i,j}(-1)^jq^i\CH_a^{i,j}[M_3;K,\CR]
\end{equation}
now, in 3d $\CN=2$ theory $T[M_3]$ with an extra 1d ``impurity'' $\Gamma_{K,\CR}$ supported on $O\times S^1$.
The analogue of the Conjecture \ref{Conjecture1} is then the following:
\begin{conj}
The WRT invariant of $M_3$ with a knot $K$ colored by representation $\CR$ can be decomposed into the following form:
\begin{equation}
	Z_{SU(2)_k}[M_3;K,\CR]=(i\sqrt{2k})^{b_1(M_3)-1}\sum_{a,b\;\in \;\atop\Tor H_1(M_3,\Z)/\Z_2}e^{2\pi ik\lk(a,a)}\,S_{ab}
	\, \hat{Z}_b(q;\Gamma_{K,\CR})|_{q\rightarrow e^{\frac{2\pi i }{k}}}
	\label{WRT-knot-decomposition}
\end{equation}
where
\begin{equation}
	\hat{Z}_b(q;\Gamma_{K,\CR}) \in \, 2^{-c} q^{\Delta_b} \Z[[q]]\qquad \Delta_b\in \Q,\qquad c\in\Z_+
	\label{Block-knot-qSeries}
\end{equation}
convergent in $|q|<1$ and $S_{ab}$ is the same as in Conjecture \ref{Conjecture1}
\label{Conjecture3}
\end{conj}
For certain 3-manifolds with an extra symmetry, preserved by the knot $K$, we have an extra grading whose value we denote by $\ell$:
\begin{equation}
	\CH_a[M_3;K,\CR]=\bigoplus_{i\in \Delta_a+\Z,\,j,\ell\in\Z}\CH_a^{i,j;\ell}[M_3;K,\CR]
\end{equation}
In practice, to compute these homological invariants, one needs a convenient ultra-violet (UV) description of
the fivebrane setup \eqref{CatGeom2} or, better yet, of the IR superconformal theory $T[M_3]$.
In general, the brane system \eqref{CatGeom2} and the 3d theory $T[M_3]$ may admit many such UV descriptions,
in various duality frames, all of which flow to the same IR physics and produce the same homological invariants (space of BPS states).

In a given UV gauge theory description of $T[M_3]$,
the objects in the tensor category $\CC$ of BPS line operators can be understood as vector spaces ---
the ``Hilbert spaces of $\CN=2$ quantum mechanics on the line, coupled to the bulk gauge theory''.
The most familiar line operators are Wilson operators $W_\lambda$, labeled by a representation $\lambda$
of the UV gauge group\footnote{In general, $\CG$ is different from $G$.
For a quiver theory, it may be given by a product $\times_i G_i$ over the vertices.} $\CG$ (which, in general, has little to do with $G$).
The corresponding object of the category $\CC$ is, tautologically, the representation $\lambda$.
The Wilson lines are expected to form a complete basis of line operators in the IR theory,
in the sense that any line operator (not necessarily associate to a knot) should have the following decomposition:
\bea
\overbrace{\rule{1.0cm}{0pt}}^{\text{IR}} &  & ~~~\overbrace{\rule{2.5cm}{0pt}}^{\text{UV}} \label{Line-UV} \\
\Gamma_{K,\CR} & ~=~ & \bigoplus_{\lambda,i,j,\ell} W_\lambda\otimes \CH_{K,\CR}^{\lambda;i,j;\ell} \,. \nonumber
\eea
The reason is very simple --- one can always decompose the ``defect Hilbert space'' into representations of $\CG$,
with the multiplicity space $\CH_{\Gamma}^{\lambda}$ associated with the representation $\lambda$.
The symmetry $\CG$ plays the role of the flavor symmetry in QM on the line, which couples to 3d theory $T[M_3]$ by gauging $\CG $ in the bulk.
If $\Gamma$ is a generic physical line operator (\textit{cf.} superposition of physical line operators),
the corresponding multiplicity spaces $\CH_{\Gamma}^{\lambda}$ are honest vector space that are at least doubly-graded,
with the $(i,j)$-grading coming from $U(1)_q\times U(1)_R$ symmetry of the brane system \eqref{CatGeom2}.
When the flavor symmetry $U(1)_{\beta}$ is a symmetry of both $T[M_3]$ and $\Gamma$, the third $\ell$-grading is also present.

As a trivial example, consider a line defect given by a quantum mechanics that is completely decoupled from the bulk. Then,
\be
\Gamma_{\text{QM}} \; = \; W_0\otimes \CH_{\text{QM}} \,,
\label{babyexmpl}
\ee
where $\CH_{\text{QM}}$ is the Hilbert space of the quantum mechanics and $W_0$ is the trivial Wilson loop. So, we see that \eqref{Line-UV} has a chance of being correct, even if there are infinitely many possibilities of engineering defects via 1d-3d coupled systems. In this sense, $\lambda$ is analogous to the label ``$a$'' of the homological blocks, which reduces infinitely many possible boundary conditions to a finite set. In fact, this isn't just an analogy --- it comes from the equivalence of two categories
\be\label{C=CB}
\CC\cong \CC_B.
\ee
Namely, a line operator can be viewed as the interface between two boundary conditions, and picking a ``reference boundary condition'' (\eg~the one giving $\hat{Z}_0$ with zero brane charge) will identify the two categories. This equivalence is a special feature of 3d quantum field theories, which associate categories to $S^1$, with the objects that can be interpreted either as line operators or as boundary conditions at spatial infinity.

At this point, $\lambda$ can still be an arbitrary representation of the UV gauge group $\CG$, and one might naively think that $\CC$ is
\be\label{CUV}
\CC_{\text{UV}}=\mathrm{Rep}(\CG),
\ee
given by the representation category of $\CG$. But in light of the equivalence \eqref{C=CB}, we expect $\CC$ to be much smaller, since $\CC_B$ is generated by finitely many simple objects (boundary conditions that give homological blocks), as we have seen in all our examples.

Indeed, \eqref{CUV} is often redundant, due to presence of Chern-Simons couplings in $T[M_3]$. There is a set $\Lambda_{M_3}$ of ``integrable representations'' of $\CG$ giving a complete basis of $W_\lambda$ in $T[M_3;G]$, and every Wilson loop labeled by a non-integrable representation can be decomposed as
\be\label{WilsonDec}
W_{\nu}=\bigoplus_{\lambda\in \Lambda_{M_3}\atop i,j,\ell\in\Z}W_{\lambda}\otimes \CH_{\nu}^{\lambda;i,j;\ell}.
\ee
All such relations generate an ideal (in the categorical sense) of $\CC_{\text{UV}}$, and $\CC$ is given by the quotient
\begin{equation}
	\CC=\CC_{UV}/\CI. 	
\end{equation}
As a vector space, $\CC$ is the same as $\Lambda_{M_3}\otimes\C$, but the algebra structure depends on $\CI$ and is determined by the theory $T[M_3]$ in a non-trivial way. Both $\CC_{\text{UV}}$ and $\CI$ depend on the duality frame (presentation of $M_3$), but it is expected that the quotient category $\CC$ is a topological invariant of $M_3$. It describes line operators / boundary conditions in the IR theory, which is what usually meant by $T[M_3]$.

We expect, for arbitrary $M_3$, the independent Wilson loops in the theory $T[M_3;G]$ are labeled by elements in\footnote{So far, we have ignored the difference between $G$ and the Langlands dual group $G^\vee$ because the quantities considered so far are not sensitive to the global structure of the group. However, the spectrum of Wilson loops is sensitive to this difference, and \eqref{WilsonCharge} is stated for the adjoint form $G=G_{\text{ad}}$. If $G$ has a center, one needs to add the representations of $G$ where the center acts non-trivially. Reader interested in this subject may consult \cite{EVFCBI} where this issue was extensively discussed in a similar brane system.\label{ft:Global}}
\be\label{WilsonCharge}
\Lambda_{M_3}=\mathrm{Tor}\,H_1(M_3,\Lambda_{\text{rt}})/W_G,
\ee
where $\Lambda_{\text{rt}}$ is the root lattice of $\frak{g}$ and $W_G$ is the Weyl group (not to be confused with Wilson line $W_{\lambda}$ in UV description of $T[M_3]$). The argument for this statement parallels our discussion of boundary conditions given by M2-brane charges ``$a$'', which take values in exactly the same set. And the fundamental Wilson loops are expected to come from knots $K$ wrapping torsion 1-cycles, colored by a representation $\CR$ of $G$ (\ie~$\CR\in \Lambda_{\text{char}}(G)/W_G$).

If $G$ is abelian, then $\Lambda_{M_3}$ is a group and one can introduce an action of line operators on boundary conditions, which we call the {\it shift} action in what follows. This action is expected because Wilson line operators $W_\lambda,\;\lambda\in \Lambda_{M_3}$, can be engineered using M2-branes\footnote{{\it i.e.} $W_\lambda=\Gamma_{K,\CR}$ when torsion homology class of $\CR$-cable of $K$ is given by $\lambda\in  \Lambda_{M_3}$} as in \eqref{CatGeom2}, and the label ``$a$'' for homological blocks also measures M2-brane charges.
In other words, a Wilson loop $W_\lambda$ relates two Hilbert spaces
\be
\CH_{T[M_3]}(D^2,W_\lambda;a) \; \sim \; \CH_{T[M_3]}(D^2,W_\lambda;a-\lambda).
\ee
One does not \textit{a priori} know whether this is an isomorphism or not, or how the gradings on the two sides are related.
However, in various examples below we observe that, at the level of homological blocks $\hat{Z}_a(q)$,
Wilson loop acts as a simple shift operator, even for $G=SU(2)$ if the Weyl group action is properly taken into account.

As before, our main tools for probing the new homological invariants are SUSY-protected quantities.
In particular, using a UV descriptions of $T[M_3]$ and a half-BPS impurity $\Gamma_{K,\CR}$,
one can compute the partition function of $T[M_3]$ on $D^2\times S^1$ with $\Gamma_{K,\CR}$ inserted along $O\times S^1$ as follows:
\begin{multline}\label{BlockKnot}
	\hat{Z}_a(q,t;\Gamma_{K,\CR})\equiv Z_{T[M_3]}(D^2\times S^1,\Gamma_{K,\CR};a)=\\ =\int \frac{dz}{2\pi i z}
	\,[\text{same as without $\Gamma_{K,\CR}$}] \sum_{\lambda\in \Lambda_{M_3}\atop i,j,\ell \in\Z}\chi_\lambda(z)q^it^\ell(-1)^{j}
	\dim\CH_{K,\CR}^{\lambda;i,j;\ell}
\end{multline}
where $z$ stands for the gauge fugacities of the UV gauge group $\CG$
and $\chi_\lambda$ is the character of the $\CG$-representation $\lambda$.

Similarly, we can consider the $S^2 \times S^1$ superconformal index with half-BPS impurities inserted at one (or both) poles of $S^2$.
The UV localization formula for the index is modified as follows (when the impurity is at the north pole):
\begin{multline}\label{IndexKnot}
	\CI(q,t;\Gamma_{K,\CR})\equiv Z_{T[M_3]}(S^2\times S^1;\Gamma_{K,\CR})= \\ =\sum_{m}\int \frac{dz}{2\pi i z}
	\,[\text{same as without $\Gamma$}] \sum_{\lambda\in\Lambda_{M_3} \atop i,j,\ell \in \Z}\chi_\lambda(zq^{m/2})q^i t^\ell (-1)^{j}
	\dim\CH_{K,\CR}^{\lambda;i,j;\ell}.
\end{multline}
As we demonstrate in a variety of examples, it obeys the modified factorization formula
\begin{equation}
\CI(q,t;\Gamma_{K,\CR}) \; = \; \sum_a|\CW_a| \hat{Z}_a(q,t;\Gamma_{K,\CR})\hat{Z}_a(q^{-1},t^{-1}) \,.
\end{equation}
Note, all of the above relations depend only on the IR data, namely on $K$ and its color $\CR$, not on the UV description of $\Gamma_{K,\CR}$.
In particular, note that $\lambda$ is always summed over in these relations, which goes back to the basic relation
between IR line operators $\Gamma_{K,\CR}$ (``impurities'') and their description \eqref{Line-UV} in the UV theory via $W_{\lambda}$.

The reader should keep in mind that there are two non-trivial maps (dualities) involved in the study of knots.\footnote{This aspect is
a surprising exception to the rule that knot homology is easier than 3-manifold homology.}
First, there is a rather non-trivial map \eqref{Line-UV} between line operators in the IR theory $T[M_3]$
and its UV description (or, rather, one of many UV descriptions).
This UV/IR map has nothing to do with applications to low-dimensional topology and is a standard phenomenon in quantum field theory.
In addition, there is a map between knots and links colored by $\CR$ in a TQFT on $M_3$
and half-BPS line operators (``impurities'') in the 3d $\CN=2$ theory $T[M_3]$.
This map is a chapter in the so-called 3d/3d correspondence.
The reason one should pay attention to both of these maps is that it is their composition which relates
homological invariants of $K \subset M_3$ to concrete calculations in $T[M_3;G]$,
$$
\boxed{{\text{UV line} \atop \text{operators}~W_{\lambda}}}
\quad \xleftrightarrow[~]{\text{UV/IR map}} \quad
\boxed{{\frac12\text{-BPS impurity} \atop \Gamma_{K,\CR} }}
\quad \xleftrightarrow[\text{~}]{\text{3d/3d correspondence}} \quad
\boxed{{\text{link}~K \subset M_3 \atop \text{colored by}~\CR}}
$$
Because each of these maps can be rather non-trivial, one should not confuse the representation $\lambda$
that labels half-BPS line operators in the UV description with the ``color'' of $K$.
The latter is denoted by $\CR$ and, whenever non-trivial, leads to the $\CR$-colored link homology \cite{Gukov:2011ry,Fuji:2012pm}
in our setup \eqref{CatGeom2}.

So far, the discussion in this section is very general and, as such, may be somewhat abstract.
We will now give some very concrete examples for $G=SU(2)$.

\subsection{Links in $S^2\times S^1$ and Rozansky's proposal}
\label{sec:Roz}

For $M_3=S^2\times S^1$, the theory $T[M_3]$ is a particular case of (\ref{TCircFib}) with $p=0$ and $g=0$, that is the theory of 3d $\CN=2$ vector multiplet with gauge group $\CG=G=SU(2)$. In this case  $\mathrm{Tor}\,H_1(M_3,\Z)=0$ and the only independent Wilson loop in $T[M_3]$ is the trivial Wilson loop. Also, there is only one homological block $\hat{Z}_0$. Therefore, for an arbitrary representation $\lambda$ of $SU(2)$,
the decomposition \eqref{WilsonDec} allows to write the Wilson loop $W_\lambda$ as \eqref{babyexmpl}:
\be
W_\lambda \; = \; W_0\otimes \CH_{\lambda}^{0;i,j;\ell} \,.
\ee
In other words, in this case we expect a factorization
\begin{equation}
\CH_a[S^2\times S^1;K^{\lambda},\CR_\lambda] \; = \; \bigoplus_{i,j,\ell\in\Z}\CH_{\lambda}^{0;i,j;\ell}\otimes {\CH}_a[S^2\times S^1]
\end{equation}
where the first factor on the right-hand side is the $S^2\times S^1$ analogue of the Khovanov-Rozansky homology
of the corresponding link $K^{\lambda}$ colored by $\CR_{\lambda}$. What is $K^{\lambda}$? And, what is $\CR_{\lambda}$?

As we shall see below, already in this relatively simple case of $M_3 = S^1 \times S^2$ the map between
colored knots and links in $M_3$ and the UV line operators $W_{\lambda}$ in $T[M_3]$ is rather non-trivial.
Since Wilson loops in $T[M_3]$ come from M2-branes wrapping the $S^1$ circle of $S^1 \times S^2$,
we expect that the precise dictionary $W_{\lambda} \leftrightarrow (K^{\lambda}, \CR_{\lambda})$
maps $\lambda$ to a certain link in $M_3$ that winds around $S^1$,
and the total winding number (counted with multiplicities given by $\CR_{\lambda}$) is equal to $|\lambda|$.
Regarding $M_3 = S^1 \times S^2$ as a special case of the Hopf fibration (namely, the trivial fibration),
in the rest of this section we present evidence for the following identification\footnote{Note,
in the unrefined / decategorified setting, $K^{\lambda}$ is simply the unknot colored by $\CR_{\lambda} = \lambda$.
Although this is no longer the case in homological world, there is still a close relation between ``cabling''
and ``color'' \cite{khovanov2005categorifications,Gukov:2005qp}.}
\be
\boxed{\phantom{\int}
W_\lambda \quad \leftrightarrow \quad (K^{\lambda}, \CR_{\lambda}) \; = \; \text{``$\lambda$-cable'' of the unknot along the Hopf fiber}
\phantom{\int}}
\label{Gamma-Lp1}
\ee
Specifically, $\lambda = {\square}^{\otimes 2n}$ corresponds to a link $L_{2n}$ made of $2n$ copies of the unknot
along the Hopf fiber, all colored by the fundamental representation of $SU(2)$.
And $(K^{\lambda}, \CR_{\lambda})$ for general $\lambda$ is a linear combination of $L_{2n}$'s
that replicates the decomposition of $\lambda$ into ${\square}^{\otimes 2n}$'s.
Such decomposition is usually referred to as the ``cabling formula'' (see Appendix~\ref{app:cabling}).

Curiously, a similar problem was already considered by Rozansky from a purely mathematical point of view in \cite{rozansky2010categorification}.
Specifically, the Conjecture~6.10 in \cite{rozansky2010categorification} gives a concrete description for the homological invariant
of the link $L_{2n}$.
We shall verify that Rozansky's proposal\footnote{The Conjecture~6.10 in \cite{rozansky2010categorification} contains a typo in the description of the ideal $I_{\text{rel}}$. Namely,
\be\label{OldRelation}
a_i p_i({\bf x})=\theta_i p_i({\bf x})=0
\ee
should be replaced by
\be\label{NewRelation}
a_i p_{n-i+1}({\bf x})=\theta_i p_{n-i+1}({\bf x})=0.
\ee
} matches exactly with the quantum field theory computation if the generators $a_i$ and $\theta_i$ are assigned flavor charges $i$, that is
\be
\mathrm{deg}_t \, a_i=\mathrm{deg}_t \, \theta_i=i.
\ee

Using the decomposition of tensor products of $SU(2)$ representations, we can write the expectation value of $2n$ Wilson loops in $T[M_3]$
as a sum over expectation values of Wilson loops labeled by irreducible representations. Only even representations with highest weight $2l=0,2,\ldots,2n$ appear in this decomposition:
\be\label{WilsonDec1}
\langle W_{{\square}^{\otimes 2n}}\rangle =\sum_{l=0}^n \left[{{2n}\choose {n-l}}-{{2n}\choose {n-l-1}}\right] \langle W_{2l}\rangle.
\ee
The expectation value of the Wilson loop $W_{2l}$ in $T[M_3]$,
which gives the graded Euler characteristic of $\CH_\lambda$,\footnote{In this normalization, we have
\be\label{WilsonNorm}
\langle W_0\rangle=1,
\ee
enabling us to directly extract the information about the space $\CH_{\lambda}^{i,j;\ell}$.}
\be
\sum_{i,j,\ell}(-1)^jq^it^\ell\dim\CH_\lambda^{0;i,j,\ell} \equiv\chi_{q,t}(\CH_\lambda)=\langle W_{2l}\rangle\equiv \frac{\hat{Z}_0(q,t;W_{2l})}{\hat{Z}_0(q,t)}
\ee
can be computed using \eqref{BlockKnot}, with $T[S^1\times S^2,SU(2)]$ now being the 3d $\CN=4$ super-Yang-Mills theory.
We obtain
\begin{equation}
	\langle W_{2l}\rangle \; = \; \frac{(qt)^{l} (1/t;q)_l}{(q^2t;q)_l},
\end{equation}
and the expectation values of $W_{2l+1}$ all vanish.

On the other hand, as a vector space, the homological invariant $H^\bullet(S^1\times S^2,L_{2n})$ proposed in Conjecture~6.10 of \cite{rozansky2010categorification} admits a similar decomposition into subspaces of pure ${\bf x}$-degree\footnote{This decomposition doesn't respect the ring structure as each subspace is not closed under multiplication.}
\be
H^\bullet_{\text{Rozansky}}(L_{2n})=\bigoplus_{l=0}^{n} V_l \otimes \Q[a_1,\theta_1,a_2,\theta_2,\ldots, a_l,\theta_l].
\ee
Here, $V_l$ is a quotient of the vector space spanned by monomials of $x_1,\ldots,x_{2n}$, with degree $n-l$, by the relations
\be
x_i^2=0 \quad \text{and} \quad \sum_{i=1}^{2n} x_i=0.
\ee
It is easy to check that\footnote{Basically, there are ${{2n}\choose {n-l}}$ non-vanishing degree $(n-l)$ monomials, and they are subject to ${{2n}\choose {n-l-1}}$ independent relations, obtained by multiplying $\sum_i x_i$ with degree $(n-l-1)$ monomials.}
\be
\dim V_l= {{2n}\choose {n-l}}-{{2n}\choose {n-l-1}}
\ee
are exactly the coefficients in \eqref{WilsonDec1}. In particular, $V_j$ with $j<0$ is empty, and the dimension of $V_0$ is given by the $n$-th Catalan number.
Then
\be\label{EquiEuler2}
\chi_t(H^\bullet)=\sum_{j=0}^n \dim V_l\cdot\fq^{2n-2j}\cdot\frac{(t;t\fq^{-2})_j}{(t\fq^{-4};t\fq^{-2})_j},
\ee
where we have used the $\fq$-degree of the generators,
\begin{align}
\mathrm{deg}_{\fq} x_i=& \quad 2,\\
\mathrm{deg}_{\fq} a_i=&-2i-2,\\
\mathrm{deg}_{\fq} \theta_i=&-2i+2.
\end{align}
If we shift the overall $\fq$-degree of $H^\bullet$ by $-2n$ units to remove the $\fq^{2n}$ factor in \eqref{EquiEuler2},
what remains is in perfect agreement with the expectation value of Wilson loops in 3d theory $T[M_3]$:
\be
\langle W_{2l}\rangle= (qt)^j\cdot \frac{(1/t;q)_j}{(tq^2;q)_j}=\fq^{-2j}\cdot\frac{(t,t\fq^{-2})_j}{(t\fq^{-4};t\fq^{-2})_j},
\ee
with a typical relation between $q$ and $\fq$ \cite{Fuji:2012pm}:
\be
\fq^2 \; = \; qt \,.
\ee
So, we have verified the conjecture by Rozansky at the level of equivariant Euler characteristics for this link $L_{2n}$,
\be
\chi_{q,t}(\CH_{\text{BPS}}) \; = \; \chi_{q,t}(H_{\text{Rozansky}})
\ee
presenting some evidence for the isomorphism between $H^\bullet_{\text{Rozansky}} (S^1\times S^2,L_{2n}) \{ -2n \}$
and $\CH_{\lambda} \; : = \; \bigoplus_{i,j,\ell\in\Z}\CH_{\lambda}^{0;i,j;\ell}$ with $\lambda = {\square}^{\otimes 2n}$.

As a side remark, positivity also features in the invariants $\hat{Z}_a(M_3; K_{\CR})$
of three manifolds with knots, in a way similar to the case without knot (\ref{block-pos}).
For example, in our case of $M_3 = S^2 \times S^1$,
\begin{multline}
	-\hat{Z}_0(q,-t;W_{6})=
	\left(2 t^3+2 t^2\right) q^3+\left(2 t^4+4 t^3+4 t^2+2 t\right) q^4+\\
	\left(2 t^5+6 t^4+10 t^3+10 t^2+4 t\right) q^5+
	\left(2 t^6+8 t^5+16
	   t^4+22 t^3+18 t^2+8 t+2\right) q^6+
	   \\
	   \left(2 t^7+8 t^6+20 t^5+36 t^4+44 t^3+34 t^2+14 t+2\right) q^7+O\left(q^8\right)
\end{multline}
has only positive coefficients.

\subsection{Links in $L(p,1)$}

For $M_3=L(p,1)$, or more generally $M_3=O(-p)\rightarrow \Sigma_g$, we have $\Lambda_{M_3}= \Z_p/\Z_2$
and the basic Wilson loops in $T[M_3]$ are given by $W_{\lambda}$ with $\lambda=0,\ldots, p$.\footnote{Notice that we have added odd representations according to description in footnote~\ref{ft:Global}.}
These correspond to the ``$\lambda$-cablings'' of the unknot \eqref{Gamma-Lp1} wrapping the Hopf fiber of $L(p,1)$.

More generally, for plumbed $M_3$ (see section \ref{sec:plumbed}), if $(K^{\lambda}, \CR_{\lambda})$
is a $\lambda$-cabling of the unknot wrapping the ``Hopf fiber'' associated with the node $v$ of the plumbing graph,
we have $\Gamma_{K,\la}{=}W_{\lambda_v}$, where $\lambda_v$ is the representation of $\CG$ obtained from
the representation $\la$ of $G$ via the ``evaluation homomorphism'' $e_v: \CG\rightarrow G$ for the node $v$.

\subsubsection{Unrefined index of ${T[L(p,1)]}$ with Wilson loops}
\label{sec:LensWilson}

The unrefined superconformal index --- with contribution of the Cartan component of the adjoint chiral removed, as in \eqref{I-UN-unref} --- with the Wilson line inserted at the north pole reads:
\begin{equation}\label{SU2Int}
\begin{aligned}
\langle W_\la \rangle\equiv 	\CI(q;W_\lambda) & =  \frac{1}{2} \sum_{m \in \mathbb{Z}} \int \frac{dz}{2\pi i z} z^{ p m}\cdot \chi_\lambda(q^{m/2}z)\\
&\ \ \ \times \left( q^{2 \left| m \right|}+q^{-2 \left| m \right|}+4 - 2 \left(q^{ \left| m \right|}+q^{- \left| m \right|} \right) \left( z^2 +\frac{1}{z^2} \right)+\left( z^4+\frac{1}{z^4} \right) \right).
\end{aligned}
\end{equation}
Assuming $\lambda$ is irreducible, we identify it with its highest weight in $\Lambda_{\text{wt},\frak{su}(2)}/\Z_2 \cong \Z_{\geq 0}$.
Our convention is such that
\be
\chi_{\lambda}(z)=\frac{z^{\lambda+1}-z^{-\lambda-1}}{z-z^{-1}}.
\ee
So, the insertion of an operator $W_\lambda$ with odd value of $\lambda$ automatically gives zero,
because of the non-trivial action of the center $\Z_2 \subset SU(2)$.

In the table below we present explicit results for $p=5$ up to $\lambda=38$:
\vspace{0.2cm}
\\
\vspace{0.2cm}
\begin{tabular}{|c|c||c|c||c|c||c|c||c|c|}\hline
$\lambda$ & $\langle W_\lambda\rangle$ & $\lambda$ & $\langle W_\lambda\rangle$ &$\lambda$ & $\langle W_\lambda\rangle$&$\lambda$ & $\langle W_\lambda\rangle$  &$\lambda$ & $\langle W_\lambda\rangle$\\ \hline
0 & 3 & 2 & $-1$ &4 & 0 & 6 &$q^{-3}$ & 8 & $-q^{-3}-2q^{-5}$ \\ \hline
10 & $2q^{-5}+q^{-7}$ & 12 &$-q^{-7} $ &14& 0 &16& $q^{-16}$ &18& $-q^{-16}-2q^{-20}$ \\ \hline
20 & $2q^{-20}+q^{-24}$ & 22 & $-q^{-24}$ &24& 0 &26& $q^{-39}$ &28& $-q^{-39}-2q^{-45}$ \\ \hline
30 & $2q^{-45}+q^{-51}$ & 32 & $-q^{-51}$ &34& 0 &36& $q^{-72}$ &38& $-q^{-72}-2q^{-80}$ \\ \hline
\end{tabular}\ .
\\
This table is almost periodic. This can be understood from the action of Wilson lines on the homological blocks.
For the $D^2\times S^1$ partition function of $T[M_3]$ with $W_\lambda$ inserted at the center of the disk, $O \in D^2$, we have:
\begin{equation}
	\hat{Z}_a(q;W_\lambda) \; = \; \frac{1}{|\CW_a|}\int\frac{dz}{2\pi i z}
	(1-z^{\pm 2})\,\chi_{\lambda}(z)\,\sum_{n\in p\Z+a}q^{n^2/p}\,z^{2n}.
	\label{block-Lp1-knot}
\end{equation}
Wilson loops act in a very simple way on the homological blocks $\hat{Z}_a(q)$. For $L(p,1)$, we have
\be
\hat{Z}_0=1,\quad \hat{Z}_1=-q^{1/p},
\ee
with the rest of the blocks being zero. Let us also formally extend $\hat{Z}_a$ for $a$ outside the range $\{0,1,\ldots,\lfloor\frac{p}{2}\rfloor\}$ by $\hat{Z}_a=0$ and define a ``shift operator''
\begin{equation}
 [j]\hat{Z}_a(q) \; := \; q^{j(2a-j)/p}\hat{Z}_{a-j}(q),\qquad a,j\in \Z.
\end{equation}
Then,
\be
\hat{Z}_a(q;W_{\lambda}) \; =  \sum_{b \, \equiv \, \pm a \pmod p}[\lambda/2] \, \hat{Z}_b(q) ,\qquad a\in \Z_p/\Z_2.
\ee
This equation is not cyclic, and we took $a$ to live in $\Z_{\ge 0}$. Then, one can verify that
\be
\langle W_\lambda \rangle \; = \sum_{a\in \Z_p/\Z_2}|\CW_a|\hat{Z}_a(q;W_\lambda)\hat{Z}_a(q^{-1}).
\ee

One can also deduce the action of $W_\lambda$ on $Z_a=\sum_{b}S_{ab}\hat{Z}_b$ by applying the $S$-transform,
but it won't be very illuminating because $Z_a(q;W_\lambda)$ will be a linear combination of all $Z_a(q)$'s. This difference between $\hat{Z}_a$ and $Z_a$ again demonstrates the advantage of working with the homological blocks $\hat{Z}_a$, rather than $Z_a(q)$.

\subsubsection{Refinement and categorification}

The refined version of (\ref{block-Lp1-knot}) reads
\begin{equation}
	(qt;q)_\infty\hat{Z}_a(q,t;W_\lambda)=
	\frac{1}{|\CW_b|}\left.\int\frac{dz}{2\pi i z}\,
	\frac{(z^2;q)_\infty(z^{-2};q)_\infty}{(z^2tq;q)_\infty(z^{-2}tq;q)_\infty}\,\chi_{\lambda}(z)\sum_{n\in p\Z+b}q^{n^2/p}\,z^{2n}\right. .
	\label{Z-hat-ref-knot}
\end{equation}
One can show that it has the positivity property (\ref{block-pos}) and, therefore, can be interpreted
as the Poincar\'e polynomial of the homological invariant that categorifies $SU(2)$ WRT invariants of
$M_3 = L(p,1)$ with knots wrapping the Hopf fiber.

For example, consider $M_3 = L(5,1)$. With our choice of $G = SU(2)$, it has three abelian flat connections and,
correspondingly, three homological blocks, all tabulated in \cite[sec.6]{Gukov:2016gkn}.
Now, we can see that incorporating knots does not spoil the positivity of \eqref{Z-hat-ref-knot}
and its interpretation as the Poincar\'e polynomial. Yet, the presence of the knot changes all homological blocks
in a non-trivial way, mixing homological invariants of the knot $K$ and the ambient 3-manifold $M_3$ in one combined entity,
{\it e.g.} for $M_3 = L(5,1)$ and $\lambda=2$:
\begin{multline}
	-(-qt;q)_\infty\hat{Z}_0(q,-t;W_2)=
	(t+1) q+\left(t^2+2 t+1\right) q^2+\left(t^3+3 t^2+4 t+2\right) q^3+\\ \left(t^4+4 t^3+8 t^2+8 t+3\right) q^4+\left(t^5+4 t^4+10 t^3+16
	   t^2+14 t+5\right) q^5+
	   \\ \left(t^6+4 t^5+11 t^4+23 t^3+31 t^2+23 t+7\right) q^6+\\
	   \left(t^7+4 t^6+12 t^5+28 t^4+47 t^3+54 t^2+37 t+11\right)
	   q^7+O\left(q^8\right),
\end{multline}
\begin{multline}
	(-qt;q)_\infty\hat{Z}_1(q,-t;W_2)=
	\sqrt[5]{q}+(t+1) q^{6/5}+\left(2 t^2+4 t+2\right) q^{11/5}+\left(2 t^3+6 t^2+8 t+4\right) q^{16/5}+\\
	\left(2 t^4+7 t^3+14 t^2+15 t+6\right)
	   q^{21/5}+\left(2 t^5+8 t^4+19 t^3+29 t^2+25 t+9\right) q^{26/5}+\\
	   \left(2 t^6+8 t^5+22 t^4+43 t^3+55 t^2+41 t+13\right)
	   q^{31/5}+O\left(q^{36/5}\right),
\end{multline}
and
\begin{multline}
	-(-qt;q)_\infty\hat{Z}_2(q,-t;W_2)=
	q^{4/5}+(t+1) q^{9/5}+\left(t^2+2 t+1\right) q^{14/5}+\left(2 t^3+4 t^2+4 t+2\right) q^{19/5}+\\
	\left(2 t^4+5 t^3+8 t^2+8 t+3\right)
	   q^{24/5}+\left(2 t^5+5 t^4+11 t^3+17 t^2+13 t+4\right) q^{29/5}+\\
	   \left(2 t^6+6 t^5+14 t^4+25 t^3+29 t^2+21 t+7\right)
	   q^{34/5}+O\left(q^{36/5}\right).
\end{multline}

Note that, when $p=1$, we have $L(p,1)\cong S^3$ and the unknot wrapping the Hopf fiber becomes
the familiar unknot in $S^3$, yet with a nontrivial framing equal to 1.
Since $H_1(S^3,\Z)=0$, as in section \ref{sec:Roz}, we expect factorization in this case:
\be
W_\lambda=W_0\otimes \CH_{\lambda}^{0;i,j;\ell},
\ee
\begin{equation}
\CH_a[S^3;K^{\lambda},\CR_\lambda]=\bigoplus_{i,j,\ell\in\Z}\CH_{\lambda}^{0;i,j;\ell}\otimes {\CH}_a[S^3]
\end{equation}
Moreover, as in section \ref{sec:Roz}, we expect that $(K^{\lambda},\CR_\lambda)$ is the cabling of the unknot determined by $\lambda$.
In appendix~\ref{app:cabling}, we present evidence for this by comparing it with the ordinary Khovanov homology
of the cabling of the unknot, {\it cf.} \cite{khovanov2005categorifications}.

\subsection{Comparison with refined CS}
\label{sec:refCS-knots}

In this section, we wish to compare the categorification of WRT invariants of knots in 3-manifolds considered
in the previous section with analogous calculations using the tools of refined CS theory and DAHA \cite{Aganagic:2011sg,Cherednik:2011nr}.
To the best of our knowledge, such calculations (that involve both knots and 3-manifolds) were not done in the literature
on refined Chern-Simons theory.

Let us consider the unknot $K$ wrapping the Hopf fiber in $M_3 = L(p,1)$ and colored by an irreducible representation $\CR = \lambda$.
Note, that now the representation $\lambda$ (whose weight we denote by the same letter) does represent the ``color'' of $K \subset M_3$.
The partition function of the refined CS theory with such insertion
can be expressed through $S$ and $T$ matrices (listed in section \ref{sec:refCS}) as follows:
\begin{multline}
	Z_{SU(2)}^\text{ref.~CS}[L(p,1);K,\lambda]=\sum_{n=1}^{k-1} g^{-1}_n\,S_{1n}S_{n,\lambda+1}(T_{nn})^{-p}=\\
	=
	\sum_{n=1}^{k-1}q^{-\frac{pn^2}{4}-\frac{pn\beta}{2}-\frac{p\beta^2}{4}}
	\prod_{j=0}^\beta (q^{-\frac{n+j+\beta}{2}}-q^{\frac{n+j+\beta}{2}})
	(q^{-\frac{n-j+\beta}{2}}-q^{\frac{n-j+\beta}{2}})\,M_{\lambda}(q^{\beta+n}).
\end{multline}
where $M_\lambda(x)$ is the $SU(2)$ Macdonald polynomial, whose coefficients are rational functions of $q$ and $t=q^\beta$.
By repeating the manipulations of section~\ref{sec:refCS}, we obtain (up to the same simple overall factor):
\begin{equation}
	Z_{SU(2)}^\text{ref. CS}[L(p,1);K,\lambda]
	=\sum_{a,b\in \Z_p /\Z_2} e^{2\pi i (k+2\beta)\text{CS}(a)}\cdot
	S_{ab}\cdot \frac{1}{|\CW_b|}\,\hat{Z}^\text{(ref.CS;$\beta$)}_b(q;K,\lambda)
\end{equation}
with
\begin{multline}
	\hat{Z}^\text{(ref.CS;$\beta$)}_a(q;K,\lambda)=\int\frac{dz}{2\pi iz}\,(z^{\pm2};q)_{\beta+1}\,M_\lambda(z^2)\,\sum_{n\in p\Z+a}q^{n^2/p}\,z^{2n}=\\
	=\left.\int\frac{dz}{2\pi iz}\,\left(\frac{(z^{\pm2};q)_\infty}{(z^{\pm2}qt;q)_\infty}\right)
	M_\lambda(z^2)
	\sum_{n\in p\Z+a}q^{n^2/p}\,z^{2n}\right|_{t=q^\beta}
	\label{Z-ref-loop}
\end{multline}
For $\lambda=1$, the Macdonald polynomial coincides with the Schur polynomial, $M_1(z^2)=\chi_1(z)$,
and the expression above agrees with the $D^2\times S^1$ partition function with a Wilson line insertion (\ref{Z-hat-ref-knot}).
However, in order to find an agreement for more general $\lambda$,
we need to make a replacement $\chi_\lambda(z)\rightarrow M_\lambda(z^2)$ in (\ref{Z-hat-ref-knot}).
At the catogorified level, this means that (\ref{Gamma-Lp1}) needs to be replaced by
\begin{equation}
W_{\lambda} \quad \leadsto \quad \bigoplus_{\mu\leq \lambda} W_\mu\otimes \CH_\lambda^\mu
\end{equation}
where $\CH_\lambda^\mu$ are appropriate categorifications of the coefificients $C^\mu_\lambda(q,t)$ in the decomposition $M_\lambda(z^2)=\sum_{\mu\leq \lambda}C^\mu_\lambda(q,t)\chi_\mu(z)$.
This suggests that the insertion of the Macdonald polynomial in the partition function of 3d $\CN=2$ theory $T[M_3]$
corresponds to a $\lambda$-colored unknot wrapping the Hopf fiber of $M_3$, {\it cf.} \eqref{Gamma-Lp1}:
\be
\boxed{\phantom{\int}
M_\lambda(z^2) \quad \leftrightarrow \quad \text{$K$ = unknot along the Hopf fiber colored by $\CR = \lambda$}
\phantom{\int}}
\ee
It would be interesting to test this against many examples of $\lambda$-colored $sl(N)$ homology for various knots
and links \cite{Gukov:2011ry,Cherednik:2011nr,Fuji:2012pm,Itoyama:2012fq,Gukov:2015gmm,Wedrich:2016smm,Kononov:2016cwp}.

Note, that the coefficients in fusion rules of Macdonald polynomials are non-trivial rational functions of $q,t$.
Therefore, the dependence on $\lambda$ in (\ref{Z-ref-loop}) does not agrees with the cabling formula,
unlike the case of Wilson line insertions.
On the other hand, the dependence on framing in the refined CS is simple and is given
by the corresponding power of the matrix element $T_{\lambda+1,\lambda+1}$.
In particular, when $p=1$ the result of (\ref{Z-ref-loop}) is
\begin{equation}
	\frac{\hat{Z}^\text{(ref.CS;$\beta$)}_0(q;K,\lambda)}{\hat{Z}^\text{(ref.CS;$\beta$)}_0(q)}=T_{\lambda+1,\lambda+1}M_\lambda(qt).
\end{equation}
Note that, for $\lambda>1$, it is a nontrivial rational function of $q$ and $t$,
which means that the underlying doubly graded vector space is infinite dimensional,
unlike the finite dimensional vector spaces produced by Wilson line insertions (see Appendix~\ref{app:cabling}).


\acknowledgments{We would like to thank J.~E.~Andersen, M.~Aganagic, F.~Benini, C.~Cordova, A.~Gadde, E.~Gorsky,
K.~Hori, H.~Kim, S.~Nawata, M.~Romo, S.~Shakirov, L.~Rozansky and K.~Ye for useful comments and discussions.
The work of S.G. and D.P. is supported in part by the U.S. Department of Energy, Office of Science, Office of
High Energy Physics, under Award Number DE-SC0011632.
In addition, the work of D.P. is supported by the center of excellence grant ``Center for Quantum Geometry of Moduli Space" from the Danish National Research Foundation (DNRF95).
P.P. gratefully acknowledges the support from Marvin L. Goldberger Fellowship and the DOE Grant DE-SC0009988.
The research of C.V. is supported in part by NSF grant PHY-1067976.
This work was performed in part (by P.P.) at Aspen Center for Physics which is supported by National Science Foundation grant PHY-1066293.
The authors would like to thank Simons Center for Geometry and Physics and the organisers of the Simons Summer Workshop 2016, where the work on the project has begun, for generous hospitality.

\appendix

\section{${\Z[[q]]}$-valued invariant for negative definite plumbed 3-manifolds}
\label{app:Zq-from-WRT}

We are interested in obtaining a certain analytic continuation of $su(2)_k$ WRT invariant \cite{reshetikhin1991invariants,witten1989quantum} of a plumbed 3-manidfold $M_3(\Gamma)$,\footnote{Everything below in principle can be generalized to $su(N)$ and $u(N)$.} where $\Gamma$ is a tree. Let us first consider the case of positive integer level, with $k\in \Z_+$. The colored Jones polynomial of link $\CL(\Gamma)$ is explicitly known (see \eg~\cite{witten1989quantum})\footnote{We use the normalization in which the colored Jones polynomial of unlink with $L$ components with canonical framing is given by
\begin{equation}
 J[\text{Unlink}]_{n_1,\ldots,n_L}=\prod_{v=1}^L \frac{\q^{n_v/2}-\q^{-n_v/2}}{\q^{1/2}-\q^{-1/2}}.
\end{equation}
}:
\begin{multline}
 J[\CL(\Gamma)]_{n_1,\ldots,n_L}=\frac{2i}{\q^{1/2}-\q^{-1/2}}\prod_{v\;\in\; \text{Vertices}\;\cong\;\{1,\ldots,L\}}
\q^{\frac{a_v(n_v^2-1)}{4}}
\,\left(\frac{2i}{\q^{n_v/2}-\q^{-n_v/2}}\right)^{\text{deg}(v)-1}
\times \\
\prod_{(v_1,v_2)\;\in\;\text{Edges}}
\frac{\q^{n_{v_1}n_{v_2}/2}-\q^{-n_{v_1}n_{v_2}/2}}{2i}
\end{multline}
where $\q=e^{2\pi i/k}$.\footnote{In this appendix, we use $\q$ to denote a $k$-th ($k\in\Z_+$) root of unity $e^{\frac{2\pi i}k}$ and reserve $q$ to denote the corresponding continuous variable which will appear in analytic continuation.} For the purposes of computing the WRT invariant of 3-manifold obtained by Dehn surgery on a link $\CL(\Gamma)$, it is useful to consider the following quantity:
\begin{equation}
 F[\CL(\Gamma)]\equiv \sum_{{n}\in \{1,\ldots,k-1\}^{L}} J[\CL(\Gamma)]_{n_1,\ldots,n_L}
\prod_{v=1}^L \frac{\q^{n_v/2}-\q^{-n_v/2}}{\q^{1/2}-\q^{-1/2}}.
\end{equation}
Then the WRT invariant of $M_3(\Gamma)$ is given by \cite{reshetikhin1991invariants}\footnote{Normalized such that
\begin{equation}
 \tau[S^3]=1.
\end{equation}
}
\begin{equation}
 \tau[M_3(\Gamma)]=\frac{F[\CL(\Gamma)]}{F[\CL(+1\bullet)]^{b_+}F[\CL(-1\bullet)]^{b_-}}
\label{WRT-surgery}
\end{equation}
where $b_\pm$ are the number of positive/negative eigenvalues of the linking matrix $M$ while $\pm1 \bullet$ denotes a plumbing graph with one vertex corresponding to an unknot with $\pm 1$ framing. The $SU(2)_k$ CS partition function, in the usual physical normalization, with
\begin{equation}
 Z_{SU(2)_k}[S^2\times S^1]=1,
\end{equation}
differs from (\ref{WRT-surgery}) by a simple factor:
\begin{equation}
 Z_{SU(2)_k}[M_3(\Gamma)]=\sqrt\frac{2}{k}\,\sin\frac{\pi}{k}\,\cdot\,\tau[M_3(\Gamma)].
\end{equation}

The following Gauss sum reciprocity formula will be very useful for us (see \eg~\cite{deloup2007reciprocity,jeffrey1992chern}):

\begin{multline}
 \sum_{n\;\in\;\Z^L/2k\Z^L}
 \exp\left(\frac{\pi i}{2k}(n,Mn)+\frac{\pi i}{k}(\ell,n)\right)
 =\\
 \frac{e^{\frac{\pi i\sigma}{4}}\,(2k)^{L/2}}{|\det M|^{1/2}}
  \sum_{a\;\in\; \Z^L/M\Z^L}
  \exp\left(-2\pi i k\left(a+\frac{\ell}{2k},M^{-1}\left(a+\frac{\ell}{2k}\right)\right)\right)
  \label{reciprocity}
\end{multline}
where $\ell \in \Z^L$, $(\cdot,\cdot)$ is the standard pairing on $\Z^L$ and $\sigma=b_+-b_-$ is the signature of the linking matrix $M$. In particular, from this formula it follows that
\begin{equation}
F[\CL(\pm 1\bullet)]=\sum_{n=1}^{k-1}\q^{\pm\frac{n^2-1}{4}}\,
\left(\frac{\q^{n/2}-\q^{-n/2}}{\q^{1/2}-\q^{-1/2}}\right)^2=
\frac{(2k)^{1/2}\,e^{\pm\frac{\pi i}{4}}\,\q^{\mp\frac{3}{4}}}{\q^{1/2}-\q^{-1/2}}.
\end{equation}
Therefore
\begin{multline}
 \tau[M_3(\Gamma)]=\frac{e^{-\frac{\pi i\sigma}{4}}\,\q^{\frac{3\sigma}{4}}}{2\,(2k)^{L/2}\,(\q^{1/2}-\q^{-1/2})}
 \times\\
 {\sum_{{n}\in \Z^L/2k\Z^L}}'\prod_{v\;\in\; \text{Vertices}}
\q^{\frac{a_v(n_v^2-1)}{4}}
\,\left(\frac{1}{\q^{n_v/2}-\q^{-n_v/2}}\right)^{\text{deg}(v)-2}
\times \\
\prod_{(v',v'')\;\in\;\text{Edges}}
\frac{\q^{n_{v'}n_{v''}/2}-\q^{-n_{v'}n_{v''}/2}}{2}
\label{WRT-Gamma-0}
\end{multline}
where we used invariance of the summand under $n_v\leftrightarrow -n_v$ and the fact that $L-|\text{Edges}|=1$. The prime $'$ in the sum means that the singular values $n_v=0,\,k$ are omitted.

Consider the following factor in the expression above:
\begin{equation}
\prod_{(v',v'')\;\in\;\text{Edges}}
(\q^{n_{v'}n_{v''}/2}-\q^{-n_{v'}n_{v''}/2})
=\sum_{{s}\in\{\pm 1\}^\text{Edges}}
\prod_{(v',v'')\;\in\;\text{Edges}}s_{(v',v'')}\q^{s_{(v',v'')}n_{v'}n_{v''}/2}.
\end{equation}
If one picks a vertex $v$ and makes a change $n_v\rightarrow -n_v$, a term in the sum with a given configuration of signs associated to edges (that is $s\in\{\pm 1\}^\text{Edges}$) will transform into a term with a different configuration times $(-1)^{\text{deg}\,v}$. Using the fact that the graph $\Gamma$ is a tree, by a sequence of such transforms, any configuration of signs can be brought to the configuration with all signs $+1$. Therefore (\ref{WRT-Gamma-0}) can be rewritten as follows:
\begin{multline}
 \tau[M_3(\Gamma)]=\frac{e^{-\frac{\pi i\sigma}{4}}\,\q^{\frac{3\sigma-\sum_v a_{v}}{4}}}{2\,(2k)^{L/2}\,(\q^{1/2}-\q^{-1/2})}
 \times\\
 {\sum_{{n}\in \Z^L/2k\Z^L}}'
\;\; \q^{\frac{(n,Mn)}{4}}
 \prod_{v\;\in\; \text{Vertices}}
\,\left(\frac{1}{\q^{n_v/2}-\q^{-n_v/2}}\right)^{\text{deg}(v)-2}.
\label{WRT-Gamma-1}
\end{multline}
Before we can apply (\ref{reciprocity}) to (\ref{WRT-Gamma-1}) we need to explicitly regularize the sum. In order to do this, let us introduce the following auxiliary quantities:
\begin{equation}
\Delta_v\in\Z_+:\; \Delta_v=\text{deg}\,v-1\mod 2,\qquad \forall v\;\in\; \text{Vertices},
\label{Delta-def}
\end{equation}
\begin{equation}
\omega\in \C:\;\;0<|\omega|<1.
\end{equation}
Then
\begin{multline}
 {\sum_{{n}\in \Z^L/2k\Z^L}}'
 \;\; \q^{\frac{(n,Mn)}{4}}
  \prod_{v\;\in\; \text{Vertices}}
 \,\left(\frac{1}{\q^{n_v/2}-\q^{-n_v/2}}\right)^{\text{deg}(v)-2}=\\
 \lim_{\omega\rightarrow 1} \frac{1}{2^L}\sum_{{n}\in \Z^L/2k\Z^L}
  \q^{\frac{(n,Mn)}{4}}
 F_\omega(x_1,\ldots,x_L)|_{x_v=\q^{n_v/2}}
 \label{WRT-omega-limit}
\end{multline}
where
\begin{multline}
F_\omega(x_1,\ldots,x_L)=\\
\prod_{v\;\in\; \text{Vertices}}\left({x_v-1/x_v}\right)^{\Delta_v}
 \,\left\{
 \left(\frac{1}{x_v-\omega/x_v}\right)^{\text{deg}(v)-2+\Delta_v}
 + \left(\frac{1}{\omega x_v-1/x_v}\right)^{\text{deg}(v)-2+\Delta_v}
 \right\}
 \\
\vspace{3em}
  \stackrel{\omega\approx 0}{=} \sum_{m\geq 0}\sum_{\ell\in \CI_m}N_{m,\ell}\,\prod_v x_v^{
\ell_v} \;\cdot\omega^m
\;\;\in\;\Z[x_1^{\pm1},\ldots,x_L^{\pm1}][[\omega]]
\label{F-omega}
\end{multline}
with $\CI_m$ being a \textit{finite} set of elements from $\Z^L$. If, after expansion in $\omega$ one collects coefficients in front of powers of $x$, the expression has the following form:
\begin{equation}
F_\omega(x_1,\ldots,x_L)\stackrel{\text{Formally}}{=\joinrel=\joinrel=\joinrel=}\sum_{\ell\in 2\Z^L +\delta} F_\omega^\ell\prod_v x_v^{
\ell_v}
\qquad \in \Z[\omega][[x_1^{\pm1},\ldots,x_1^{\pm L}]]
\end{equation}
where
\begin{equation}
 F_\omega^\ell=\sum_{m:\,\ell\in \CI_m}N_{m,\ell}\,\omega^m \;\;\in \Z[\omega]
\end{equation}
and
\begin{equation}
 \delta\in \Z^L/2\Z^L,\qquad \delta_v \equiv \text{deg}\,v \pmod 2.
\end{equation}
Note that $F_1^\ell=\sum_{m}N_{m,\ell}\in \Z$ does not depend on the choice of $\Delta\in \Z^L$ in (\ref{Delta-def}). Formally,
\begin{multline}
F_1(x_1,\ldots,x_L)=\sum_{\ell\in 2\Z^L +\delta} F_1^\ell\prod_v x_v^{\ell_v}=\\
\prod_{v\,\in\,\text{Vertices}}\left\{
{\scriptsize \begin{array}{c} \text{Expansion} \\ \text{at } x\rightarrow 0 \end{array} }
\frac{1}{(x_v-1/x_v)^{\text{deg}\,v-2}}
+
{\scriptsize \begin{array}{c} \text{Expansion} \\ \text{at } x\rightarrow \infty \end{array} }
\frac{1}{(x_v-1/x_v)^{\text{deg}\,v-2}}\right\}.
\end{multline}
Now let us assume that the quadratic form $M:\Z^L\times \Z^L\rightarrow \Z$ is negative definite, \footnote{In principle, this condition can be relaxed. It is only necessary that $M$ is negative on a certain subspace of $\Z^L$.}\ie~$\sigma=-L$). Then we can define the following $q$-series, convergent for $|q|<1$:
\begin{equation}
\hat Z_b(q)\stackrel{\text{Def}}{=\joinrel=}
2^{-L} q^{-\frac{3L+\sum_v a_{v}}{4}}
\sum_{\ell \in 2M\Z^L+b}F^\ell_1\,q^{-\frac{(\ell,M^{-1}\ell)}{4}}
\;\in\; 2^{-c}q^{\Delta_b}\Z[[q]]
\label{Z-hat-def}
\end{equation}
where
\begin{equation}
 c\in \Z_+,\qquad c\leq L,
\end{equation}
\begin{equation}
  b\in (2\Z^L+\delta)/2M\Z^L\,/\Z_2 \cong (2\text{Coker}\,M+\delta)\,/\Z_2
\stackrel{\text{Set}}{\cong} H_1(M_3,\Z)\,/\Z_2,
\end{equation}
\begin{equation}
 \Delta_b=-\frac{3L+\sum_v a_{v}}{4}-\max_{\ell \in 2M\Z^L+b}\frac{(\ell,M^{-1}\ell)}{4}\,\in \Q
\end{equation}
where $\Z_2$ acts as follows:
\begin{equation}
 b\rightarrow -b
\end{equation}
which is a symmetry of (\ref{Z-hat-def}).

Using relation (\ref{WRT-omega-limit}) and applying Gauss reciprocity formula (\ref{reciprocity}) we arrive at the following expression for the WRT invariant:
\begin{multline}
 \tau_k[M_3(\Gamma)]=\frac{e^{-\frac{\pi i\sigma}{4}}\,\q^{-\frac{3L+\sum_v a_{v}}{4}}}{2\,(2k)^{L/2}\,(\q^{1/2}-\q^{-1/2})}\,
\lim_{\omega\rightarrow 1}\sum_{{n}\in \Z^L/2k\Z^L}
  \q^{\frac{(n,Mn)}{4}}
 F_\omega(x_1,\ldots,x_L)|_{x_v=\q^{n_v/2}}
\label{WRT-Gamma-2}=\\
\frac{2^{-L} \q^{-\frac{3L+\sum_v a_{v}}{4}}}{2\,(\q^{1/2}-\q^{-1/2})\,|\det M|^{1/2}}
\sum_{\scriptsize\begin{array}{c}a\in \mathrm{Coker}\,M \\ b \in 2\mathrm{Coker}\, M+\delta \end{array}}
 e^{-2\pi i(a,M^{-1}b)} e^{-2\pi ik(a,M^{-1}a)}\times\\
\lim_{\omega\rightarrow 1}
\sum_{\ell \in 2M\Z^L+b}F^\ell_\omega\,\q^{-\frac{(\ell,M^{-1}\ell)}{4}}.
\end{multline}

Now, assume that the limit $\lim_{q\rightarrow \q}\hat{Z}_b(q)$, where $\hat{Z}_b(q)$ is defined in (\ref{Z-hat-def}) and $q$ approaches $k$-th primitive root of unity from inside of the unit disc $|q|<1$, exists and moreover,
\begin{equation}
	\lim_{\omega\rightarrow 1}
	\sum_{\ell \in 2M\Z^L+b}F^\ell_\omega\,\q^{-\frac{(\ell,M^{-1}\ell)}{4}}
	=\lim_{q\rightarrow \q}
	\sum_{\ell \in 2M\Z^L+b}F^\ell_1\,q^{-\frac{(\ell,M^{-1}\ell)}{4}}.
	\label{limit-exchange}
\end{equation}
Then
\begin{multline}
 \tau_k[M_3(\Gamma)]=\frac{1}{2\,(\q^{1/2}-\q^{-1/2})\,|\det M|^{1/2}}
 \,\times\\
\sum_{a\in \mathrm{Coker}\,M}e^{-2\pi ik(a,M^{-1}a)}
\sum_{b \in 2\mathrm{Coker}\, M+\delta} e^{-2\pi i(a,M^{-1}b)}\lim_{q\rightarrow \q} \hat{Z}_b(q).
\label{WRT-prop-1}
\end{multline}
We do not provide a proof of (\ref{limit-exchange}), and therefore (\ref{WRT-prop-1}), but believe it should be similar to the proof of the similar statement in \cite{lawrence1999modular}. However, even without the relation (\ref{WRT-prop-1}), the formula (\ref{Z-hat-def}) provides an invariant of negative-definite plumbed 3-manifolds valued in $q$-series with integer coefficients. As was already mentioned before, the resulting $q$-series do not depend on the auxiliary choices of $\Delta_v$ in (\ref{F-omega}). One can show that they are also invariant under the action of Kirby moves (see Figure~\ref{fig:moves}) acting on $\Gamma$, and therefore indeed depend only on homeomorphism class of $M_3(\Gamma)$.

The formula (\ref{Z-hat-def}) for the homological blocks $\hat{Z}_a$ can also be rewritten as a contour integral:
\begin{equation}
	\hat{Z}_b(q)=q^{-\frac{3L+\sum_v a_{v}}{4}}\cdot\text{v.p.}\int\limits_{|z_v|=1}
\prod_{v\;\in\; \text{Vertices}}
\frac{dz_v}{2\pi iz_v}\,
\left({z_v-1/z_v}\right)^{2-\text{deg}(v)}\cdot\Theta^{-M}_b(z)
\end{equation}
where $\Theta^{-M}_b(x)$ is the theta function of the lattice corresponding to minus the linking form $M$:
\begin{equation}
	\Theta^{-M}_b(x)=\sum_{\ell \in 2M\Z^L+b}q^{-\frac{(\ell,M^{-1}\ell)}{4}}
	\prod_{i=1}^Lx_i^{\ell_i}
\end{equation}
and ``v.p.'' again means that we take principle value integral (\ie~take half-sum of contours $|z_v|=1\pm \epsilon$). Such prescription corresponds to regularization by $\omega$ made in (\ref{F-omega}).


\section{Cabling in $M_3$ versus color in ${T[M_3]}$}
\label{app:cabling}

Here, we illustrate a delicate feature of 3d/3d correspondence that shows up at the categorical level when knots are introduced in $M_3$.
(The issue goes away if one looks at the problem without knots or at the decategorified level.)
Namely, we compare 3d $\CN=2$ theory $T[M_3]$ on $D^2 \times S^1$ to quantum group invariants of $M_3$ and their categorification.
As explained in section~\ref{sec:knots} and illustrated in Figure~\ref{fig:impurity}, introducing knots and links in $M_3$
corresponds to adding a line operator (``impurity'') of a suitable type in $T[M_3]$.

A curious aspect of this correspondence is that the representation $\lambda$ (``color'') of the line operator in $T[M_3]$
most naturally maps to the cabling data of the original knot $K$ in $M_3$.
At the decategorified level, that data would be the same as the color of $K$ itself, but in the homological world the story is more interesting.

\begin{figure}[ht]
\centering
\includegraphics[scale=0.5]{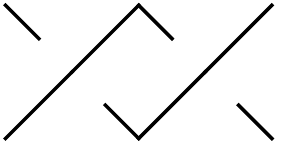}\qquad
\includegraphics[scale=0.5]{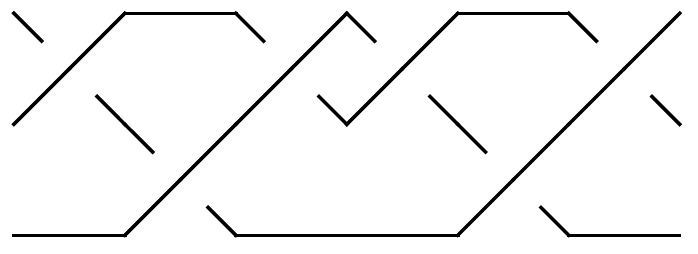}\qquad
\includegraphics[scale=0.5]{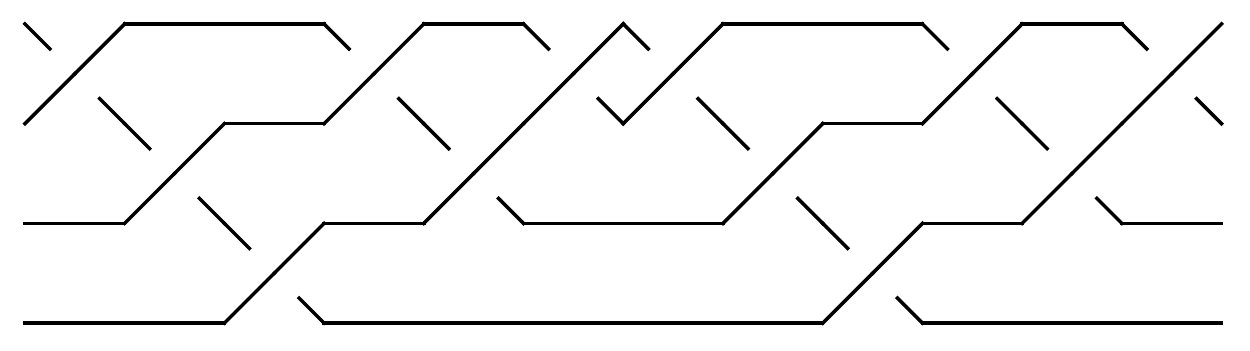}
\caption{Braid representations of 2-, 3- and 4-cables of unknot with framing 1.}
\label{fig:cables}
\end{figure}

Since this phenomenon can be seen already in the basic case of the unknot, we illustrate it with the simplest choice
of $G=SU(2)$, $M_3 = S^3$ and $K = \text{unknot}$ with framing 1.
Its colored Jones polynomials can be computed via the standard cabling formula
that isolates an $n$-dimensional irreducible representation in the tensor product of 2-dimensional fundamental representations.
As in \cite{khovanov2005categorifications},
applying the same formula to the ordinary Khovanov homology of the $n$-cabling $K^n$ (illustrated in Figure~\ref{fig:cables}):
\begin{equation}
	 \begin{array}{rl}
	 \CP_2(\fq,t;K)=&\CP_{1}(\fq,t;K^2)-1 \\
	 \CP_3(\fq,t;K)=&\CP_{1}(\fq,t;K^3)-2\CP_1(\fq,t;K) \\
	 \CP_4(\fq,t;K)=&\CP_{1}(\fq,t;K^4)-3\CP_{1}(\fq,t;K^2)+1 \\
	 &\ldots
	 \end{array}
\label{cablingfla}
\end{equation}
we find\footnote{using \texttt{KnotTheory} package for Mathematica}
\begin{equation}
	 \begin{array}{rl}
     \CP_1(\fq,-t;K)=&\mathfrak{q}^{5/2}+\sqrt{\mathfrak{q}} \\
	 \CP_2(\fq,-t;K)=&\mathfrak{q}^2+\mathfrak{q}^6 t^2+\mathfrak{q}^4 t^2  \\ \CP_3(\fq,-t;K)=&\mathfrak{q}^{5/2}+\mathfrak{q}^{9/2}+\mathfrak{q}^{17/2} t^4+\mathfrak{q}^{21/2} t^4+\mathfrak{q}^{13/2} t^2+\mathfrak{q}^{5/2} (-t) \\
	 \CP_4(\fq,-t;K)=&\mathfrak{q}^{16} t^8+\mathfrak{q}^{14} t^8+\mathfrak{q}^{12} t^6+\mathfrak{q}^{10} t^4+\mathfrak{q}^4 \left(1-t^3\right)+\mathfrak{q}^6 \left(t^2-t^3\right)+\mathfrak{q}^8 \left(-t^5+t^4+t^2\right) \\
	 &\ldots
	 \end{array}
	 \label{cabled-unknot-Kh}
\end{equation}
where we changed $t\rightarrow -t$ in order to show explicitly the $\Z_2$ grading (by fermion number) via $\pm$ signs.
Note that the choice of framing played an important role in this calculation.
In particular, if it was trivial, the result would be very different (even at the level of total dimensions)
because the cables would be just disconnected unions of unknots, instead of non-trivial links shown in Figure~\ref{fig:cables}.

Now, let us compare \eqref{cabled-unknot-Kh} with the expectation values (\ref{Z-hat-ref-knot}) of the Wilson line operators
in 3d $\CN=2$ theory $T[S^3]$:
\begin{equation}
 	 \begin{array}{rl}
 	 \langle W_1\rangle=&t^{-1/4}(\mathfrak{q}^{5/2}+\sqrt{\mathfrak{q}}) \\
 	 \langle W_2\rangle=&\mathfrak{q}^2+\frac{\mathfrak{q}^6}{t}+\frac{\mathfrak{q}^4}{t} \\
 	 \langle W_3\rangle=&\mathfrak{q}^{5/2}+\mathfrak{q}^{9/2}+\frac{\mathfrak{q}^{17/2}}{t^2}+\frac{\mathfrak{q}^{21/2}}{t^2}-\frac{\mathfrak{q}^{5/2}}{t}+\frac{\mathfrak{q}^{13/2}}{t} \\
 	 \langle W_4\rangle=& \frac{\mathfrak{q}^{16}}{t^4}+\frac{\mathfrak{q}^{14}}{t^4}+\frac{\mathfrak{q}^{12}}{t^3}+\frac{\mathfrak{q}^{10}}{t^2}+\mathfrak{q}^6 \left(\frac{1}{t}-\frac{1}{t^2}\right)+\mathfrak{q}^4 \left(1-\frac{1}{t^2}\right)+\mathfrak{q}^8 \left(-\frac{1}{t^3}+\frac{1}{t^2}+\frac{1}{t}\right)\\
 	 &\ldots
 	 \end{array}
	 \label{cabled-unknot-Wilson}
 \end{equation}
where, as in section~\ref{sec:Roz}, we made the replacement $q=\fq^2/t$.
We find perfect agreement at the level of $\Z\times \Z_2$ graded spaces,
where the first factor is the $\fq$-grading and the second factor is the fermion number grading.
The $t$-gradings do not exactly agree (although they are somewhat correlated).
This is a standard feature of the relation between colored knot homology
and homology of cablings \cite{khovanov2005categorifications,Gukov:2005qp}.

The comparison between (\ref{cabled-unknot-Kh}) and (\ref{cabled-unknot-Wilson}) clearly shows that the representation (``color'')
of the Wilson line operator in $T[S^3]$ appears to map to the cabling data of the original unknot in $M_3 = S^3$.
Note, {\it a priori} neither (\ref{cabled-unknot-Kh}) nor (\ref{cabled-unknot-Wilson}) have any direct relation
to the homology of the unknot colored by representation $\lambda$.
Indeed, $\lambda$-colored homology of the unknot \cite{Gukov:2011ry,Gorsky:2013jxa}
should be obtained by using the categorified Jones-Wenzl projectors \cite{MR2901969,MR2826934}
instead of the naive cabling relations \eqref{cablingfla}.

Similarly, (\ref{cabled-unknot-Wilson}) may miss the mark for two reasons.
First, the variable $t$ that appears in (\ref{cabled-unknot-Wilson}) keeps track of the flavor symmetry grading,
not the homological (R-symmetry) grading. This issue is not a serious objection, especially in view of
the ``homological-flavor locking'' that allowed to identify these two gradings in many examples in this paper.
A more serious issue with (\ref{cabled-unknot-Wilson}) has to do with a non-trivial map of line operators in 3d/3d correspondence.
In particular, the $n$-th cabling\footnote{more precisely, the linear combination of cablings up to $n$-th, given by \eqref{cablingfla}}
of a knot $K$ and its $n$-colored version may be indistinguishable at the decategorified level, but they are completely
different operators in the physical setup \eqref{CatGeom2} and, in general, have different homological invariants (spaces of BPS states).


\section{Categorification of the Turaev-Viro invariants}
\label{sec:TV}

The aim of this paper is to introduce and study new three-manifold invariants. Along the way, we have seen the important roles played by boundary conditions, abelian flat connections, homological blocks, refinement, and line operators. Here, we outline how these ideas can be applied to categorification of the Turaev-Viro invariants.

Defined via a state-sum model, the Turaev-Viro (TV) invariants \cite{TURAEV1992865} involve a ``counting'' problem from the start.
This motivates a series of questions: do TV invariants of $M_3$ have a natural home in physics? Could it be that they are counting BPS states associated with the theory $T[M_3]$ and can be categorified in a way similar to the WRT invariant? How are they related to the new invariants discussed in this paper? Can they also be decomposed into the ``atoms'' of three-manifold invariants --- the homological blocks and abelian connections? Can they be refined?

{}From the perspective of the state-sum model, categorification of the Turaev-Viro invariants may potentially seem even more natural than that of Chern-Simons (WRT) invariants, though the two problems are closely connected. Thus, the TV invariant of a closed 3-manifold $M_3$ is simply the square of the corresponding $SU(2)$ WRT/CS invariants \cite{turaev1992shadow,walker1991witten,turaev1994quantum}:
\be
\mathrm{TV}(M_3, q) = \left|Z^{\text{CS}}(M_3, q)\right|^2.
\ee
As explained in detail in this paper, the WRT invariants can be written as a linear sum of homological blocks and,
when there is only one block,
\be
\mathrm{TV}(M_3, q) = \hat{Z}_0(M_3,q)\hat{Z}_0(M_3,q^{-1})=\CI_{T[M_3]}(q)
\ee
is exactly given by the index of the theory $T[M_3]$. So, for homological spheres, we have already categorified their TV invariants!

For general $M_3$, the story will be more interesting.\footnote{To streamline the presentation, we will ignore normalization factors, Weyl group actions, \textit{etc}.} If we use $V_B$ to denote the vector space of boundary conditions on $S^1$ of the theory $T[M_3]$ (a de-categorification of $\CC_B$), then there are two natural bases $\{|\hat{a}\rangle\}$ and $\{|a\rangle\}$ corresponding to the homological blocks and (connected components of) abelian flat connections. There are several distinct elements in $V_B$, one is $|0\rangle$ --- the boundary condition at the origin of $D^2$, and another is $|\text{CS}\rangle$ --- the boundary condition at $\partial D^2$ that is used to reproduce the WRT invariant. Partition function of the theory $T[M_3]$ on $S^1\times I\times S^1$ defines the inner product on $V_B$ and a map $V_B\rightarrow V_B^*$. Then we have\footnote{To obtain homological invariants, one just needs to decompactify the time $S^1$ circle, and $V_B$ will be come the category $\CC_B$ that the 3d theory $T[M_3]$ associates to $S^1$. Then the homological invariants are identified with $\mathrm{Hom}(B_1,B_2)$ between two boundary conditions $B_1$, $B_2$ in $\CC_B$.}
\bea
Z^{\text{CS}}(q,t)=&\langle 0 | \text{CS} \rangle,\\
Z_a(q,t) =&\langle 0 | a \rangle,\\
\hat{Z}_a(q,t) =&\langle 0 | \hat{a} \rangle,\\
\CI(q,t) =&\langle 0|0 \rangle,
\eea
and also, in the unrefined limit,
\be
\mathrm{TV}(q)=Z^{\text{CS}}(q)Z^{\text{CS}}(q^{-1})=\langle 0 | \text{CS}\rangle \langle \text{CS}|0\rangle.
\ee
Using
\be
| \text{CS}\rangle=\sum_a e^{2\pi ik\text{CS}(a)}|a \rangle,
\ee
we have
\be
| \text{CS}\rangle \langle \text{CS}|=\sum_{a,b}e^{2\pi ik\left(\text{CS}(a)-\text{CS}(b)\right)}|a \rangle\langle b|=\sum_{j}\CO_j,
\ee
where
\be
\CO_j=\sum_{a}\tilde{q}^{\,\lk(j,2a-j)}|a-j \rangle\langle a|,
\ee
is the shift operator that acts on $Z_a$ by
\be
Z_a\mapsto \tilde{q}^{\,\lk(j,2a-j)} Z_{a-j}
\ee
with
\be
\tilde{q}=e^{-2\pi i k}
\ee
being the modular transform of $q$. The shift operators are the S-dual of Wilson loops, whose action on the homological block for $M_3=L(k,1)$, as we have seen in section~\ref{sec:LensWilson}, is
\be
\hat{Z}_a\mapsto {q}^{j(2a-j)} \hat{Z}_{a-j}.
\ee

The S-duality of type IIB string theory becomes 3d mirror symmetry of $T[M_3]$ that exchanges Wilson loops and vortex loops (see \eg~\cite{Assel:2015oxa} for a detailed description of the map in a very large class of 3d $\CN=4$ theories), and one expects that $\CO_j$ corresponds to the insertion of a vortex line. Similar to a Wilson loop, which carries an electric charge, a vortex loop carries a magnetic flux, creating a holonomy along its meridian --- no surprise that it will shift $a$ which is characterized by holonomies. The relation between vortex loops and abelian flat connections can be understood more precisely using the M-theory geometry, as we now explain.

Vortex loops come from codimension-two defects in 6d (2,0) theory and can be engineered by a stacks of ``defect M5-branes'' as follows:
\beq
\begin{matrix}
{\mbox {\textrm{M5-branes:}}}~~~\qquad & \R & \times & M_3& \times & D^2  \\ \qquad & &
&   \cap &  & \cap \\
{\mbox{\rm space-time:}}~~~\qquad & \R&  \times  & T^*M_3 & \times & TN   \\ \qquad & &
&   \cup &  & \cup \\
{\mbox{\rm ``defect" M5$'$-branes:}}~~~\qquad &\R & \times & M_3 & \times  & T^*|_O   \\
\end{matrix},
\label{CatGeom3}
\eeq
where $T^*|_O$ is the cotangent space at $O$. The defect fivebranes have a flat background 2-form connection (``gerbe connection'') on its world volume. The vortex lines will carry fluxes, labeled by elements in $H_1(M_3,\Z)^*$.
Analogous to the BPS state counting analysis with M2-branes, the non-torsion part of $H_1(M_3,\Z)^*$ results in $Q$-exact deformations of the system, enabling us to consider only the torsion part $(\mathrm{Tor}\,H_1(M_3,\Z))^*$.
Taking into account multiple fivebranes gives $(\mathrm{Tor}\,H_1(M_3,\Lambda_{\text{wt},\frak{g}}))^*/W_G$.
In other words, non-trivial BPS vortex loop are labeled by components of abelian flat connections.
For M2-branes propagating in this system \eqref{CatGeom3}, this generates the above mentioned shift action of vortex loops on flat connections.

We have now identified the system relevant for the Turaev-Viro invariant: $T[M_3]$ on $\R\times S^2$ with the insertion of a exotic defect, given by a linear combination of vortex loops,\footnote{This defect can be viewed as the S-dual of a Wilson loop labeled by a reducible representation, but it will be interesting to find an alternative way to characterize this defect.} whose BPS Hilbert space should categorify the TV invariant. This Hilbert space will be doubly-graded for generic $M_3$ and at least triply-graded for Seifert $M_3$. In the latter case, we will have refinement for both TV invariants and the ``TV homologies.'' It would be interesting to test this proposal in concrete examples, which we leave for future work.


\bibliographystyle{JHEP}
\bibliography{3Man}

\end{document}